\documentclass[prb,showpacs,preprintnumbers,amsmath,amssymb]{revtex4}
\usepackage{graphicx}% Include figure files
\usepackage{dcolumn}% Align table columns on decimal point
\usepackage{bm}% bold math
\usepackage{subfigure}

\begin{document}
\title{Chiral skyrmion states in non-centrosymmetric magnets \label{BHPhase}}

\author{A. O. Leonov}
\thanks
{Corresponding author} 
\email{leon-off@mail.ru}

\affiliation{Zernike Institute for Advanced Materials, University of Groningen, Groningen, 9700AB, The Netherlands}

% %%% PACS numbers

%\keywords{surface-induced anisotropy, ferromagnetic nanowires, nanotubes, vortex states of the magnetization}%Use showkeys class option if keyword
                              %display desired
         
\maketitle

%%%%%%%%%%%%%%%%%%%%%%%%%%%%%%%%%%%%%%%%%%%%%%%%%%%%%%%%%%%%%%%%%%%%%%%%%%%%%%%%%%%%%%%%%%%%%%%%%%%%%%%%%%%%%%%%%%%%%%%
\section{Introduction \label{BHIntro}}
%%%%%%%%%%%%%%%%%%%%%%%%%%%%%%%%%%%%%%%%%%%%%%%%%%%%%%%%%%%%%%%%%%%%%%%%%%%%%%%%%%%%%%%%%%%%%%%%%%%%%%%%%%%%%%%%%%%%%%%

%\textbf{As stated in the manuscript and again in our previous response and tutorial attachements, we study in present manuscript the low-temperature phenomenological theory with fixed modulus of magnetization, i.e. we are using a micromagnetic approximation, |M|=const, to the magnetic states in chiral magnets.  In [Rossler06] and in Ref.[Muhlbauer09] and others studies, Landau-Ginzburg functionals are used that lift this hard constrained on the modulus of magnetization and include a Landau term. Thus the modulus of magnetization becomes soft. As already shown in Ref. [Rossler06] this leads to important inhomogeneities of the magnetization that can be beneficial to the stabilization of Skyrmionic condensed states, and a modified gradient model yields thermodynamic stabilization of the Skyrmion states owing to the coupling of the angular/directional degrees of freedom and the longitudinal primary order parameter of the magnetization distribution. In present manuscript, we are studying the low-temperature approximation and not these precursor states. }

%\textbf{The quest for Skyrmionic  states in chiral magnets is currently a very topical subject and attracts great attention and controversies among  specialists working on chiral bulk systems and their colleagues from nanomagnetic community.}

Chiral skyrmion states exist in non - centrosymmetric magnetic crystals \cite{JETP89,JMMM94,Nature06,pss94} as a consequence
of the asymmetric exchange Dzyaloshinskii-Moriya interactions that destroy the homogeneous magnetic state and generally lead to twisted incommensurate magnetic spin-structures \cite{Dz64}).
%
%These couplings can be induced also by the broken inversion  symmetry at surfaces of  magnetic nanolayers \cite{Crepieux98,PRL01,Bode07,Honolka08}.
%
%The chiral skyrmions in such systems are particle-like multiply twisted states with radial symmetry and  a fixed physical size that result from the competition  between direct and Dzyaloshinskii-Moriya  exchange.

Recently, microscopic observation of skyrmion lattices and 
free skyrmions in magnetic layers of the chiral helimagnets 
with the noncentrosymmetric cubic B20 crystal structure 
confirm the existence of these chiral topological 
spin-textures in otherwise rather simple magnetic metals
\cite{Yu10,Yu10a}.
The  stabilization of these states and 
their transformation properties impressively illustrate
the theoretically predicted solitonic nature of these 
chiral two-dimensionally localized spin-states 
\cite{JETP89,JMMM94,Nature06,pss94}.
In particular, the experiments clearly show
the ability of skyrmions to form densely 
packed two-dimensional arrangements
and how the field-driven transformation process can 
decompose these lattices by setting free 
the constituent  skyrmions 
as excitations.
The stabilization of such skyrmion lattices against
one-dimensionally modulated helices in these cubic helimagnets
at low temperatures requires
a subtle effect possibly combining uniaxial 
magnetic anisotropy (or cubic and exchange anisotropy) with the magnetic field
[XI,XII].
However, both in magnetic films and in magnetic crystals,
symmetry imposed restrictions on the chiral Dzyaloshinskii-Moriya interactions 
may allow to create skyrmion lattices with high perfection 
in applied fields because the competing conical helix state
does not exist \cite{JETP89,JMMM94}.

This chapter is mainly devoted to numerically rigorous
solutions of hexagonal skyrmion lattices  for cubic helimagnets.
It justifies and extends previous approximate solutions
that used a circular cell approximation (CCA) for the calculation 
of the free energy of skyrmion lattices \cite{JETP89,JMMM94}[XI].
%
%In the present chapter we resort to numerical techniques to obtain rigorous skyrmion solutions.
 The theoretical results of the present chapter provide a comprehensive description of skyrmion
lattice evolution in an applied magnetic field and/or in the presence of uniaxial, cubic, and exchange anisotropy. %It is shown that skyrmion textures are composed of countable localized magnetic units that can be set free and manipulated  by tuning the competition between direct and chiral Dzyaloshinskii-Moriya exchange.
%skyrmion lattices are stable with respect to one-dimensional helices in applied fields for crystals from Laue classes D$_{2d}$ and C$_{nv}$ in the approximation of constant magnetization length, i.e., far from the ordering temperatures.
%
The low-temperature phenomenological theory with fixed modulus of magnetization, M=const, is applied to the magnetic states in chiral magnets. 
%èäåÿ ñíà÷àëà ââåñòè âçàèìîäåéñòâèÿ ÄÌ.

%\textit{DM interactions.}

%à) Ñíà÷àëà ðàññêàçàòü, ÷òî èçíà÷àëüíî ýòî âçàèìîäåéñòâèå áûëî ââåäåíî, ÷òîáû îáúÿñíèòü ñëàáûé ôåððèìàãíåòèçì â ÀÔÌ.

%In the present chapter  I study  the low-temperature phenomenological theory with fixed modulus of magnetization, i.e. I use a micromagnetic approximation, |M|=const, to the magnetic states in chiral magnets.  In chapter \ref{confinement} Landau-Ginzburg functionals are used that lift this hard constraint on the modulus of magnetization and include a Landau term. Thus the modulus of magnetization becomes soft. As already shown in Ref.\cite{Nature06}[XII,XIV,XV] this leads to important inhomogeneities of the magnetization that can be beneficial to the stabilization of Skyrmionic condensed states, and a modified gradient model yields thermodynamic stabilization of the Skyrmion states owing to the coupling of the angular/directional degrees of freedom and the longitudinal primary order parameter of the magnetization distribution. 
%
%In present chapter, however, I study the low-temperature approximation and not these precursor states. 

%%%%%%%%%%%%%%%%%%%%%%%%%%%%%%%%%%%%%%%%%%%%%%%%%%%%%%%%%%%%%%%%%%%%%%%%%%%%%%%%%%%%%%%%%%%%%%%%%%%%%%%%%%%%%%%%%%%%%%%%%%%%
\section{Phenomenological theory of modulated states in chiral helimagnets \label{PTBH}}

\label{PT1}
%%%%%%%%%%%%%%%%%%%%%%%%%%%%%%%%%%%%%%%%%%%%%%%%%%%%%%%%%%%%%%%%%%%%%%%%%%%%%%%%%%%%%%%%%%%%%%%%%%%%%%%%%%%%%%%%%%%%%%%%%%%%

\subsection{Dzyaloshinsky-Moriya  interaction}

\vspace{3mm}

In many magnetic crystals the magnetic properties may be strongly influenced by the asymmetric exchange interaction known also as the Dzyaloshinsky-Moriya  interaction (DMI). When acting between two spins $\mathbf{S}_i$ and $\mathbf{S}_j$, it leads to a term in the Hamiltonian which is generally described by a spin vector product:
\begin{equation}
H_{\mathrm{DM}}=\mathbf{D}_{ij}\cdot(\mathbf{S}_i\times\mathbf{S}_j).
\label{DMvector}
\end{equation} 
where $\mathbf{D}_{ij}$ is the Dzyaloshinskii vector. 

Dzyaloshinskii-Moriya interactions arise in certain groups of magnetic crystals with low symmetry where the effects of coupling (\ref{DMvector}) do not cancel. Their effect is  to cant (i.e. slightly rotate) the spins by a small angle. In general, $\mathbf{D}_{ij}$ may not vanish even in centrosymmetric crystalls. 
% and lies parallel or perpendicular to the line connecting the two spins, depending on the symmetry. 
Anisotropic exchange interaction occurs commonly  in antiferromagnets and then results in a small ferromagnetic component of the moments which is produced perpendicularly to the spin-axis of the antiferromagnet. The effect is known as \textit{weak ferromagnetism}. It is found, for example, in $\alpha$-Fe$_2$O$_3$, MnCO$_3$, and CoCO$_3$. To explain the phenomenon of weak ferromagnetism the interaction (\ref{DMvector}) was phenomenologically introduced by Dzyaloshinskii \cite{Dz57}. %based on the phenomenological considerations. 
Moriya found a microscopic mechanism due to the spin-orbit coupling responsible for the interactions (\ref{DMvector}) \cite{Moriya60}. %The spin-orbit interactions play a role in a similar manner to that of the oxygen atom in superexchange \cite{Blundell}.

Another fundamental macroscopic manifestation of the antisymmetric coupling (Eq. \ref{DMvector}) takes place in non-centrosymmetric magnetic crystals. Dzyaloshinskii showed that in this case the interaction (\ref{DMvector}) stabilizes long-periodic spatially modulated structures with fixed sense of rotation of the vectors $\mathbf{S}_i$. Within a continuum approximation for magnetic properties, the interactions responsible for these modulations are expressed by inhomogeneous invariants. One calls these contributions to the free magnetic energy, involving first derivatives of magnetization or staggered magnetization with respect to spatial coordinates, inhomogeneous Dzyaloshinskii-Moriya interactions. They are linear with respect to the first spatial derivatives of a magnetization $\mathbf{M}$ of type \cite{Dz64}
\begin{eqnarray}
\mathcal{L}_{ij}^{(k)} = M_i\left(\frac{\partial M_j}{\partial x_k}\right)
-M_j\left(\frac{\partial M_i}{\partial x_k}\right).
\label{LI}
\end{eqnarray}
where $M_i$ and $M_j$ are components of magnetization vectors that arise in certain combinations in expressions (\ref{LI}) depending on crystal symmetry, and $x_k$ are spatial coordinates. Such antisymmetric mathematical forms were studied in the theory of phase transitions by E. M. Lifshitz and are known as \textit{Lifshitz invariants} \cite{LandauLifshitz}.

Depending on the crystal symmetry \cite{Dz64,JETP89}, the Dzyaloshinskii-Moriya energy $W_{\mathrm{DM}}(\mathbf{M})$ includes certain combinations of Lifshitz invariants (\ref{LI}).
%
%Particularly, for important uniaxial crystallographic classes, $n$mm ($C_{nv}$), $\bar{4}$2m ($D_{2d}$), and $n$22 ($D_{n}$) DM energy contributions have the following form \cite{JETP89}:
Particularly, for important uniaxial crystallographic classes, $(\mathit{nmm})$($C_{nv}$), $\overline{4}2m$($D_{2d}$), and $n22$ ($D_n$) functional $W_{\mathrm{DM}}$ can be written as 
\begin{equation}
%(nmm): W_D=D\,(\mathcal{L}_{xz}^{(x)}+\mathcal{L}_{yz}^{(y)}),
(nmm): W_{\mathrm{DM}}=\int w_{\mathrm{DM}}dV=\int[D\,(\mathcal{L}_{xz}^{(x)}+\mathcal{L}_{yz}^{(y)})]dV,
\label{Cnv}
\end{equation}
\begin{equation}
%(\overline{4}2m): W_D=D\,(\mathcal{L}_{xz}^{(y)}+\mathcal{L}_{yz}^{(x)}),
(\overline{4}2m): W_{\mathrm{DM}}=\int[D\,(\mathcal{L}_{xz}^{(y)}+\mathcal{L}_{yz}^{(x)})]dV,
\label{D2d}
\end{equation}
\begin{equation}
%(n22): W_D=D_1\,(\mathcal{L}_{xz}^{(y)}-\mathcal{L}_{yz}^{(x)})+D_2\mathcal{L}_{xy}^{(z)}.
(n22): W_{\mathrm{DM}}=\int[D_1\,(\mathcal{L}_{xz}^{(y)}-\mathcal{L}_{yz}^{(x)})+D_2\mathcal{L}_{xy}^{(z)}]dV.
\label{Dn}
\end{equation}
where $n=3,4,6$, and $D_1$, $D_2$, $D$ are Dzyaloshinskii constants.

Lifshitz invariants for n ($C_n$) and $\overline{4}$ ($S_4$) classes consist of terms with simultaneous presence of two Dzyaloshinskii constants related to directions $x,\,y$ in the basal plane:
\begin{equation}
(n): W_{\mathrm{DM}}=\int[D_3\,(\mathcal{L}_{xz}^{(x)}+\mathcal{L}_{yz}^{(y)})+D_4\,(\mathcal{L}_{xz}^{(y)}-\mathcal{L}_{yz}^{(x)})]dV,
\label{Cn}
\end{equation}
\begin{align}
%(\overline{4}): \quad  w_D=D_5\,(\mathcal{L}_{xz}^{(x)}-\mathcal{L}_{yz}^{(y)}) +
%D_6\,(\mathcal{L}_{xz}^{(y)}+\mathcal{L}_{yz}^{(x)}).
(\overline{4}):  W_{\mathrm{DM}}=\int[D_5\,(\mathcal{L}_{xz}^{(x)}-\mathcal{L}_{yz}^{(y)}) +
D_6\,(\mathcal{L}_{xz}^{(y)}+\mathcal{L}_{yz}^{(x)})]dV.
\label{S4}
\end{align}
For cubic helimagnets belonging to 23 (T) (as MnSi, FeGe, and other B20 compounds) and 432 (O) crystallographic classes Dzyaloshinskii-Moriya interactions are reduced to the following form:
\vspace{3mm}
\begin{equation}
%W_D=D\,(\mathcal{L}_{yx}^{(z)}+\mathcal{L}_{xz}^{(y)}+\mathcal{L}_{zy}^{(x)})=D\,\mathbf{M}\cdot\mathrm{rot}\mathbf{M}.
W_{\mathrm{DM}}=\int[D\,(\mathcal{L}_{yx}^{(z)}+\mathcal{L}_{xz}^{(y)}+\mathcal{L}_{zy}^{(x)})]dV=\int[D\,\mathbf{M}\cdot\mathrm{rot}\mathbf{M}]dV.
\label{LifshitzCubic}
\end{equation}

Dzyaloshinskii-Moriya interactions stabilizing chiral magnetic states may be also induced by the symmetry breaking at the surface in confined systems as magnetic nanolayers, nanowires, and nanodots. As a genuine consequence of surface-induced DM couplings different types of chiral modulations have been observed \cite{Bode07,Heinze11}.  Therefore, thin film systems are appropriate candidate structures to study chiral magnetic skyrmions.  In particular, micromagnetic analysis of the chirality selection for the vortex ground states of magnetic nanodisks shows that the sign and the strength of the DM coupling strongly influence their structures, magnetization profiles and core sizes \cite{Butenko09}. The calculated relations between strength of the DM interactions and vortex-core sizes provide a method to determine the magnitude of surface-induced DM couplings in ultrathin magnetic films/film elements.

\subsection{The general micromagnetic energy functional}

%we study in present manuscript the low-temperature phenomenological theory with fixed modulus of magnetization, i.e. we are using a micromagnetic approximation, |M|=const, to the magnetic states in chiral magnets.  (intro for CHapter 4)

Within the phenomenological theory introduced by Dzyaloshinskii \cite{Dz64} the magnetic energy density of a  non-centrosymmetric ferromagnet with spatially dependent magnetization $\mathbf{M}$ can be written as 
%
%\begin{equation}
%w=A \sum_{i,j}(\partial_i M_j)^2-\mathbf{M}\cdot\mathbf{H}-\mathbf{M}\cdot\mathbf{H^{(m)}}+w_0(\mathbf{M})+w_D(\mathbf{M})
%\label{DMdens1}
%\end{equation}
%
%\begin{equation}
%w=A \sum_{i,j}\left(\frac{\partial M_j}{\partial %x_i}\right)^2-\mathbf{M}\cdot\mathbf{H}-\frac{1}{2}\mathbf{M}\cdot\mathbf{H}_d+w_0(\mathbf{M})+w_D(\mathbf{M}),
%\label{DMdens1}
%\end{equation}

\begin{equation}
%W(\mathbf{M})=\underbrace{A \sum_{i,j}\left(\frac{\partial m_j}{\partial x_i}\right)^2
%-D\,\mathbf{m}\cdot\mathrm{rot}\mathbf{m}
%-\mathbf{M}\cdot\mathbf{H}}_{W_0(\mathbf{M})}
%+W_a(\mathbf{m})
W(\mathbf{M})=\underbrace{A \sum_{i,j}\left(\frac{\partial m_j}{\partial x_i}\right)^2
+D\,w_D(\mathbf{M})
-\mathbf{M}\cdot\mathbf{H}}_{W_0(\mathbf{M})}
+W_a(\mathbf{m})
\label{DMdens1}
\end{equation}
%
%\begin{equation}
%w=\underbrace{A  \left( \mathbf{grad} \mathbf{M} \right)^2}_{\partial^2}
%+\underbrace{w_D (\mathbf{M})}_{\partial^1}
%+\underbrace{w_0 (\mathbf{M})}_{\partial^0}\label{density}
%\end{equation}
%
where $A>0$ and $D$ are coefficients  of exchange and Dzyaloshinskii-Moriya interactions; $\mathbf{H}$ is an  applied magnetic field; $x_i$ are the Cartesian components of the spatial variable. 
$w_D$ is composed of Lifshitz invariants. Almost all calculations of the present chapter have been done for cubic helimagnets with $w_D=\mathbf{m}\cdot\mathrm{rot}\mathbf{m}$.

$W_a(\mathbf{m})$ includes short-range anisotropic energies: 
%
%\begin{equation}
%w_a(\mathbf{m})=-\sum_{i=1}^{3}\left[B  \left(\frac{\partial M_i}{\partial x_i} \right)^2
%+K_c M_i^4\right]-K_u M_3^2 
%\end{equation}
%
\begin{equation}
W_a(\mathbf{m})=-\sum_{i=1}^{3}\left[B_{EA}  \left(\frac{\partial m_i}{\partial x_i} \right)^2
+K_c (\mathbf{m}\cdot\mathbf{n}_i)^4\right]-K_u (\mathbf{m}\cdot\mathbf{a})^2 
\label{additional}
\end{equation}
where $B_{EA}$,  $K_c$, and $K_u$ are coefficients of exchange, cubic, and uniaxial magnetic anisotropies, correspondingly;
 $ \mathbf{a}$ and $\mathbf{n}_i$ are unit vectors along easy uniaxial and cubic magnetizaton axes, respectively.

%\textit{words about isotropic model.}
%In the next section we present 

%Functional $W_0(\mathbf{M})$ in (\ref{DMdens1}) includes only primary interactions \textit{essensial} essential to stabilize skyrmionic and helical phases.   % and attributes to the solutions for chiral modulations in all chiral ferromagnets. 
%
%On the other hand, weaker energy contributions $w_a(\mathbf{m})$ impose distortions on the solutions and determine thermodynamical stability of individual phases and conditions of phase transformations between them.
%

Functional $W_0(\mathbf{M})$  includes only basic interactions essential to stabilize skyrmion and helical states. Solutions for chiral modulated phases and their most general features attributed to all chiral ferromagnets are determined by this functional. Generically, there are only small energy differences between various modulated states. On the other hand, weaker energy contributions (as magnetic anisotropies (\ref{additional})) impose distortions on solutions of model (\ref{DMdens1}) which reflect crystallographic symmetry and values of magnetic interactions in individual chiral magnets. It is essential to recognize that these weaker interactions determine the stability limits of the different modulated states. The fact that thermodynamical stability of  individual phases and conditions of phase transfomations between them are determined by magnetocrystalline anisotropy and other relativistic or weaker interactions  means  that (i) the basic theory only determines a set  of different and unusual modulated phases, while (ii) the transitions between these modulated states,  and their evolution in magnetization processes depends on symmetry and details of magnetic secondary effects in chiral magnets, in particular the strengths of relativistic magnetic interactions. Thus functional (\ref{DMdens1}) is  the \textit{generic} model for a manifold
of interaction functionals describing different groups of noncentrosymmetric magnetic crystals, because it allows to identify the basic modulated structures that may be found in them.

Dzyaloshinskii's phenomenology (\ref{DMdens1}) is a main theoretical tool to analyze and interprete experimental results on chiral magnets. During last three decades of intensive investigations of chiral modulations in different classes of non-centrosymmetric magnetic systems a huge empirical material has been organized and systematized within the framework of this theory (see, for example, a review \cite{Izyumov84} and bibliography in papers \cite{Bogdanov02k,Nature06}). The Dzyaloshinskii interaction functional (\ref{DMdens1}) plays in chiral magnetism a similar role as the Frank functional in liquid crystals \cite{books} or Ginzburg-Landau functionals in physics of superconductivity \cite{Brandt95,Brandt03}.

%includes all necessary (primary) interactions essential to stabilize skyrmion and helical phases.  On the contrary, weaker energy contributions $W_a(\mathbf{m})$  determine thermodynamical stability of individual phases and conditions of phase transformations between them. 
%
%Two-dimensional solutions  for skyrmion lattices considered in the present chapter are supposed to be homogeneously extended into the third direction as skyrmion filements. This means that the influence of Lifshitz invariants with derivatives along $z$-direction is excluded from the consideration.

\subsection{Reduced variables and characteristic lengths \label{reducedUnits}}

For the forthcoming calculations I will use two ways of indroducing non-dimensional variables. 

In the first method, the length scales are reduced by the characteristic width of the Bloch domain wall. This method is valuable in the situations where anisotropic magnetic materials are considered and the influence of "tunable" DM interactions on the solutions of micromagnetic equations is investigated. %In particular in section \ref{SkMagnetostatic}, I consider the solutions for magnetic bubble domains stabilized by dipole-dipole interactions in the presence of induced DMI. 

In the second method, the lengths are expressed in units of $L_D$, i.e. the length scales are related to the period of the spiral state in zero field. Such a method is suitable for the calculations of the present chapter, as first I consider different modulated states as solutions of the isotropic energy functional $W_0(\mathbf{M})$ and then "activate" additional small anisotropic contributions $W_a(\mathbf{m})$.

\vspace{3mm}
\textit{A. Reduced variables with the length scales in units of the width of the Bloch domain wall}
\vspace{3mm}

Following Refs. \cite{JMMM99} I introduce the non-dimensional variables based on the domain wall width
\begin{equation}
L_B=\sqrt{\frac{A}{K_u}}.
\end{equation}
Then the energy functional (\ref{DMdens1}) can be written in the reduced form as
\begin{align}
w(\mathbf{m})=\sum_{i,j}\left(\frac{\partial m_j}{\partial \widetilde{x}_i}\right)^2
&-\frac{4\varkappa}{\pi}\mathbf{m}\cdot\mathrm{rot}\mathbf{m}-2\mathbf{m}\cdot\mathbf{h}-\nonumber\\
&-\sum_{i=1}^{3}\left[\frac{B_{EA}}{K_u}  \left(\frac{\partial m_i}{\partial \widetilde{x}_i} \right)^2+\frac{K_c}{K_u} (\mathbf{m}\cdot\mathbf{n}_i)^4\right]-(\mathbf{m}\cdot\mathbf{a})^2 
\end{align}
where
\begin{equation}
\mathbf{h}=\frac{\mathbf{H}}{H_a},\,\widetilde{\mathbf{r}}=\frac{\mathbf{r}}{L_B},\,w(\mathbf{m})=\frac{W(\mathbf{M})}{H_aM}
\end{equation}
and 
\begin{equation}
H_a=\frac{2K_u}{M} 
\end{equation}
is the anisotropy field.

The parameter 
\begin{equation}
\varkappa=\frac{\pi D}{4\sqrt{A\,K_u}}
\end{equation}
plays a similar role as the Ginzburg-Landau parameter in the theory of superconductivity. It describes the relative contribution
of the Dzyaloshinsky energy term. In Refs. \cite{pss94,JMMM99} it was shown that modulated structures can be realized as
thermodynamically stable states only if $\varkappa$ exceeds the value of 1.

\vspace{3mm}
\textit{B. Reduced variables with the length scales in units of $L_D$}
\vspace{3mm}

Following Refs. \cite{JMMM94} I introduce the non-dimensional variables based on the period of the helical state in zero magnetic field. Then the energy functional (\ref{DMdens1}) can be written in the reduced form as
\begin{equation}
w(\mathbf{m})=\sum_{i,j}\left(\frac{\partial m_j}{\partial \widetilde{x}_i}\right)^2-\mathbf{m}\cdot\mathrm{rot}\mathbf{m}-\mathbf{m}\cdot\mathbf{h}-\sum_{i=1}^{3}\left[b_{EA}  \left(\frac{\partial m_i}{\partial \widetilde{x}_i} \right)^2+k_c (\mathbf{m}\cdot\mathbf{n}_i)^4\right]-\beta_u(\mathbf{m}\cdot\mathbf{a})^2 
\end{equation}
where
\begin{equation}
\mathbf{h}=\frac{\mathbf{H}}{H_D},\,\widetilde{\mathbf{r}}=\frac{\mathbf{r}}{L_D},\,w(\mathbf{m})=\frac{W(\mathbf{M})}{H_DM}
\end{equation}
and 
\begin{equation}
H_D=\frac{D^2}{AM}.
\end{equation}

The reduced constants of exchange  $b_{EA}$, cubic $k_c$, and uniaxial $\beta_u$ anisotropies  are defined as
\begin{equation}
b_{EA}=\frac{B_{EA}A}{D^2},\,k_c=\frac{K_cA}{D^2},\,\beta_u=\frac{K_uA}{D^2}.
\end{equation}

%\vspace{3mm}
%\textit{C. The relation between two ways of introducing non-dimensional variables}
%\vspace{3mm}

%%%%%%%%%%%%%%%%%%%%%%%%%%%%%%%%%%%%%%%%%%%%%%%%%%%%%%%%%%%%%%%%%%%%%%%%%%%%%%%%%%%%%%%%%%%%%%%%%%%%%%%%%%%%%%%%%%%%%%%%%%%
%\section{Dzyaloshinskii theory of modulated states for cubic helimagnets}
%%%%%%%%%%%%%%%%%%%%%%%%%%%%%%%%%%%%%%%%%%%%%%%%%%%%%%%%%%%%%%%%%%%%%%%%%%%%%%%%%%%%%%%%%%%%%%%%%%%%%%%%%%%%%%%%%%%%%%%%%%%

%%%%%%%%%%%%%%%%%%%%%%%%%%%%%%%%%%%%%%%%%%%%%%%%%%%%%%%%%%%%%%%%%%%%%%%%%%%%%%%%%%%%%%%%%%%%%%%%%%%%%%%%%%%%%%%%%%%%%%%%%%%
%\subsection{Helical modulations}
%%%%%%%%%%%%%%%%%%%%%%%%%%%%%%%%%%%%%%%%%%%%%%%%%%%%%%%%%%%%%%%%%%%%%%%%%%%%%%%%%%%%%%%%%%%%%%%%%%%%%%%%%%%%%%%%%%%%%%%%%%%
\begin{figure}
%\vspace{7 cm}
\centering
\includegraphics[width=18cm]{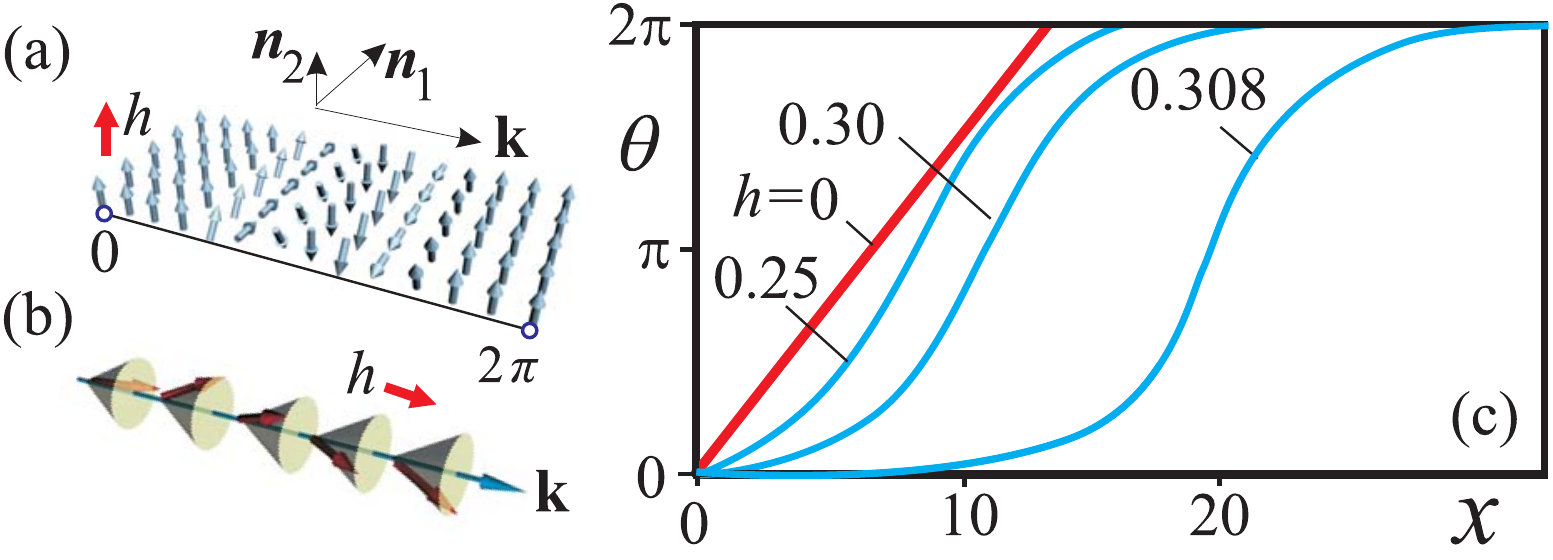}
\caption{
\label{spirals1}
One-dimensional chiral modulations in cubic helimagnets. In a helical "array" (a) the magnetization
rotates in the plane spanned by the orthogonal unity vectors $\mathbf{n}_1$ and $\mathbf{n}_2$ and the rotation sense is determined by the sign of Dzyaloshinskii constant $D$. Under the influence of the magnetic field applied perpendicularly to the propagation direction the helix is transformed into a transversally distorted  \textit{helicoid} with non-linear profiles $\theta(x)$ (c). Magnetic field applied along propagation direction stabilizes single-harmonic conical phase (b). 
}
\end{figure}

\section{One-dimensional chiral modulations \label{1DBH}}

The Dzyaloshinskii-Moriya interactions arising in non-centrosymmetric magnets play a crucial role in destabilizing the homogeneous ferromagnetic arrangement and twisting it into a helix (Fig. \ref{spirals1}). At zero magnetic field such helices are single-harmonic modes forming the global minimum of the functional $W_0(\mathbf{M})$\cite{Dz64}: %induce long-range one-dimensional modulations of the magnetization  - helices \cite{Dz64}.
\begin{equation}
\mathbf{M}= M_s \left[ \mathbf{n}_1 \cos 
\left(\mathbf{k} \cdot \mathbf{r} \right)
+ \mathbf{n}_2 \sin  \left(\mathbf{k}
\cdot \mathbf{r} \right) \right], 
\quad  |\mathbf{k}| = \frac{1}{2L_D}
\label{helix0}
\end{equation}
where $\mathbf{n}_1$, $\mathbf{n}_2$  are the unit vectors in the plane of the magnetization rotation orthogonal to the wave vector $\mathbf{k}$ ($\mathbf{n}_1\perp\mathbf{n}_2; \mathbf{n}_1\perp\mathbf{k};\,\mathbf{n}_2\perp\mathbf{k}$). %$L_D$ is proportional to the ratio of the counter-acting exchange and Dzyaloshinskii constants (\ref{helix0}) and introduces a fundamental \textit{length} characterizing a magnitude of chiral modulations in non-centrosymmetric magnets.

The modulations (\ref{helix0}) have a fixed rotation sense determined by the sign of Dzyaloshinskii-Moriya constant $D$ and are continuously degenerate with respect to propagation directions in the space.

An applied magnetic field lifts the degeneracy of the helices (\ref{helix0}) and stabilizes two types of one-dimensional modulations: cones and helicoids (Fig. \ref{spirals1} (a), (b)).

\subsection{Helicoids \label{helicoidsBH}} 

If the propagation vector $\textbf{k}$ of a spiral state is perpendicular to an applied magnetic field, I will call such a state \textit{helicoid} (Fig. \ref{spirals1} (a)). 

\vspace{3mm}
\textit{A. Solutions for the polar angle $\theta$ in the helicoid}
\vspace{3mm}

Analytical solutions for the polar angle $\theta(x)$ of the magnetization written in spherical coordinates,
\begin{equation}
\mathbf{M} = M_s\left( \sin \theta(x) \cos \psi, \sin \theta(x) \sin \psi, \cos \theta(x) \right),
\end{equation}
 are derived by solving a \textit{pendulum} equation
\begin{equation}
A \frac{d^2 \theta} {d x^2} -H\cos \theta =0.
\end{equation}
Such solutions are expressed as a set of elliptical functions \cite{Dz64} and describe a gradual expansion of the helicoid period with increased magnetic field  (see the set of angular profiles $\theta(x)$ in Fig. \ref{spirals1} (c)).
In a sufficiently high magnetic field $H_H$ \cite{JMMM94} [XI] the helicoid  infinitely expands and transforms into a system of isolated non-interacting 2$\pi$-domain walls (kinks) separating domains with the magnetization along the applied field \cite{Dz64,JMMM94}. Non-dimensional value of this critical field is 
\begin{equation}
h_H=\frac{H_H}{H_D}=\frac{\pi^2}{8}=0.30843.
\label{Hhspiral}
\end{equation}
%
 %the helicoid  infinitely expands and transforms into a system of isolated non-interacting 2$\pi$-domain walls (kinks) separating domains with the magnetization along the applied field \cite{Dz64,JMMM94}.

\vspace{3mm}
\textit{B. Solutions for the azimuthal angle $\psi$ in the helicoid}
\vspace{3mm}

Distribution of the polar angle $\theta(x) $ in magnetic field is common for helimagnets of all crystallographic classes. Azimuthal angle $\psi$, on the contrary, is fixed by the different forms of the Lifshitz invariants. 
%
%Dependences of polar 
%
%Angle $\psi$ in the helical states is defined by the crystal symmetry, i.e. by the different functional form of the Lifshitz invariants (\ref{LifshitzU}-\ref{S4}). 

\begin{figure}
%\vspace{7 cm}
\centering
\includegraphics[width=18cm]{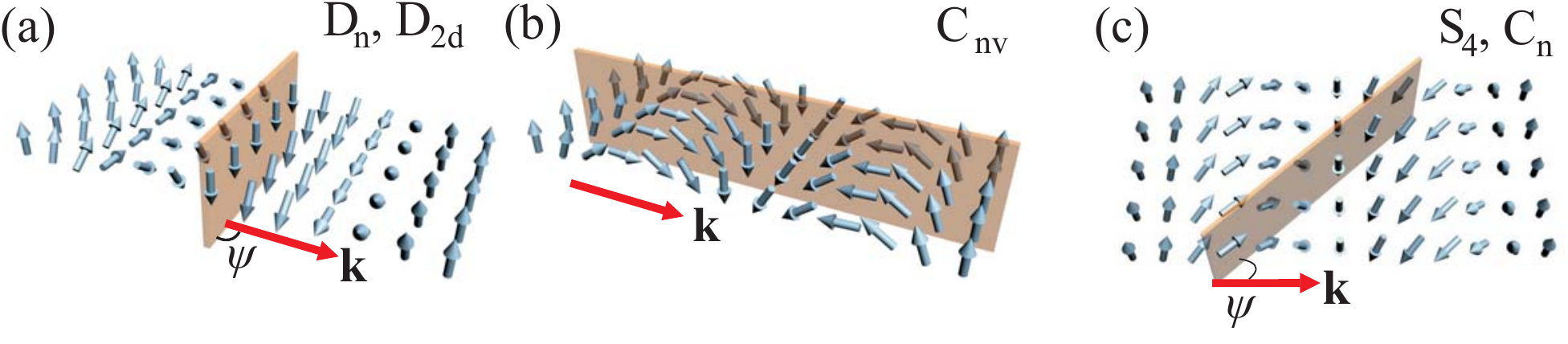}
\caption{
\label{spirals2}
Basic modulated structures: a) helicoid for systems with D$_{2d}$ and D$_{n}$ symmetry; b) cycloid for helimagnets with C$_{nv}$ symmetry. The plane of magnetization rotation (shown by red color) makes angle $\psi=\pi/2$ (a) and $\psi=0$ with the propagation direction $\mathbf{k}$. For the magnets of S$_4$ and C$_n$ crystallographic classes (c) angle $\psi$ is specified by the ratio of Dzyaloshinsky constants (see text for details). %: $\psi=\arctan(-D_{\mu}/D_{\nu})$.
}
\end{figure}

For cubic helimagnets  as well as for magnets belonging to the crystallographic classes D$_{2d}$  and D$_n$  the magnetization $\mathbf{M}$ rotates in the plane perpendicular to the propagation direction like in a common Bloch wall (Fig. \ref{spirals2} (a)), i.e. $\psi=\pi/2$.

For helimagnets of C$_{nv}$ symmetry, the magnetization vector undertakes N\'eel-type rotation along the propagation direction and  comprises  cycloid (Fig. \ref{spirals2} (b)), i.e. $\psi=0$.

For helicoids with competing DM interactions, angle $\psi$ is determined by the ratio of DM constants: $\psi=\arctan(-D_{\mu}/D_{\nu}),\,\nu=3,5;\,\mu=4,6$ (Fig. \ref{spirals2} (c)).

\subsection{Cone \label{ConeBHPhase}}

A conical spiral is a solution of the functional $W_0(\mathbf{M})$ with propagation direction along the magnetic field in which the magnetization rotation retains single-harmonic character:
\begin{equation}
\psi = \frac{z}{2L_D},\quad \cos \theta = \frac{|\mathbf{H}|}{2H_D}.
\label{cone1}
\end{equation}
In such a helix the magnetization component along the applied field has a fixed value
\begin{equation}
M_{\bot} = M \cos \theta = \frac{MH}{2H_D},
\end{equation}
and the magnetization vector $\mathbf{M}$ rotates within a cone surface. The critical value 
\begin{equation}
h_d=2H_D 
\end{equation}
marks the saturation field of the cone phase.

The conical state combines properties of the homogeneous state and the flat spiral as a compromise between Zeeman and DM energies. This conical phase is the global minimum of functional $W_0(\mathbf{M})$ (\ref{DMdens1}).

Note, that a conical spiral will propagate along direction of an applied magnetic field, if corresponding Lifshitz invariants  are present along this direction.

\section{Chiral localized skyrmions: the building blocks for skyrmionic textures \label{ISBH}}

\subsection{Equations}

\begin{figure*}[th]
\centering
\includegraphics[width=15cm]{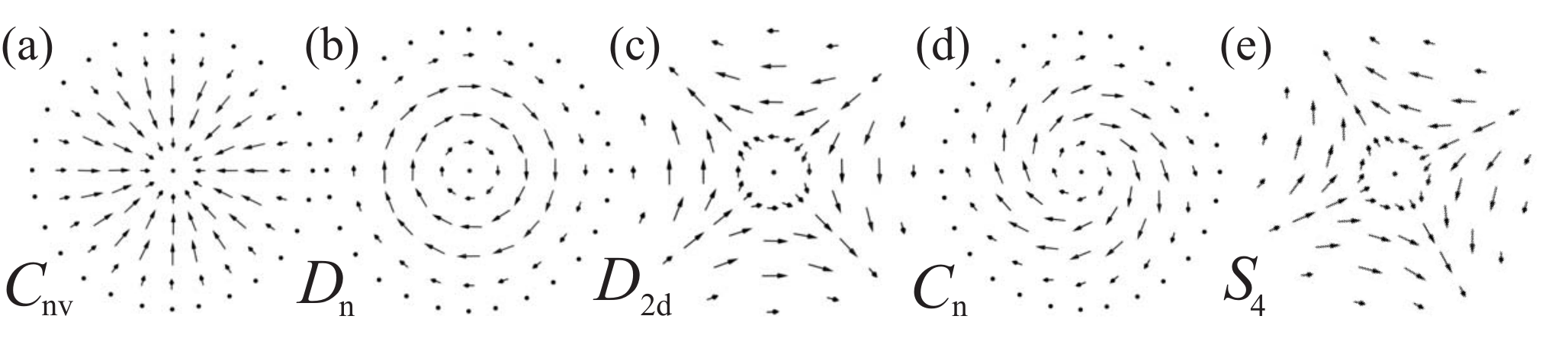}
\caption{
\label{Fig2}
Projections of the magnetization onto the basal plane for localized skyrmions of non-centrosymmetric magnets with $C_{nv}$ (a),
$D_n$ (b), $D_{2d}$ (c), $C_n$ (d), and $S_4$ (e) symmetry
}
\end{figure*}

The equations minimizing functional $W_0(\mathbf{M})$ in (\ref{DMdens1}) include solutions not only for one-dimensional helical states (section \ref{1DBH}), but also  for two-dimensional isolated skyrmions (IS) with magnetization written in spherical coordinates $\mathbf{M}= M (\sin \theta(\rho) \cos \psi(\varphi),\sin \theta(\rho) \sin \psi(\varphi), \cos \theta(\rho))$ and cylindrical coordinates used for the spatial variable $\mathbf{r}=(\rho \cos \varphi, \rho \sin \varphi,z)$.%, $\psi(\varphi)$ and $\theta(\rho)$, where spherical coordinates are used for the magnetization 
%

%\vspace{3mm}
%\textit{A. Solutions for the polar angle $\theta=\theta(\rho)$ in IS}
%\vspace{3mm}
%Dependence of the polar angle $\theta$ on the polar coordinate $\rho$ defined from (\ref{OneD}) is common for helimagnets of all crystallographic classes considered before.

The equilibrium solutions $\theta=\theta(\rho)$ for isolated Skyrmions are common for helimagnets of all crystallographic classes. The dependences $\theta=\theta(\rho)$ are derived from the Euler equation \cite{JETP89,JMMM94,pss94,JMMM99}: %for the polar angle $\theta(\rho)$  
\begin{eqnarray}
 \frac{d^2 \theta}{d \rho ^2} 
+ \frac{1}{\rho}\frac{d \theta}{d \rho}
 -\frac{\sin2 \theta}{2\rho ^2} 
 -\frac{\sin^2 \theta}{\rho} 
- \frac{h}{2}\sin \theta =0
\label{OneD}
\end{eqnarray}
with the boundary conditions 
\begin{equation}
\theta(0)=\pi,\, \theta(\infty)=0.
\label{boundaryBH}
\end{equation}
The Euler equation (\ref{OneD}) has been obtained by variation of $W_0(\mathbf{M})$. % after substituting (\ref{SphericalM}) and (\ref{Cylindrical}). 
 The non-dimensional units have been introduced in accordance with the section \ref{reducedUnits} \textit{B}.  In the case of DM interactions with competing counterparts, however, the spatial coordinates have to be normalized by 
\begin{equation}
L_D=\frac{A}{\sqrt{D_{\mu}^2+D_{\nu}^2}},\,\nu=3,5;\,\mu=4,6.
\end{equation}
Then DM energy contributions in reduced form can be parametrized by the relative ratios 
\begin{equation}
d_{\nu}=\frac{D_{\nu}}{\sqrt{D_{\nu}^2+D_{\mu}^2}},\, d_{\mu}=\frac{D_{\mu}}{\sqrt{D_{\nu}^2+D_{\mu}^2}},\,\sqrt{d_{\nu}^2+d_{\mu}^2}=1.
\end{equation}
Azimuthal angle $\psi$, as in the case of helicoids (see Fig. \ref{spirals2}), depends on the symmetry class of the corresponding helimagnet (Fig. \ref{Fig2}): 
\begin{align}
& \mathrm{C}_{nv}:\,\psi=\varphi,\nonumber\\ 
& \mathrm{D}_n:\,\psi=\varphi-\pi/2,\nonumber\\
& \mathrm{D}_{2d}:\,\psi=-\varphi+\pi/2.
\label{psiBHPhase}
\end{align}
For classes with competing DM interactions the functions $\psi(\varphi)$ are specified by the ratio of DM constants \cite{JETP89}:
\begin{equation}
 \psi(\varphi)=\varphi+\arctan{(-\frac{d_{\mu}}{d_{\nu}})}.
 \end{equation}

The total energy of an isolated skyrmion with respect to the homogeneous state can be written as
\begin{equation}
E = \int\limits_0^{\infty} \varepsilon (\theta, \rho) d \rho,\,
\varepsilon (\theta, \rho) =  2\pi\rho\left[\left( \frac{d \theta}{d \rho } \right)^2 +\frac{\sin^2 \theta}{\rho^2} 
+ h\,(1- \cos \theta) + \frac{d \theta}{d \rho } + \frac{\sin 2 \theta}{2\rho}\right]
\label{energy1}
\end{equation}
where $\varepsilon (\theta, \rho)$ is an  energy density.

%%%%%%%%%%%%%%%%%%%%%%%%%%%%%%%%%%%%%%%%%%%%%%%%%%%%%%%%%%%%%%%%%%%%%%%%%%%%%%%%%%%%%%%%%%%%%%%%%%%%%%%%%%%%%%%%%%%%%%%%%%%%%%%%
\subsection{Methods \label{MethodsBHPhase}}
%%%%%%%%%%%%%%%%%%%%%%%%%%%%%%%%%%%%%%%%%%%%%%%%%%%%%%%%%%%%%%%%%%%%%%%%%%%%%%%%%%%%%%%%%%%%%%%%%%%%%%%%%%%%%%%%%%%%%%%%%%%%%%%%

\begin{figure}
\centering
\includegraphics[width=18cm]{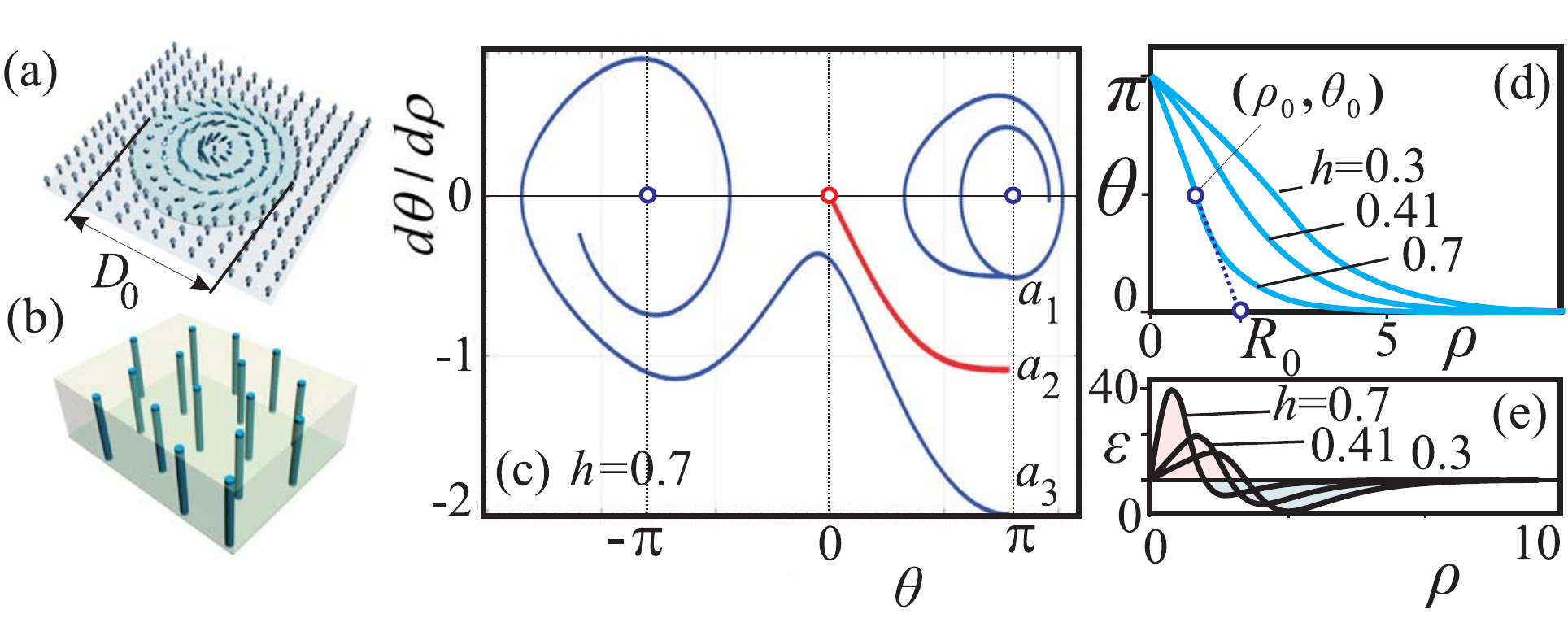}
\caption{
\label{Fig22}
Isolated skyrmions: (a) cross-section through an isolated skyrmion shows axisymmetric distribution of the
magnetization (shaded area indicates the core with the diameter $D_0$); (b) isolated skyrmions are homogeneously extended  into the third dimension as skyrmionic filements; typical solutions of Eq. (\ref{OneD}) for isolated skyrmions are shown as phase portraits (c) on the plane ($(\theta,\theta_{\rho})$)  and  magnetization profiles $\theta(\rho)$ (d). (e)  Energy densities $\varepsilon(\theta,\rho)$ for different values of the applied magnetic field $h$.
}
\end{figure}

The most appropriate method to obtain solutions of (\ref{OneD}) for isolated skyrmions is to solve the auxiliary Cauchy problems for these equations  with the initial conditions \cite{JMMM94}:
\begin{equation}
\theta(0)=\pi, \frac{d \theta } { d \rho}(0)=a_i.
\label{a}
\end{equation} 
For arbitrary values of $a_i$ the lines $\theta_{\rho}(\theta)$ normally end by spiraling around one of the attractors ($\theta_i,0$) where $\theta_i$ are specified by the magnetic field $h$. As an example in Fig. \ref{Fig22} (b) two lines with $a_1=0.5$ and $a_3=2$ are plotted.  

The curves end in the points ($2k \pi, 0$) with $k=1,2...$ only for certain discrete values of initial derivatives $a_i$. Then these particular trajectories chosen among all possible trajectories in \textit{phase space} $(\theta, d \theta / d \rho)$ represent localized solutions of the boundary value problem (\ref{OneD}). 

%Among possible trajectories in \textit{phase space} $(\theta, d \theta / d \rho)$ only for certain discrete values of initial derivatives $a_i$ the curves end in the points ($2k \pi, 0$) with $k=1,2...$ representing localized solutions of the boundary value problem (\ref{OneD}). 
%
In Fig. \ref{Fig22} (b) such a \textit{separatrix} solution corresponds to $(d \theta / d \rho)(0) =a_2= 1.088$.
Note, that in magnetic fields applied opposite to the magnetization in the center of an isolated skyrmion, besides the ordinary skyrmions with  $\Delta \theta = \theta(0)-\theta (\infty) = \pi$, also skyrmions with any odd number of half- turns $\Delta \theta $ = $3 \pi$, $5 \pi$  can exist \cite{JMMM99}.

The set of profiles $\theta(\rho)$ for different values of the applied magnetic field is plotted in Fig. \ref{Fig22} (d).
As these profiles bear strongly localized character, a skyrmion core diameter $D_0$ can be defined in analogy to definitions for domain wall width \cite{Hubert98}, i.e.  as two times the value of $R_0$, which is the coordinate  of the point where the tangent at the inflection point ($\rho_0, \theta_0$) intersects the $\rho$-axis (Fig. \ref{Fig22} (a), (d)):
\begin{equation}
D_0=2 (\rho_0-\theta_0(d\theta/d\rho)^{-1}_{\rho=\rho_0}).
\label{D0}
\end{equation}
According to conventions of Refs. \cite{JMMM94,pss94,JETP89} such  arrow-like solutions  will be decomposed into skyrmionic cores with linear dependence 
\begin{equation}
\theta(\rho)=\pi(1-\frac{\rho}{R}),\, \rho \leq L_D 
\end{equation}
and exponential "tails" with 
\begin{equation}
\theta\propto \exp{[-\rho\sqrt{\frac{h}{2}}]},\,\rho \gg L_D .
\label{exponenta}
\end{equation}
The exponential character of skyrmion asymptotics has been derived by solving the Euler equation (\ref{OneD}) for $\rho\rightarrow\infty$:
\begin{equation}
\frac{d^2\theta}{d\rho^2}-\frac{h\theta}{2}=0.
\end{equation}

Therefore, a "nucleus" with a diameter $2R$ can be considered as a two-dimensional particle-like state as it accumulates almost all energy of the isolated skyrmion.  At the same time the asymptotic exponential tails will be viewed as the "field" generated by the particle \cite{JETP95}.

From subdivision of the skyrmion structure general features of two-dimensional localized skyrmions can be revealed.

\subsection{Analytical results for the linear ansatz}
%
%A "nucleus" with a diameter $2R$ can be considered as a two-dimensional elementary particle as it accumulates almost all energy of Isolated skyrmion \cite{JETP95}. One can find equilibrium radius $R$ from  substituting linear ansatz into (\ref{energy1}) and minimizing with respect to $R$.
%%
Equilibrium radius $R$ of the skyrmion core can be found from  substituting the linear ansatz into (\ref{energy1}) and minimizing with respect to $R$.
The skyrmion energy (\ref{energy1}) is reduced to a quadratic potential
\begin{eqnarray}
\label{ansatz2}
E (R)  = E_0 + \alpha R^2  - \frac{\pi}{2} R,
\quad
R_{min} = \frac{2.641}{h}, \quad E_{min}=E_0-\frac{2.074}{h}
\label{lineEnergy}
\end{eqnarray}
where 
\begin{equation}
E_0 = 6.154
\end{equation}
is the "internal" energy of the skyrmions, 
\begin{equation}
\alpha = 0.297\,h,
\end{equation}
and the parabola vertex point $(R_{min},E_{min})$ determines the minimum of energy (\ref{lineEnergy}).

This simplified model offers an important insight into physical mechanisms underlying the formation of the chiral skyrmions.
The exchange energy $E_0$ does not depend on the skyrmion size and presents an amount of positive energy "trapped" within the skyrmion (see red-shaded positive peak of energy for solutions $\theta(\rho)$ in Fig. \ref{Fig22} (e)) . 
The equilibrium skyrmion size arises as a result of the competition between chiral and Zeeman energies:
\begin{equation}
 R_{min}\propto \frac{|D|}{H}.
 \end{equation}
In centrosymmetric systems with $D=0$ localized solutions are radially unstable and collapse spontaneously under the influence of applied magnetic field \cite{JMMM94}. % or do not exist at all \cite{JMMM94}.

\subsection{Inter-skyrmion interaction and condensation of isolated skyrmions into the lattice \label{condensationBH}}
%
%the assimptotic exponential tails can be viewed as the "field" generated by the particle \cite{JETP95}.

Asymptotic behaviour of the skyrmion solutions with $\theta \propto  \exp{[-\rho\sqrt{h/2}]},\, \rho \rightarrow \infty$ is determined by the Dzyaloshinskii-Moriya interactions. It can be considered as a specific "field" generated by the particle \cite{JETP95} which causes the repulsive character of the inter-skyrmion potential:
\begin{equation}
U(L)\propto\,\sqrt{L}\,\exp{[-L\sqrt{\frac{h}{2}}]}
\label{interactionBHIS}
\end{equation}
where $L>>1$ is the distance between skyrmion cores.

The ensemble of repulsive particle-like isolated skyrmions can condense into a lattice if the value of an applied magnetic field is smaller than the critical value $h_{S}$.  % with decreasing values of applied magnetic field $h$. 
In this case negative energy density associated with DM interactions (blue-shaded area of energy distribution $\varepsilon(\theta,\rho)$, Fig. \ref{Fig22} (e)) outweights the positive  exchange contribution (red-shaded area), and the skyrmion strings tend to fill the whole space with some equilibrium radius $R_{min}$. %(the line for $h=0.3$ in Fig.\ref{Fig22}(c) was calculated according to CCA).
For equation (\ref{OneD}),
\begin{equation}
h_{S}=0.400659.
\end{equation}
%
 %(see also section \ref{competitionSkHel}).
%For $h<h_{cr}$
%
The mechanism of lattice formation through nucleation and condensation of isolated skyrmions follows a classification introduced by DeGennes \cite{DeGennes75} for (continuous) transitions into incommensurate modulated phases.

\subsection{Distinction of solutions for localized skyrmions from Belavin-Polyakov solitons}

Note, that solitonic solutions 
with the same boundary conditions  
$\theta(0)=\pi,\,\theta(\infty)=0$   
as those for isolated skyrmions can be obtained 
also for isotropic centrosymmetric ferromagnets 
(well-known Belavin-Polyakov solutions 
for the nonlinear SO(3) $\sigma$-model \cite{BP75}).
In this case for $h=\beta=0$ 
differential equation (\ref{OneD}) has a manifold of analytical solutions:
\begin{equation}
\theta(\rho)=2 \arctan{\left(\frac{\rho}{\rho_0}\right)^N},\, \psi(\varphi)=N\varphi+\alpha.
\label{BP}
\end{equation}
described by the angle $\alpha$ and the parameter of integration $\rho_0$:
\begin{equation}
\alpha \in [0,\pi],\,\rho_0\in [0,\infty)
\end{equation} 
In spite of the seeming similarity with isolated skyrmions considered before, Belavin-Polyakov (BP) solitons represent a distinct branch of solutions.

First, the solutions of Eq. (\ref{BP}) are \textit{achiral} localized structures with the energy
\begin{equation}
E_0=4\pi N 
\end{equation}
 independent of the sense of rotation, 
i.e. angle $\alpha$. 
% and on the parameter $\rho_0$.
%
On the contrary, the sense of rotation and the exact value of angle $\alpha$ in isolated skyrmions is dictated by the crystallographic symmetry and corresponding DM interactions (see formulas in section \ref{ISBH}) and Fig. \ref{Fig2}).

Second, solutions (\ref{BP}) have no 
definite size.  Their energy is invariant 
under scale transformation of the profiles
\begin{equation}
\theta(\rho)\rightarrow\theta\left(\frac{\rho}{\lambda}\right),\, \lambda>0. 
\end{equation}
The solutions (\ref{BP}) represent always separatrix lines 
in the phase portraits (Fig. \ref{Fig22} (c)) which for 
any value of initial derivatives hit the point $\theta(\infty)=0$.
Applied magnetic field and/or uniaxial anisotropy force Belavin-Polyakov solutions to end by spiraling 
around pole $(\pi/2,0)$ so that they never reach point $(0,0)$.
From the analysis of energy (\ref{lineEnergy}) it is 
seen that it has a parabolic dependence on size of the soliton 
with minimum for zero radius $R$.
Thus applied magnetic field or internal anisotropic interactions lead to the spontaneous collapse of Belavin-Polyakov solutions.
In chiral skyrmions the influence of DM interactions shifts the vertex of  parabola describing the skyrmion energy (\ref{lineEnergy}) into the region of finite skyrmion radii.
On the phase plane $(\theta,d\theta/d\rho)$ only 
curves with appropriate initial derivatives will 
end in the point $(0,0)$ (Fig. \ref{Fig22} (c)).

Third, asymptotic behaviour of Belavin-Polyakov solutions 
has a $1/\rho$-character defined by the exchange energy.
In isolated chiral skyrmions $\theta\propto\exp[-\rho]$ which is caused by DM interactions. 
Moreover, energy density 
distributions $\varepsilon(\rho)$ (Fig. \ref{Fig22} (e)) 
reveal two distinct regions: positive exchange-energy "bags" 
concentrated in the skyrmion center and extended areas 
with negative DM-energy density stretching up to infinity.

%%%%%%%%%%%%%%%%%%%%%%%%%%%%%%%%%%%%%%%%%%%%%%%%%%%%%%%%%%%%%%%%%%%%%%%%%%%%%%%%%%%%%%%%%%%%%%%%%%%%%%%%%%%%%%%%%%%%%%%%%%%%%%%%
\section{Properties of ideal skyrmion lattices: double twist versus compatibility \label{PropertiesBHPhase}}
%%%%%%%%%%%%%%%%%%%%%%%%%%%%%%%%%%%%%%%%%%%%%%%%%%%%%%%%%%%%%%%%%%%%%%%%%%%%%%%%%%%%%%%%%%%%%%%%%%%%%%%%%%%%%%%%%%%%%%%%%%%%%%%%

\begin{figure}
\centering
\includegraphics[width=18cm]{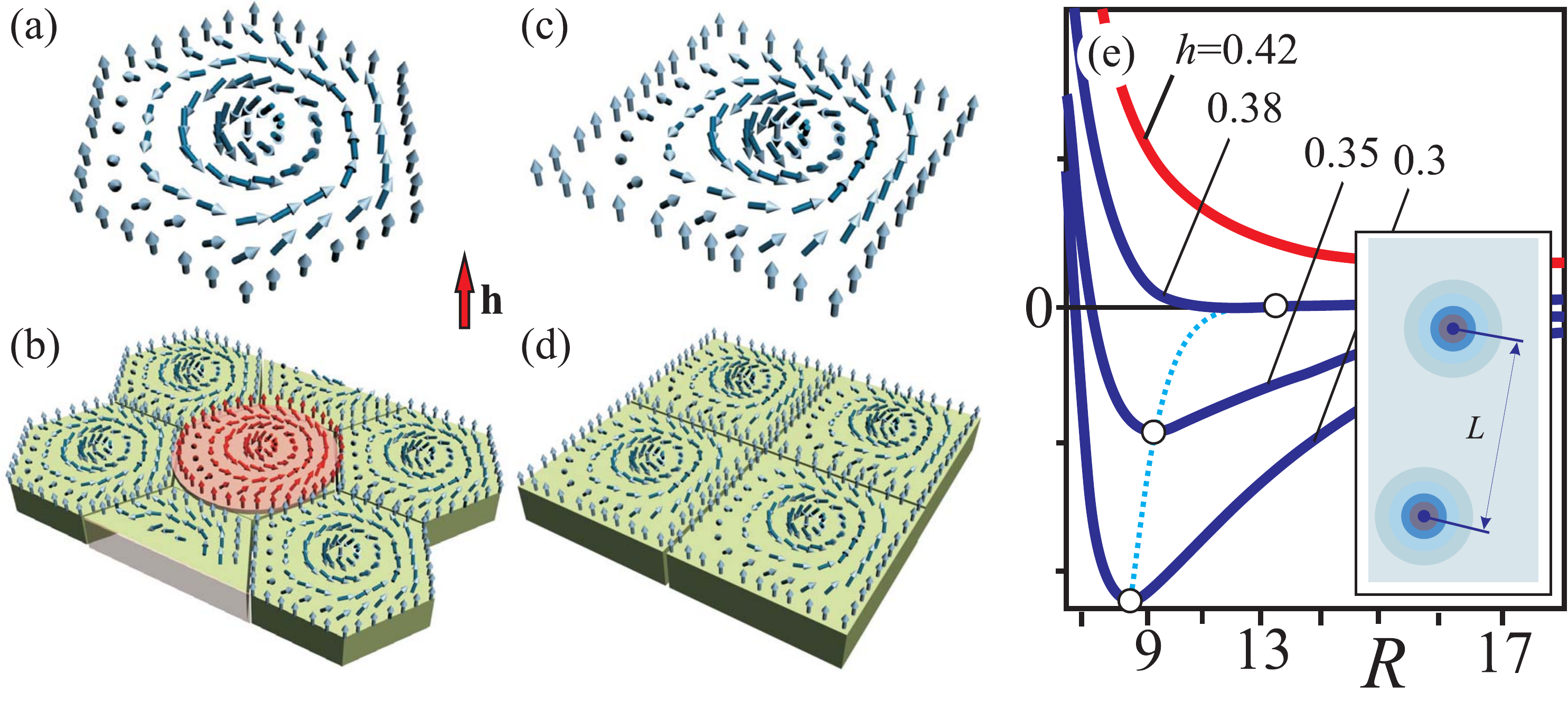}
\caption{
\label{Condensation}
Hexagonal (a),(b) and square (c),(d) skyrmion lattices: (a) and (c) are unit cells with axisymmetric distribution of the magnetization near the center; (b) and (d) are fragments of the lattices. In (b) the   replacement of the cell in the skyrmion lattice by the circle according to the method of circular cell approximation is shown as outline. (e) Below the critical field $h_{S}$ the energy of a skyrmion lattice has a minimum for some equilibrium cell size $R_{min}$.
}
\end{figure}

In  early numerical approaches used in Refs. \cite{JMMM94,JMMM99,pss94} the circular cell approximation (CCA) had been used to derive equilibrium parameters of skyrmion lattices.
In this method the lattice cell is replaced by a circle (Fig. \ref{Condensation} (b)), and then Eq. (\ref{OneD}) is integrated with boundary conditions
\begin{equation}
 \theta(0)=\pi,\, \theta(R)=0.
 \end{equation}
After that, the energy density of the lattice 
\begin{eqnarray}
W_{CCA} = \frac{1}{\pi R^2}\int_0^{R} \varepsilon (\theta, \rho)  d\rho
\label{energyCCA}
\end{eqnarray}
is minimized with respect to the cell radius $R$ (Fig. \ref{Condensation} (e)) and the equilibrium size $R_{min}$ is found.

In real hexagonal (Fig. \ref{Condensation} (a)) and/or square skyrmion lattices (Fig. \ref{Condensation} (c), (d)), the axisymmetric distribution of the magnetization is preserved only near the center of lattice cell while the overlappping solutions $\theta(\rho)$ in the inter-skyrmion regions are distorted. 
Therefore, it is worthwhile to compare  corresponding numerically rigorous solutions  with those obtained from the circular-cell approximation.

\subsection{Methods: numerical recipes \label{NumericalRecipes}}

For two-dimensional skyrmions the Euler-Lagrange equations derived from the energy functional (\ref{DMdens1}) are non-linear partial differential equations.
These equations have been solved by numerical energy minimization procedure using finite-difference discretization on rectangular grids with adjustable grid spacings and periodic boundary conditions.
Components $(m_x,m_y,m_z)$ of the magnetization vector $\mathbf{m}$ have been evaluated in the knots of the grid, and for the calculation of the energy density (\ref{DMdens1}) I used finite-difference approximation of derivatives with different precision up to eight points as neighbours.
To check the stability of the numerical routines I refined and coarsened the grids from $42\times72$ points up to $168\times288$.  % using circular cell solutions as initial guess for the relaxation procedure.
To avoid elliptical instability of the hexagonal skyrmion lattice I used grid spacings $\Delta_y\approx\Delta_x$ so that grids are approximately square in order to reduce the artificial anisotropy incurred by the discretization. %square grid with equal spacings $\Delta_y,\, \Delta_x$  along perpendicular directions $x$ and $y$.
The final equilibrium structure for the 2D baby-skyrmion hexagonal lattice was obtained  according to the following iterative procedure of the energy minimization using simulated annealing and a single- step Monte- Carlo dynamics with the Metropolis algorithm \cite{Metropolis}: 

(i) The initial configuration of magnetization vectors in the grid knots for Monte-Carlo annealing is specified by the solutions from circular-cell approximation.

(ii) A point $(x_n,y_n)$ on a grid is chosen randomly. Then, the magnetization vector in the point is rotated without change of its length. If the energy change $\Delta H_k$ associated with such a rotation is negative, the action is immediately accepted.

(iii) However, if the new state's energy is higher than the last, it is accepted probabilistically. The probability $P$ depends upon the energy and a kinetic cycle temperature $T_k$: 
\begin{equation}
P=\exp{\left[-\frac{\Delta H_k}{k_BT_k}\right]}, 
\end{equation}
where $k_B$ is Boltzmann constant.
Together with probability $P$ a random number $R_k\in [0,1]$ is generated. If $R_k<P$ new configuration accepted otherwise discarded (see, for example, \cite{MC}).
Generally speaking, at high temperatures $T_k$, many states will be accepted, while at low temperatures, the majority of these probabilistic moves will be rejected.
Therefore, one has to choose appropriate starting temperature for heating cycles to avoid transformation of metastable skyrmion textures into globally stable spiral states.

(iv) The characteristic spacings $\Delta_x$ and $\Delta_y$ are also adjusted to lead to the energy relaxation.  The procedure is stopped when no further reduction of energy is observed.

%%%%%%%%%%%%%%%%%%%%%%%%%%%%%%%%%%%%%%%%%%%%%%%%%%%%%%%%%%%%%%%%%%%%%%%%%%%%%%%%%%%%%%%%%%%%%%%%%%%%%%%%%%%%%%%%%%%%%%%%%%%%%%%%
\subsection{Features of ideal skyrmion lattices}
%%%%%%%%%%%%%%%%%%%%%%%%%%%%%%%%%%%%%%%%%%%%%%%%%%%%%%%%%%%%%%%%%%%%%%%%%%%%%%%%%%%%%%%%%%%%%%%%%%%%%%%%%%%%%%%%%%%%%%%%%%%%%%%%

\begin{figure}
\centering
\includegraphics[width=15cm]{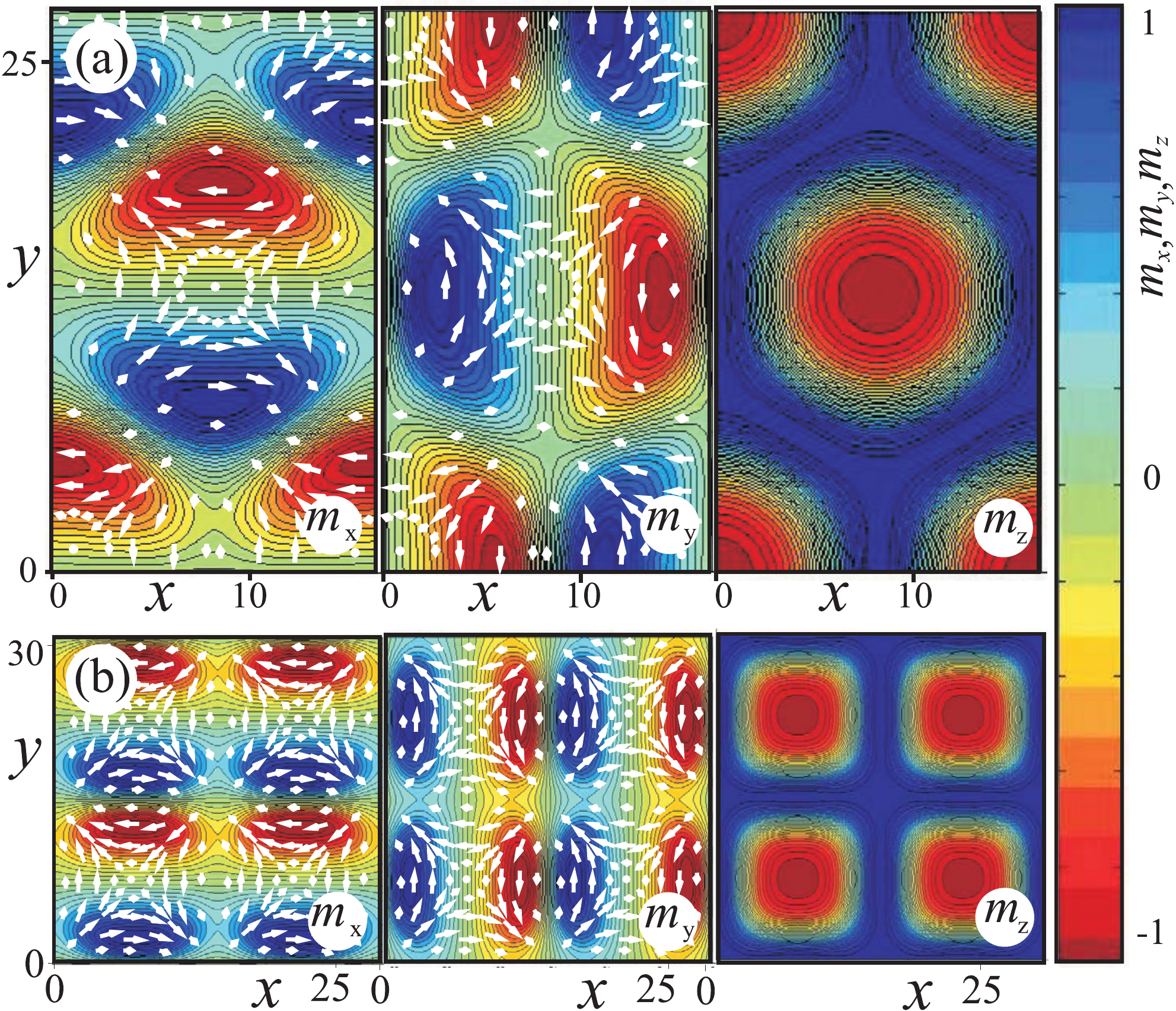}
\caption{
\label{Fig3}
Contour plots for $m_x$, $m_y$, and $m_z$ components of the magnetization  on the plane $(x,y)$ for the hexagonal (a) and square (b) skyrmion lattices of  a helimagnet with $D_{2d}$ symmetry. The white arrows show the corresponding distribution of the magnetization. 
}
\end{figure}

While condensing into the lattice, isolated skyrmions can form either hexagonal or square skyrmion order (Fig. \ref{Condensation} (a)-(d)).
Contour plots for the components $m_x,\, m_y$, and $m_z$ of the magnetization vector $\mathbf{m}$ in both  lattices are shown in Fig. \ref{Fig3} (a), (b).
Separate isolated skyrmions preserve axisymmetric distribution of the magnetization near the cell center while the overlap of solutions $\theta(\rho)$ (Fig. \ref{Fig22} (d)) distorts the inter-skyrmion regions.

\vspace{3mm}
\textit{A. Comparison of energy densities and surface areas of the lattice cells from circular-cell approximation and numerical simulations.}
\vspace{3mm}

\begin{figure}
\centering
\includegraphics[width=12cm]{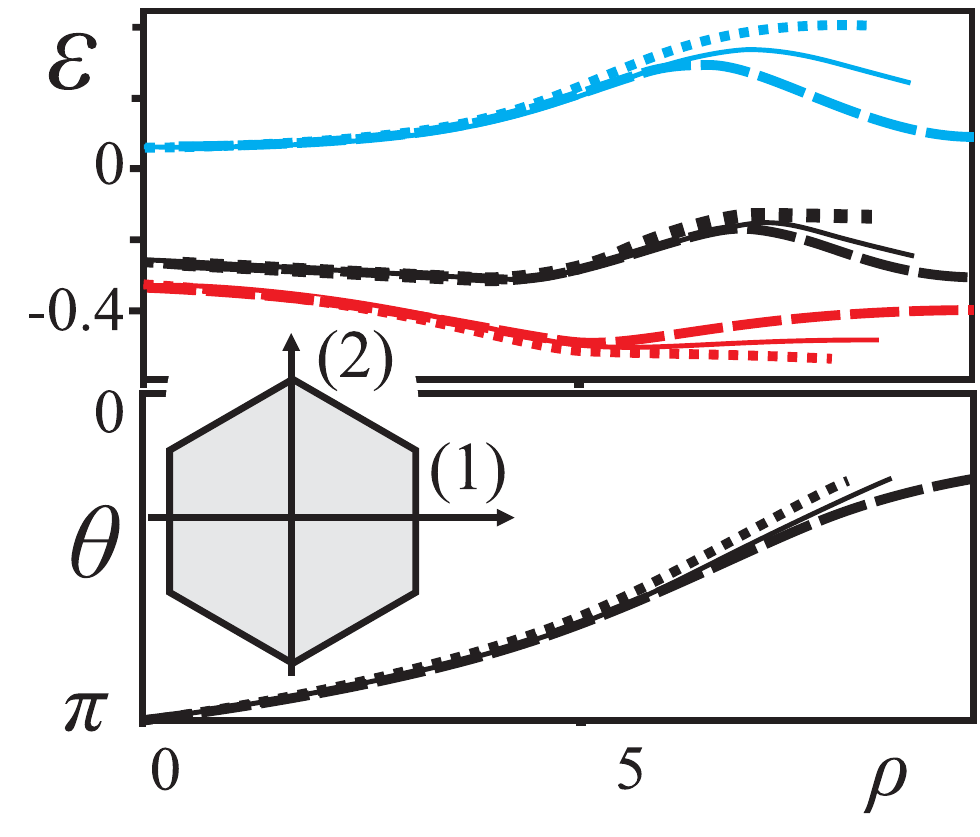}
\caption{
\label{HT}
Distributions of exchange (blue lines), DM (red lines), and $\varepsilon_i(\rho)$ total energy densities in a hexagonal lattice for two particular directions $i$ through the cell (dotted line along apothem, dashed line along the diagonal of the hexagon) plotted together with the corresponding dependences for circular-cell approximation (solid thin lines); profiles $\theta_i(\rho)$ for circular-cell approximation (solid thin line) and numerical hexagon (dotted and dashed lines).
}
\end{figure}

Figure \ref{HT} shows the distribution of the free-energy densities and magnetization profiles $\theta(\rho)$  for equilibrium hexagonal skyrmion lattice in the circular-cell approximation and from numerical simulations.
Due to the denser packing of individual skyrmions, hexagonal lattices provides smaller energy density in comparison with square lattice. % (not shown in Fig. \ref{Fig3} (c)).

The difference of energy densities in hexagonal cell and CCA cell for $h=0$ is 
\begin{equation}
\Delta W=\frac{W_{CCA}-W_{hexagon}}{W_{CCA}}=\frac{0.234-0.2312}{0.234}=0.012.
\end{equation}
For the square cell the difference is larger,
\begin{equation}
\Delta W=\frac{W_{CCA}-W_{square}}{W_{CCA}}=\frac{0.234-0.2235}{0.234}=0.0449.
 \end{equation}
The surface area of the cell in CCA is larger than the surface of the corresponding numerical hexagon, 
\begin{equation}
\Delta S=\frac{S_{CCA}-S_{hexagon}}{S_{CCA}}=0.0167, 
\end{equation}
whereas the surface area of square lattice cell is larger than the circle, 
\begin{equation}
\Delta S=\frac{S_{square}-S_{CCA}}{S_{CCA}}=0.0234.
\end{equation}
Hence,  the statement of the circular- cell approximation  \cite{JMMM94}, that surface areas of a circle and a hexagonal cell must coincide, is basically erroneous.
However, the smallness of all the differences between CCA and rigorous numerical simulations for model (\ref{DMdens1}) allows to consider circular-cell approximation as an excellent approach for the global properties of the hexagonal skyrmion lattice.
In particular, CCA yields an exact value of the upper critical field $h_{S}$ as the skyrmions are located at big distances from each other and are independent on the detailed arrangement of individual filaments: $h_{S}$ is the same for  square and hexagonal lattices.
%Figure \ref{Fig3}(b) shows magnetization profiles $\theta(\rho)$ for both CCA and hexagonal skyrmion cell.

The distortions of angular solutions near the border of hexagon lead to corresponding redistribution of exchange and DM energy density (Fig. \ref{HT}): due to the increase of exchange energy density along the apothem of the hexagon (dotted blue line), the total energy density (dotted black line) has also higher value than corresponding CCA energy density (thin black line).

%
%This remarkable property is due to specific energetics of the skyrmions. 
%
%"Double-twist" rotation of the magnetization near the skyrmion core leads to larger energy reduction than in "single-twisted" helical phases while edge areas of the cell have larger energy density than the helical states \cite{JP10}.
 % 

\vspace{3mm}
\textit{B. Expansion into the Fourier series of the $m_z$-component of the magnetization for the lattice from the rigorous calculations}
\vspace{3mm}

\begin{figure}
\centering
\includegraphics[width=9cm]{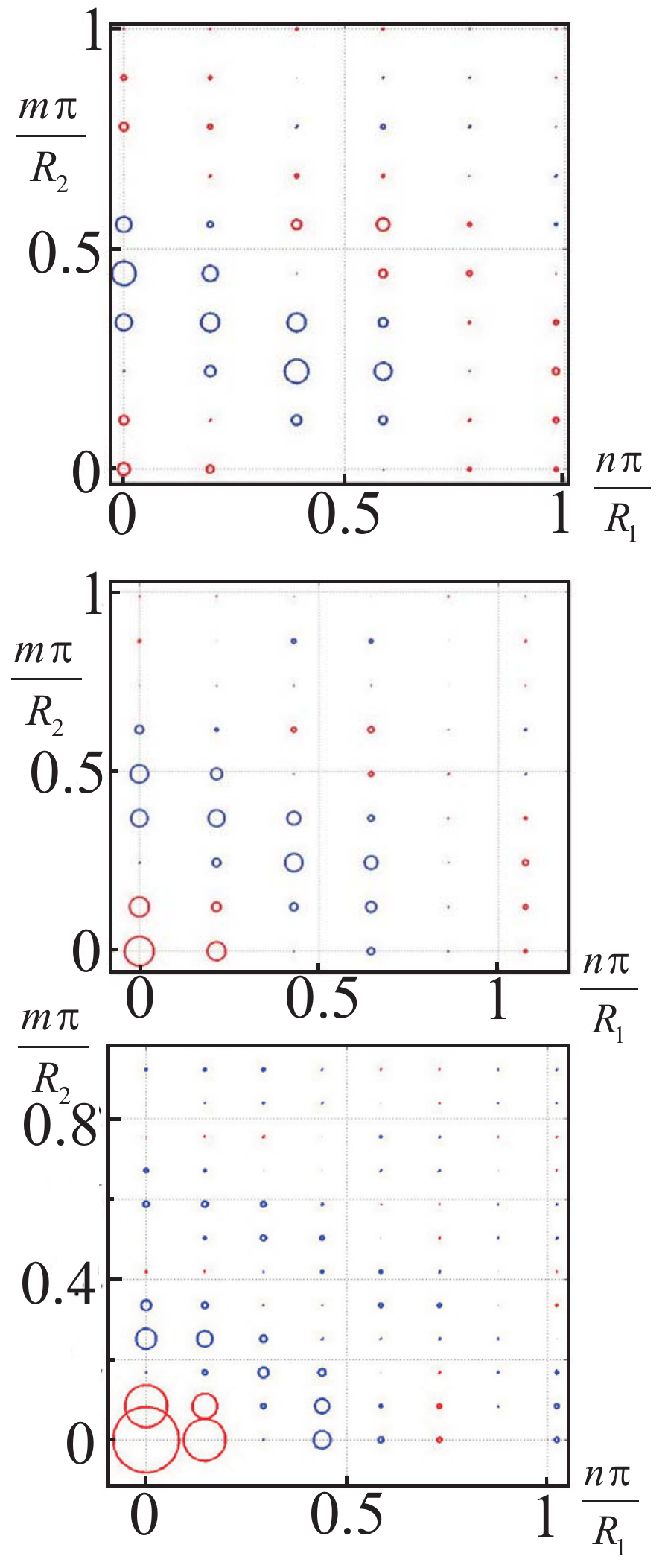}
\caption{
\label{Fourier}
Graphical representation of the coefficients of the Fourier expansion for $z$-component of the magnetization in the hexagonal skyrmion lattice  for different values of the applied magnetic field $h=0$ (a), $h=0.2$ (b), $h=0.4$ (c). Positive coefficients are marked by red color, whereas negative - by blue.  
}
\end{figure}

The Fourier expansion for $z$-component of the magnetization may be written as
\begin{align}
m_z=\sum_{i,j =0}^{\infty}\lambda_{ij}[&a_{ij}\cos(\frac{2\pi ix}{R_1})\cos(\frac{2\pi jy}{R_2})+b_{ij}\sin(\frac{2\pi ix}{R_1})\cos(\frac{2\pi jy}{R_2})+\nonumber\\
+&c_{ij}\cos(\frac{2\pi ix}{R_1})\sin(\frac{2\pi jy}{R_2})+d_{ij}\sin(\frac{2\pi ix}{R_1})\sin(\frac{2\pi jy}{R_2})]
\label{FourierMz}
\end{align}
where 
\begin{equation}
\lambda_{00}=0.25,\, \lambda_{i0}=\lambda_{0j}=0.5, \, \lambda_{ij}=1,
\end{equation}
$R_1$ and $R_2$ are characteristic sizes of the elementary lattice cell (Fig. \ref{Fig3} (a)).
With the present choice of origin of coordinates, coefficients
\begin{equation}
b_{ij}=c_{ij}=d_{ij}=0. 
\end{equation}
The coefficients $a_{ij}$ may be represented graphically  for different values of the applied magnetic field (Fig. \ref{Fourier}).
Due to axial arrangement of the core the amplitudes of higher harmonics  have comparable values with those of leading  lattice harmonics. Positive coefficients $c_{ij}$ of the expansion (\ref{FourierMz}) are marked by red color, whereas negative coefficients - by blue. Multiplying the diameter of each circle by ten one can extract the value of the underlying coefficient. 

%For example, for $h=0.2$ (Fig. \ref{Fourier} (b)) $c_{11} = -0.071$, $c_{22} = 0.047$, $c_{50} = 0.047$.

In Ref. \cite{Muhlbauer09} a triple spin-spiral crystal is presented as a skeleton for such a skyrmion lattice in cubic chiral magnets. While the topology and rough geometry of these states is the same, this theoretical interpretation of the Skyrmion states of chiral magnets assumes that the skyrmionic states can be described by the first few harmonics of a hexagonal lattice. Skyrmions in this approach have triangular cores instead of radial cores.  This point of view does not agree with the exact solutions and detailed demonstration of radial and localized solutions for skyrmions in the present chapter. The approach of \cite{Muhlbauer09} discounts the existence and relevance of the localized and radial nature of the skyrmion solutions. 
By virtue of the localized character of the skyrmion cores and its axial symmetry  such an approximation by a number of Fourier modes is very poor as the convergence of the Fourier series is slow. Owing to the localization of the skyrmions their properties cannot be modeled, nor understood from a multi-Q ansatz with a finite number of Fourier components. In particular, the important transformation process of a Skyrmion lattice into an assembly of isolated skyrmion lines under an applied field  cannot be described by the picture of a triple spin-spiral crystal.

Thus, the theoretical interpretation proposed in Ref. \cite{Muhlbauer09} is considered to be not correct.

\vspace{3mm}
\textit{C. Rigorous solutions for skyrmion lattices with Dzyaloshinskii-Moriya interactions representing the weighted sum of Lifshitz invariants}
\vspace{3mm}

\begin{figure}[tb]
\includegraphics[width=15cm]{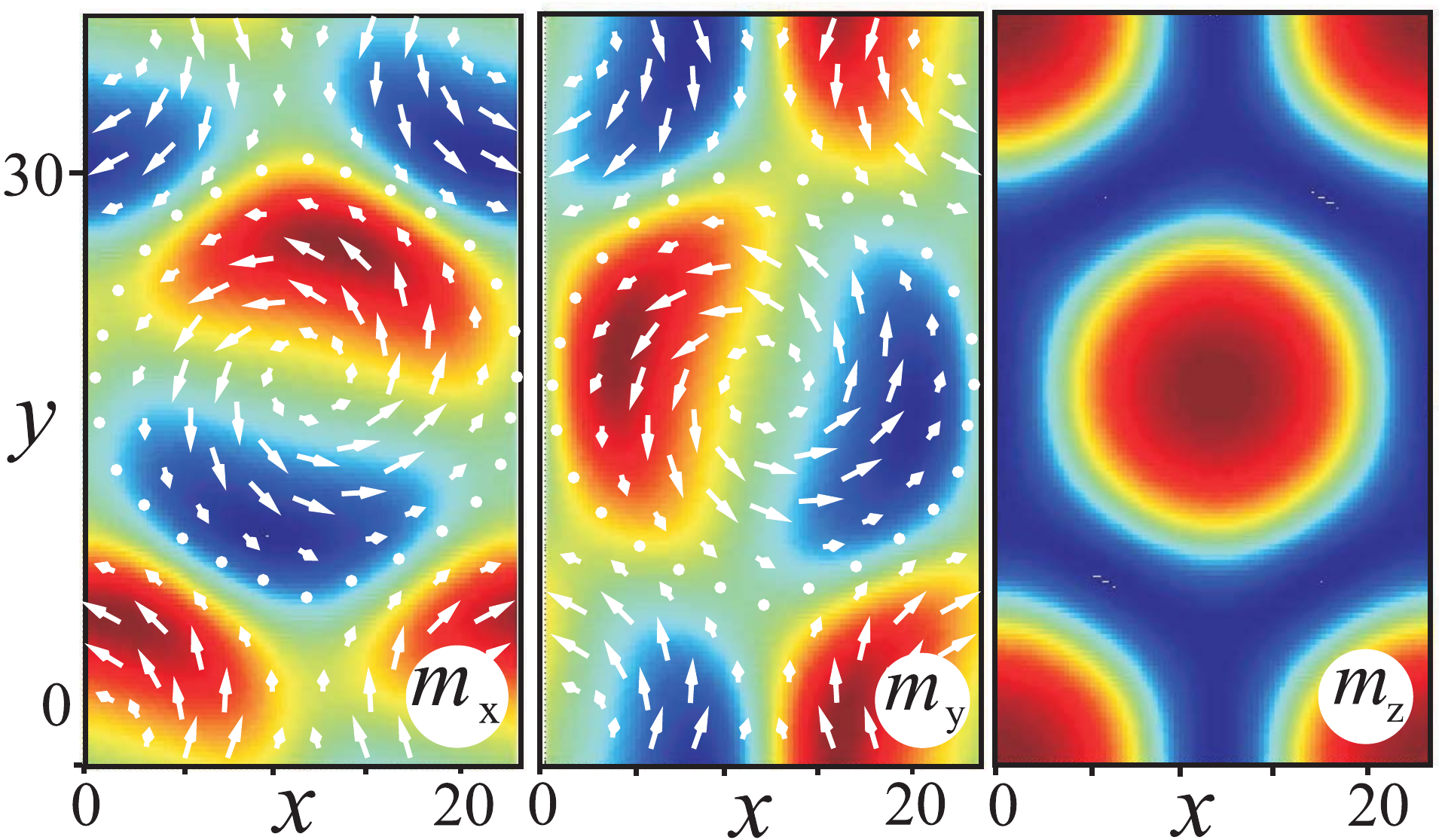}
\centering
\caption{
\label{Fig6}
Contour plots of $m_x,\,m_y$, and $m_z$ components of the magnetization for helimagnets with $C_n$ symmetry ($h=\beta=0,\,d_1=0.2,\,d_2=0.9798$).   
 }
\end{figure}

In the case of DM interactions (\ref{Cn}), (\ref{S4}) with competing counterparts, the skyrmions have a more complicated structure as shown in Fig. \ref{Fig6} for particular case of $\mathrm{C}_n$ symmetry.
For $d_1=0.2,\,d_2=0.9798$ angle $\psi=\phi+78^{\mathrm{o}}$, and the structure of skyrmions is slightly different from the "Bloch"-type skyrmion with D$_{n}$ symmetry (Fig. \ref{Fig2} (b)).
Note, that the cases $d_1=1,\,d_2=0$ and $d_1=0,\,d_2=1$ denote skyrmions with C$_{nv}$ (Fig. \ref{Fig2} (a)) and D$_n$ (Fig. \ref{Fig2} (b)) symmetry, correspondingly.

%%%%%%%%%%%%%%%%%%%%%%%%%%%%%%%%%%%%%%%%%%%%%%%%%%%%%%%%%%%%%%%%%%%%%%%%%%%%%%%%%%%%%%%%%%%%%%%%%%%%%%%%%%%%%%%%%%%%%%%%%%%%%%
%old
%\section{Competition of skyrmions with helicoids within the isotropic phenomenological model \label{competitionSkHel}}

%new
\section[Competition of skyrmions with helicoids]{Competition of skyrmions with helicoids within the isotropic phenomenological model \label{competitionSkHel}}
%%%%%%%%%%%%%%%%%%%%%%%%%%%%%%%%%%%%%%%%%%%%%%%%%%%%%%%%%%%%%%%%%%%%%%%%%%%%%%%%%%%%%%%%%%%%%%%%%%%%%%%%%%%%%%%%%%%%%%%%%%%%%%

\begin{figure}
\centering
\includegraphics[width=18cm]{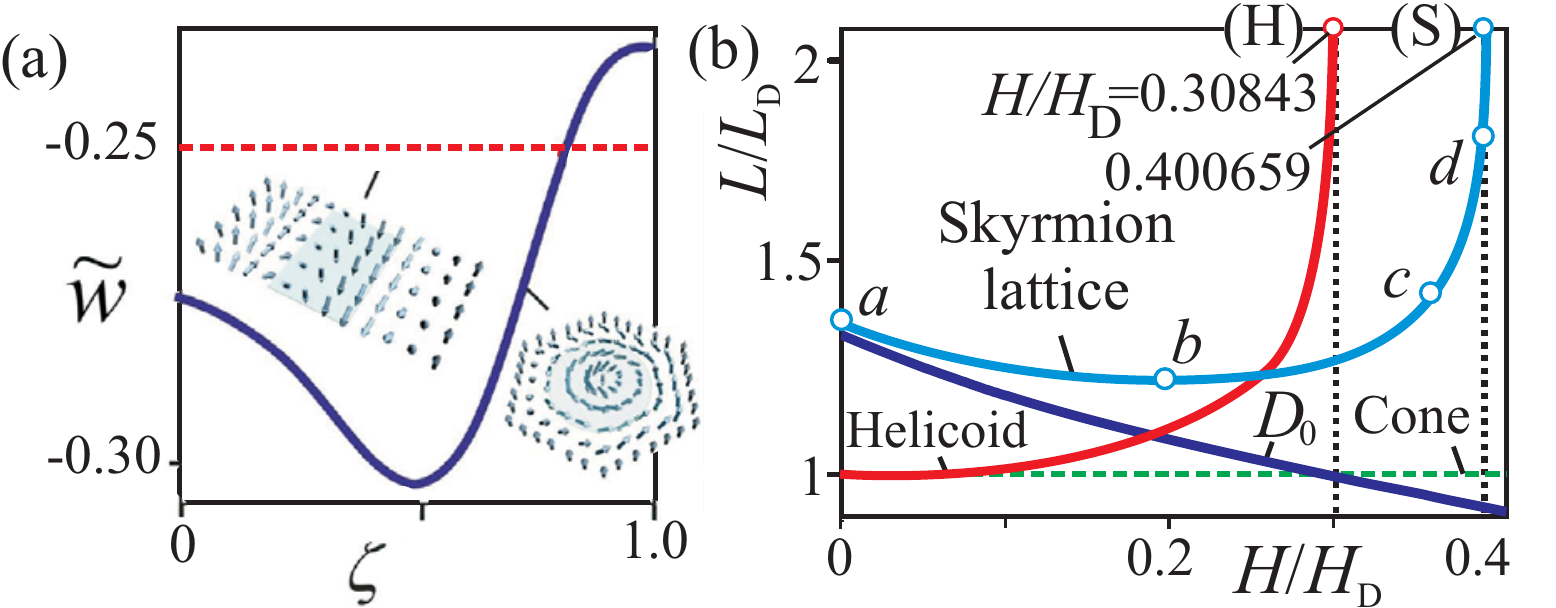}
\caption{
\label{field}
(a) Local energies $\widetilde{w}(\zeta)$ of the skyrmion lattice and helicoid at zero field (reproduced from \cite{Nature06}); (b) equilibrium sizes of the cell core ($D_0$, Eq. (\ref{D0})) and lattice period $R$ compared to helicoid and cone periods.
}
\end{figure}

From the previous calculations it is known \cite{Nature06} [XI,XIV,XV], that "double-twisted" rotation of the magnetization as in skyrmions yields an energetic advantage only at small distances from the skyrmion axis in comparison with  "single-twisted" spiral phases \cite{Nature06}.
Conversely, the energy density is larger at the outskirt of the skyrmion which is the consequence of an inherent frustration built into models with chiral couplings: the system cannot fill the whole space with the ideal, energetically most favoured double-twisted motifs. 
The equilibrium energy of the skyrmion cell at zero field 
\begin{equation}
\widetilde{w}_{S}(\zeta )=\frac{2}{\zeta^2}\int^{\zeta}_{0} \varepsilon(\rho)\rho d\rho
\end{equation}
plotted as a function of the distance from the center $\zeta$ (Fig. \ref{field} (a)) shows that an energy excess near the border outweighs the energy gain at the skyrmion center. 
As a result, the skyrmion states are metastable states in comparison with lower-energy helical phases.

At higher magnetic fields, however, the skyrmion lattice has lower energy than the helicoid. The first order transition 
between these two modulated states occurs at \cite{JMMM94}
\begin{equation}
H_1 = 0.1084 H_D.
\end{equation}
Properties of the skyrmion lattice solutions are collected in Fig.~\ref{field} and in Table \ref{table2}.
With increasing magnetic field, a gradual localization of the skyrmion core $D_0$ is accompanied  by the expansion of the lattice period.
The lattice transforms into the homogeneous state by infinite expansion of the period at the critical field
\begin{equation}
H_S = 0.40066 H_D.
\end{equation}
Remarkably, the skyrmion core retains a finite size, $D_0 (H_S) = 0.920 L_D$ and the lattice releases a set of \textit{repulsive} isolated skyrmions at the transition field $H_S$, owing to their topological stability.
These free skyrmions can exist  far above $H_S$. 
On decreasing the field again below $H_S$, they can re-condense into a skyrmion lattice (Fig. \ref{field} (b)).
A similar type of sublimation and resublimation of particle-like textures occurs in helicoids at the critical field 
$h_H$ (Eq. (\ref{Hhspiral})): the period infinitely expands and the helicoid splits into a set of isolated 2$\pi$ domain
walls or kinks \cite{Dz64,JMMM94}.

\begin{table}[h]
\caption{
\label{table2}
Critical fields and characteristic parameters of the hexagonal skyrmion lattice:
$H_1$ transition field between the helicoid and skyrmion lattice; 
$H_S$ saturation field of the skyrmion lattice; last column gives properties of IS as excitations of the saturated state for an (arbitrary) high field $H/H_D=0.7$ %for isolated skyrmions as excitations of the saturated state.
}
\begin{center}
\begin{tabular}{|p{3.5cm}||p{2cm}|p{2cm}|p{2cm}|p{2cm}|} %{llllll}
\hline
&& $H_1$& $H_{S}$& \\ \hline
 Reduced magnetic field, $H/H_D \quad \quad $&$0$&$0.1084 \quad \quad$&$0.40066 \quad \quad$&0.7\\ \hline
% \mr
Lattice cell period, $L/L_D$&$1.376 \quad \quad$&1.270& $\infty$ & - \\\hline
Core diameter, $D_0/L_D$&1.362&1.226& 0.920 & 0.461 \\ \hline
Averaged magnetization, $m_{\small{S}}$ &0.124& 0.278&1&1  \\
\hline
\end{tabular}
\end{center}
% {\small{$H_1$, transition field between the helicoid and skyrmion lattice; 
% $H_S$ is the saturation field of the skyrmion lattice; last column stands for 
% isolated skyrmions as excitations of the saturated state.}}
\end{table}

\begin{figure}
\centering
\includegraphics[width=18cm]{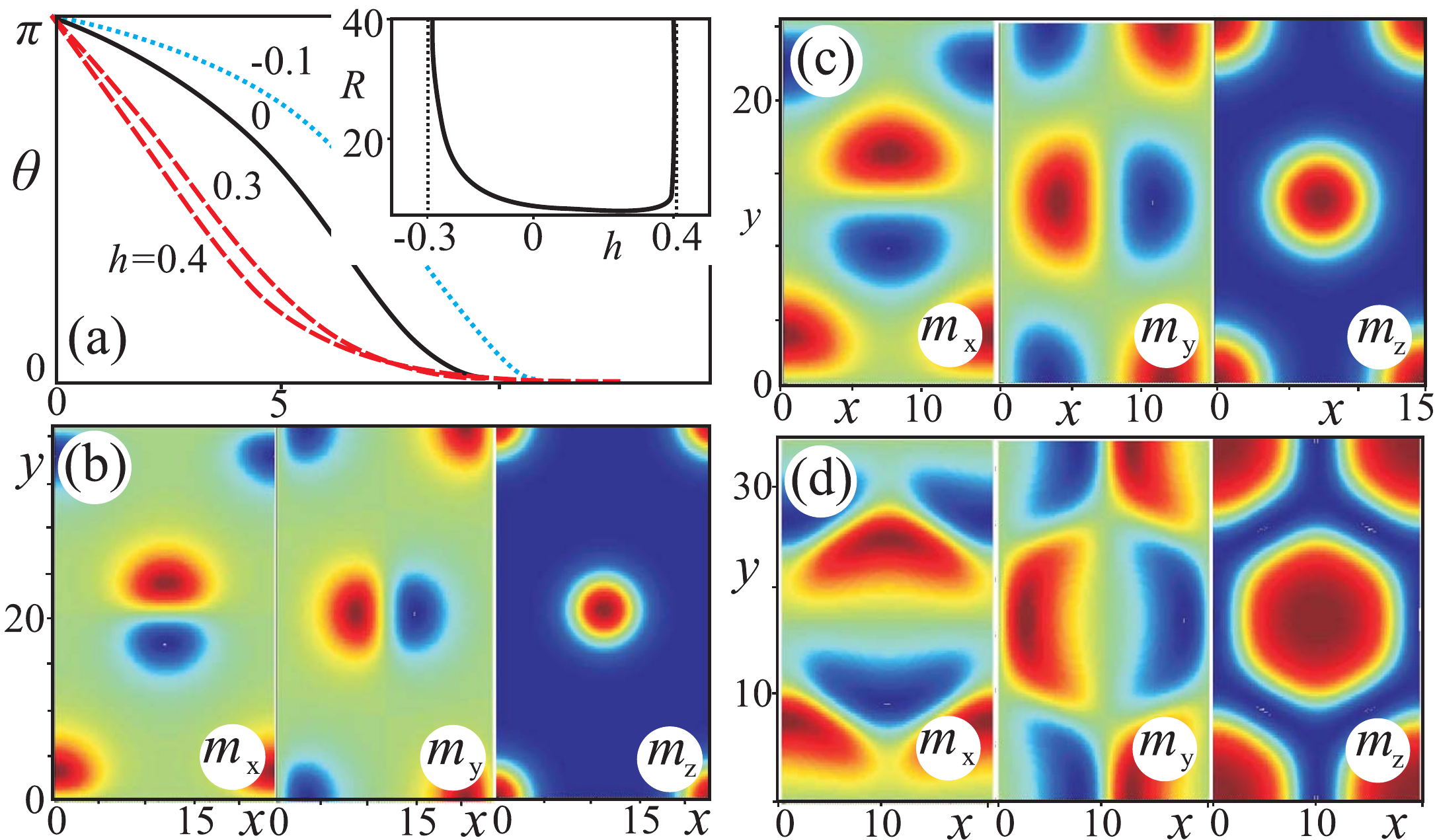}
\caption{
\label{Fig4}
Evolution of the hexagonal skyrmion lattice in magnetic field applied either opposite to the magnetization in the skyrmion center (b), (c) or parallel to it (d). The solutions are presented as angular profiles along diagonals of the hexagons (a), and contour plots for all components of the magnetization on the plane $(x,y)$: (b) $h=0.4$, (c) $h=0.3$, (d) $h=-0.2$. Inset of (a) shows the equilibrium characteristic size of the hexagonal lattice for both directions of the magnetic field: for positive values of the magnetic field the skyrmion lattice transforms into a system of isolated skyrmions with repulsive potential between them, whereas for negative magnetic field it turns into the homogeneous phase through a honeycomb structure with increasing lattice period (d).
}
\end{figure}

For a negative magnetic field applied along the magnetization in the center of skyrmion strings, both the skyrmion cores and the lattice cell size expand.
Near the critical field $h_H=-0.30843$  the vortex lattice consists of honeycomb-shaped cells separated from each other by narrow $360^{\circ}$ domain walls (Fig. \ref{Fig4} (d)). 
Note, that for negative fields the honey-comb lattice is highly instable.
It is hardly accessible and easily elongates into spiral state.
For negative magnetic fields, isolated skyrmions do not exist.

Thus, it can be concluded that for functional $W_0(\mathbf{M})$ (\ref{DMdens1}) the cone phase is the global minimum in the whole range of the applied fields where the modulated states exist ($ 0 < h < h_d$). The helicoids and skyrmion lattices can exist only as metastable states. One has to look for additional energy contributions capable to stabilize skyrmion phase. In the next sections I consider some successful candidates for this role: uniaxial, cubic, and exchange anisotropy.

%%%%%%%%%%%%%%%%%%%%%%%%%%%%%%%%%%%%%%%%%%%%%%%%%%%%%%%%%%%%%%%%%%%%%%%%%%%%%%%%%%%%%%%%%%%%%%%%%%%%%%%%%%%%%%%%%%%%%%%%%%%%%%
%old
%\section{Stabilization of skyrmion textures by uniaxial distortions in non - centrosymmetric cubic helimagnets. \label{DistortionsBHphase}}

%new
\section[Stabilization of skyrmion textures by uniaxial distortions ]{Stabilization of skyrmion textures by uniaxial distortions in non - centrosymmetric cubic helimagnets \label{DistortionsBHphase}}
%%%%%%%%%%%%%%%%%%%%%%%%%%%%%%%%%%%%%%%%%%%%%%%%%%%%%%%%%%%%%%%%%%%%%%%%%%%%%%%%%%%%%%%%%%%%%%%%%%%%%%%%%%%%%%%%%%%%%%%%%%%%%%%

From the numerical investigation of Eq.~(\ref{DMdens1}), I  show now that a sufficiently strong magnetic anisotropy $K_u$ (\ref{additional}) stabilizes skyrmionic textures in applied magnetic fields.
The uniaxial anisotropy $K_u$ in cubic helimagnets can be imposed, for example,  by surface/interface interactions in thin films or nanolayers and tuned by covering the surface with different non-magnetic materials.

In sufficiently thick magnetic layers, such induced anisotropy can be considered as a pure surface effect which distorts the uniform prolongation of skyrmion filaments perpendicularly to the surface and transforms them into convex shaped spherulites.
In thin magnetic nanolayers surface-induced uniaxial anisotropy is uniformly distributed through the layer and can be considered as homogeneous uniaxial anisotropy with constant $K_u$.
On the other side, the uniaxial anisotropy in cubic helimagnets may be induced by uniaxial strains in bulk systems. % or by surface/interface interactions in thin films or nanolayers. 

\begin{figure}
\centering
\includegraphics[width=15cm]{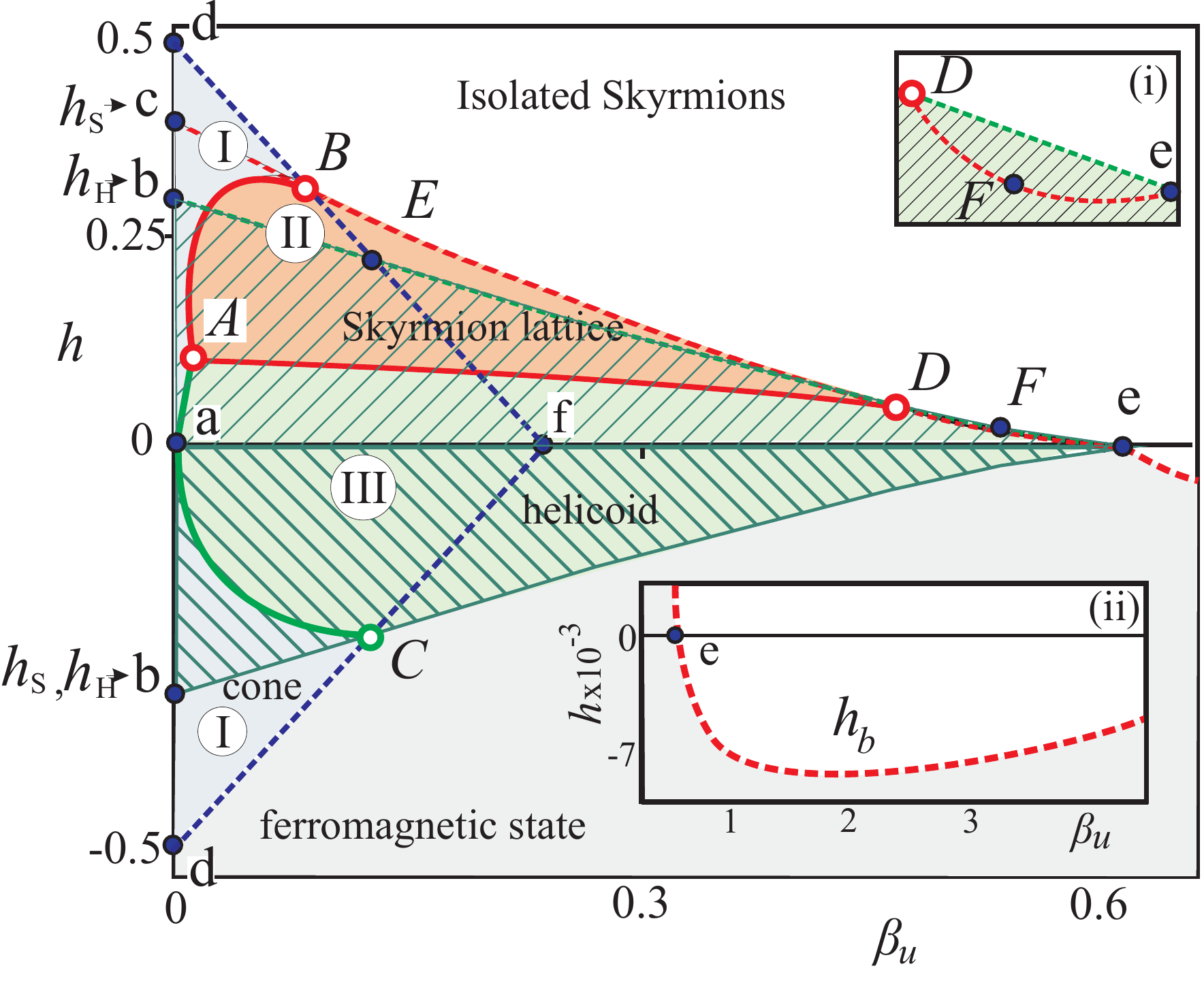}
\caption{
\label{PDUA}
Magnetic phase diagram of the solutions for model (\ref{DMdens1}) including uniaxial anisotropy $\beta_u$. Filled areas designate the regions of thermodynamical stability of corresponding modulated phases: I - conical phase (blue shading); II - skyrmion phase (red shading); III - helicoid (green shading). White shading stands for the region of isolated skyrmions and kinks. In the region with grey shading no modulated states are available. Hatching shows the existence region of helicoids. The conical phase exists within the area (a-d-B-f). For $\beta_u>0.0166$ corresponding to the point $A$ a skyrmion lattice can be stabilized in high magnetic fields.  For $\beta_u>0.25$ corresponding to the point $f$ only helicoids and skyrmions can be realized as thermodynamic phases. Two insets show the magnifications of particular parts of the phase diagram: inset (i) exhibits the region $D-F-e$ where spiral state as only one modulated phase can exist; the inset (ii) shows the line (red dashed line) of skyrmion bursting $h_b$ in negative fields (see text for details). 
}
\end{figure}

By comparing the equilibrium energies of the conical phase, the helicoids, and the rigorous solutions for hexagonal skyrmion lattice, I have constructed the phase diagram of solutions (Fig. \ref{PDUA}).

As in section \ref{ISBH}, I start analysis of the phase diagram from isolated skyrmions.

\subsection{Isolated skyrmions in chiral helimagnets with uniaxial anisotropy}

\begin{figure}
\centering
\includegraphics[width=12cm]{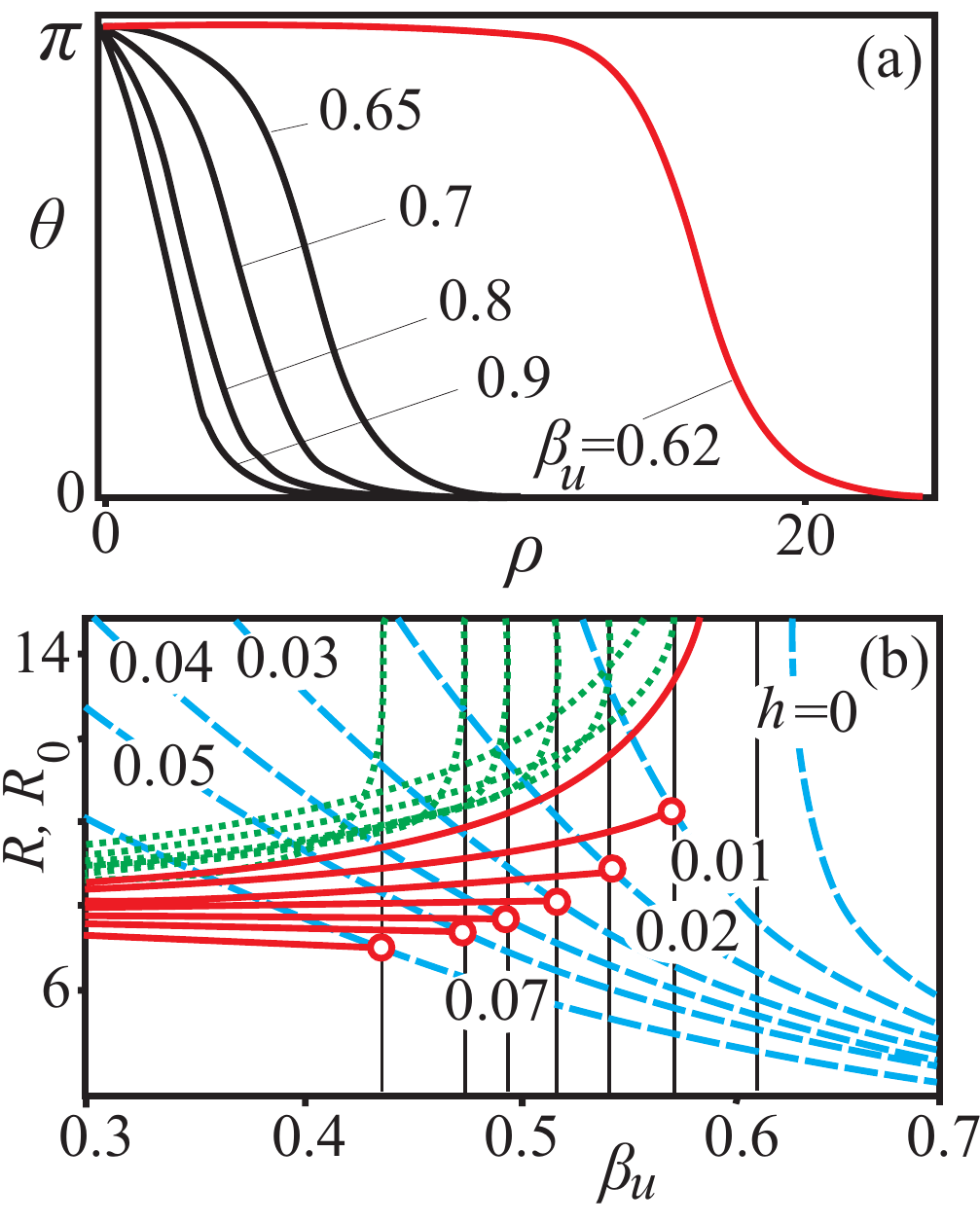}
\caption{
\label{ISUA}
(a) Angular profiles $\theta(\rho)$ for isolated skyrmions in zero magnetic field and for different values of  easy-axis uniaxial anisotropy $\beta_u$ show the expansion of skyrmion cores while approaching the critical value of uniaxial anisotropy $\beta_{cr}=0.61685$. In (b) the size of the core of isolated skyrmions determined according to the Lilley's definition (see section \ref{ISBH}) is plotted in dependence on the uniaxial anisotropy constant $\beta_u$ for different values of the applied magnetic field (dashed blue line). Red solid lines in (b) show the size of the skyrmion core in a skyrmion lattice, green dotted  lines - the size of the lattice cell. For constant value of the applied magnetic field and variable constant of uniaxial anisotropy the lattice releases the isolated skyrmions for some critical value of $\beta_u$. This corresponds to the intersection point of red and blue lines; the green lines tend to infinity; the characteristic size of the core of the  skyrmions undergoes a sudden change. For $h=0$ there is no connection between the skyrmion lattice and isolated skyrmions.
}
\end{figure}

In cubic helimagnets with uniaxial anisotropy, isolated skyrmions are solutions of the Euler equation written in the reduced form:
\begin{eqnarray}
\label{eq}
 \frac{d^2 \theta}{d \rho ^2} 
+ \frac{1}{\rho}\frac{d \theta}{d \rho}
 -\frac{\sin2 \theta}{2\rho ^2} 
 -\frac{\sin^2 \theta}{\rho} 
- \frac{h}{2}\sin \theta-\frac{\beta_u}{2}\sin 2\theta =0
\label{OneD2}
\end{eqnarray}
with the boundary conditions 
\begin{equation}
\theta(0)=\pi,\, \theta(\infty)=0.
\end{equation}

The region of metastable existence of simple $\pi$-skyrmions was calculated in Ref.\cite{pss94}. On the phase diagram (Fig. \ref{PDUA}) it is marked by white color and expands over large values of positive magnetic field and easy-axis uniaxial anisotropy.

For $h=0$ and 
\begin{equation}
\beta_u>\beta_{cr}=\frac{\pi^2}{16} 
\end{equation}
isolated skyrmions exist  as a separate branch of skyrmion solutions \cite{pss94}. %(blue dashed line in Fig.\ref{Fig5}(d)).
With decreasing constant $\beta_u$ the cores of isolated skyrmions expand, and the localized skyrmions dissappear as a solution for the critical value $\beta_{cr}$ (Fig. \ref{ISUA} (a)). The characteristic size $R_0$ of the Skyrmion core determined according to the Lilley definition (blue dashed lines in Fig. \ref{ISUA} (b)) expands to infinity for $\beta_u=\beta_{cr}$.
%
%With decreasing constant $\beta$ the characteristic size $R_0$ of the skyrmion cores  increases; IS are blown (inset of Fig.\ref{Fig5}(c)), and dissappear for the critical value $\beta_{cr}$. 

In the applied magnetic field $h>0$, the isolated skyrmions can condense into the lattice with decreasing constant of uniaxial anisotropy $\beta_u$.
%
%The applied magnetic field $h>0$ allows to join both branches of skyrmion solutions and to condense isolated skyrmions into the lattice with decreasing constant of uniaxial anisotropy $\beta$.
%
The solid red and dotted green lines in Fig. \ref{ISUA} (b) show dependences of the characteristic core and lattice cell sizes  on the changing constant of uniaxial anisotropy $\beta_u$.
In the point of intersection of red and blue lines, i.e. in the point of condensation of isolated skyrmions into the lattice, the skyrmion core undergoes a sudden leap, while the equilibrium lattice period expands unlimitedly.

For large values of uniaxial anisotropy $\pi$-skyrmions can exist even at negative fields (see inset (ii) of Fig. \ref{PDUA}). The magnetization in the skyrmion core is then oriented along the field, while the surrounding matrix is magnetized in the opposite direction. Thus, the skyrmion size increases with increasing magnetic field. Finally, when $h$ reaches a certain critical value $h_{b}(\beta_u)$ (inset (ii) of Fig. \ref{PDUA}) the skyrmion "bursts" into the homogeneous state with the magnetization parallel to the applied field. First such a behaviour of isolated skyrmions in a negative magnetic field was described in Ref.\cite{JMMM99}.
Also the technique to explore skyrmion stability was elaborated. 

In the following I exploit the methods of Ref. \cite{JMMM99} and present a comprehensive analysis of the structure and stability of all types of isolated skyrmions of the model (\ref{OneD2}).

\subsection{Localized skyrmions and the manifold of solutions of micromagnetic equations: the question of radial stability \label{StabilityISBH}}

\begin{figure}
\centering
\includegraphics[width=18cm]{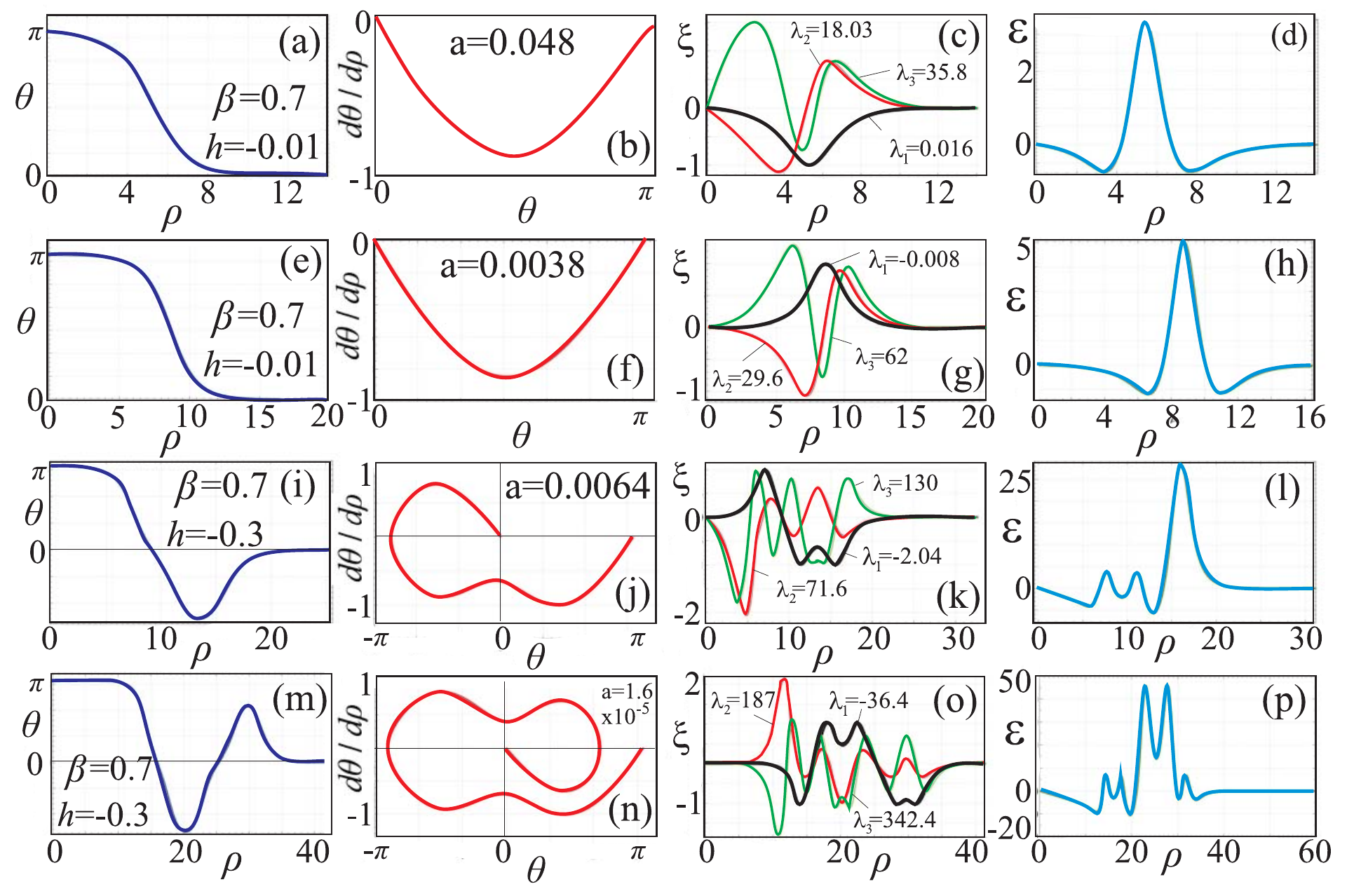}
\caption{
\label{all}
 Different types of isolated skyrmions which can be found among the solutions of the Euler equation (\ref{OneD2}) for negative magnetic field and easy axis uniaxial anisotropy. The first column (a), (e), (i), (m) shows the angular profiles $\theta(\rho)$. The second column (b), (f), (j), (n) exhibits the phase portraits. The eigenmodes plotted in the third column (c), (g), (k), (o) allow to deduce that all the solutions except skyrmions (c) are unstable. The distributions of the energy density for different types of localized solutions are plotted in the fourth column (d), (h), (l), (p).
}
\end{figure}

In addition to skyrmion solution of Eq. (\ref{OneD2}) (Fig. \ref{all} (a)) a family of specific vortex states with small values of derivative in the center $d\theta/d\rho(\rho=0)$ can be found.

The first vortex of this family is also of $\pi$-type, but has a larger core size (Fig. \ref{all} (e)). The energy distribution in such a vortex (Fig. \ref{all} (h)) looks qualitatively  the same as for the common skyrmion (Fig. \ref{all} (d)).
This vortex can exist even for zero values of Dzyaloshinskii-Moriya interaction. 

All other members of the vortex family (Fig. \ref{all} (i), (m)) are characterized by the parts with a reverse rotation of the magnetization vector - nodes. Each sequential vortex has more nodes than preceding one and exhibits oscillations of the magnetization in the tail (Fig. \ref{all} (i), (m)). The phase portraits for such vortices before hitting the point (0,0) round by turns the attractors in points (0,$\pm\pi/2$) (Fig. \ref{all} (j), (n)).

The analysis of stability for all solutions of equation (\ref{OneD2}) shows that only the skyrmion solution (Fig. \ref{all} (a)) is stable with respect to small perturbations of the structure.

To check the stability of obtained skyrmion solutions  I consider radial distortions of type $\xi(\rho)$ with constraint
\begin{equation}
\xi(0)=\xi(\pi)=0.
\label{constraint}
\end{equation}
Such distortions are the relevant leading instabilities of radial skyrmion structures $\theta(\rho)$. 
By inserting $\widetilde{\theta}(\rho)=\theta(\rho)+\xi(\rho)$ into the energy functional (\ref{energy1}) with uniaxial anisotropy I obtain the perturbation energy
\begin{equation}
E^{(2)}=\int^{\infty}_0\left[\left(\frac{d\xi}{d\rho}\right)^2+G(\rho)\xi^2\right]\rho d\rho
\label{PertEnergy}
\end{equation}
with
 \begin{equation}
G(\rho)=\cos(2\theta)\left(\frac{1}{\rho^2}+\beta\right)+\frac{h}{2}\cos\theta-\frac{\sin(2\theta)}{\rho}.
\end{equation}
Radial stability of the function $\theta(\rho)$ means that the functional $E^{(2)}$ is positive for all functions $\xi(\rho)$
which obey condition (\ref{constraint}). Correspondingly, the solutions will be unstable, if there is a function $\xi(\rho)$
that leads to a negative energy (\ref{PertEnergy}). Thus, the problem of radial stability is reduced to the solution of the spectral problem for functional (\ref{PertEnergy}). I solve it by expanding $\xi(\rho)$ in a Fourier series:
\begin{equation}
\xi(\rho)=\sum_{k=1}^{\infty}b_k\sin(k\theta(\rho))
\end{equation}
Inserting this into Eq. (\ref{PertEnergy}) reduces the perturbation energy  to the following quadratic form:
\begin{equation}
E^{(2)}=\sum_{l,k=1}^{\infty}A_{kl}b_kb_l
\end{equation}
where
\begin{equation}
A_{kl}=\int_0^{\infty}\left[kl\left(\frac{d\theta}{d\rho}\right)^2\cos(k\theta)\cos(l\theta)+G(\rho)\sin(k\theta)\sin(l\theta)\right]\rho d\rho.
\label{matrix}
\end{equation}
To establish radial stability of a solution, one has to determine the smallest eigenvalue $\lambda_1$ of the symmetric
matrix $\mathbf{A}$ (\ref{matrix}). If $\lambda_1$ is positive, the solution $\theta(\rho)$ is stable with respect to small radial perturbations. Otherwise it is unstable.

For our skyrmion solutions (Fig. \ref{all} (c)) the eigenmode $\xi_n(\rho)$ corresponding to the \textit{n}th eigenvalue ($\lambda_n$) consists mainly of the function $\sin(n\theta(\rho))$, with small admixtures of other harmonics. In particular, the eigenmode corresponding to the smallest eigenvalue $\lambda_1$ can be written as
\begin{equation}
\xi_1(\rho)=\sin(\theta(\rho))+\sum_{k=2}^{\infty}\varepsilon_k\sin(k\theta(\rho)),
\end{equation}
where $\varepsilon_k<<1$ in most cases. The function $\xi_1(\rho)$ describes a displacement of the vortex front. Thus the
lowest perturbation of the structure is connected with an expansion or compression of the profile. The calculations showed that in the region of existence of skyrmion solutions matrix (\ref{matrix}) has only positive eigenvalues, and thus these solutions are radially stable.

The smallest eigenvalues of large $\pi$-vortices (Fig. \ref{all} (e)) are always negative (Fig. \ref{all} (g)). These vortices are unstable either with respect to infinite expansion of the core, or to a contraction into a common skyrmion \cite{JMMM99}. 
The solutions of the spectral problem for vortices with nodes (Fig. \ref{all} (k), (o)) reveals their instability with respect to perturbations that remove the energetically disadvantageous humps.

\subsection{Transformation of hexagonal skyrmion lattice under influence of uniaxial anisotropy}

\begin{figure}
\centering
\includegraphics[width=10cm]{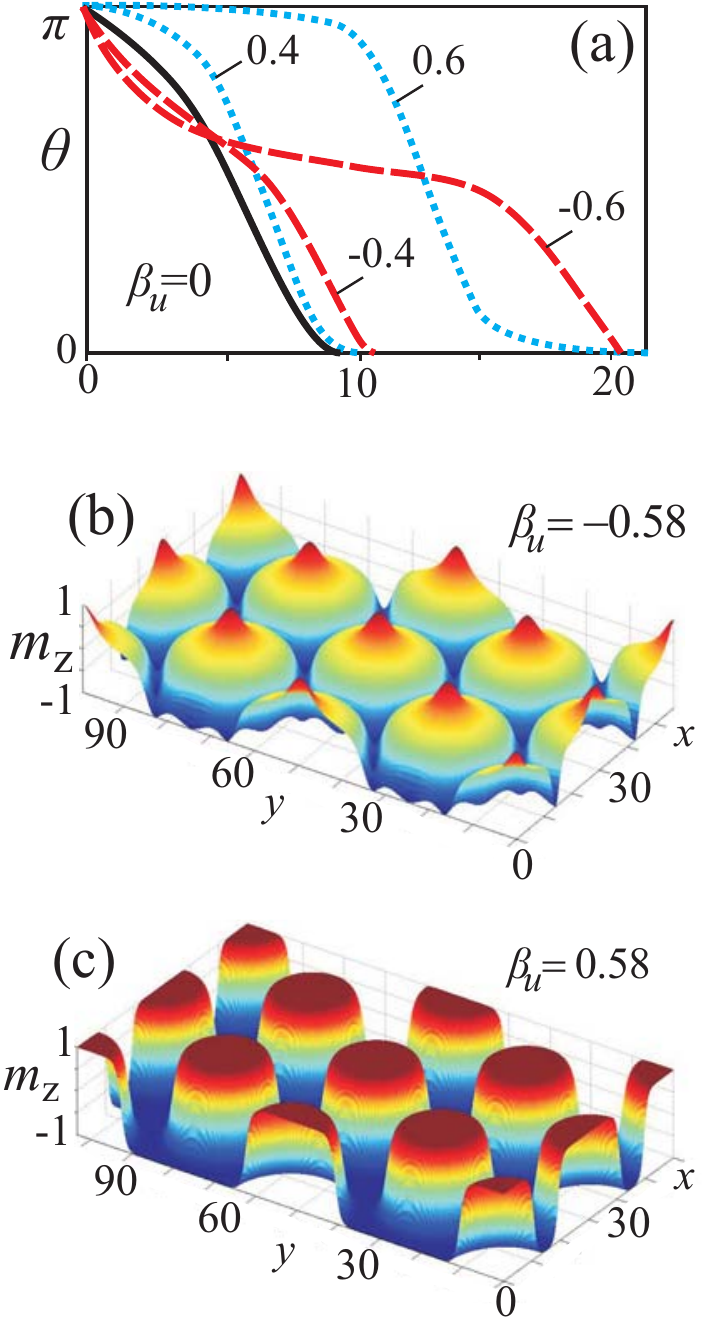}
\caption{
\label{plane}
Zero-field solutions of skyrmion lattices for easy-plane (b) and easy axis (c) uniaxial anisotropy shown as radial profiles $\theta(\rho)$ (a) and surface plots of $m_z$-component of the magnetization.
}
\end{figure} 

%The same field-transformation is observed for helimagnets with uniaxial anisotropy of easy-axis ($\beta>0$) or easy-plane ($\beta_u<0$) type (Fig.\ref{Fig5}).
%
For $h=0$ the influence of uniaxial anisotropy on the skyrmion structure is rather weak for $|\beta_u|<0.25$, but then it becomes pronounced up to the critical value $|\beta_{cr}|=\pi^2/16$ where a second-order phase transition into the homogeneous state occurs.

Easy-plane type anisotropy, $\beta_u<0$,   leads to the compression of the regions close to the skyrmion core and boundary (Fig. \ref{plane} (a), red dashed lines in (c) and surface plot (b)).
The easy-plane region of the lattice cell with $\theta(x,y)=\pi/2$ grows rapidly approaching critical value $\beta_{cr}$.

Easy axis anisotropy, $\beta_u>0$, on the contrary, expands the near-core region ($\theta=0$) and the skyrmionic outskirt with $\theta=\pi$ (Fig. \ref{plane} (c), blue dotted lines in (a)).

%%%%%%%%%%%%%%%%%%%%%%%%%%%%%%%%%%%%%%%%%%%%%%%%%%%%%%%%%%%%%%%%%%%%%%%%%%%%%%%%%%%%%%%%%%%%%%%%%%%%%%%%%%%%%%%%%%%%%%%%%%%%%%%
\subsection{Stabilization effect of uniaxial anisotropy on skyrmion states \label{StabilizationUA}}
%%%%%%%%%%%%%%%%%%%%%%%%%%%%%%%%%%%%%%%%%%%%%%%%%%%%%%%%%%%%%%%%%%%%%%%%%%%%%%%%%%%%%%%%%%%%%%%%%%%%%%%%%%%%%%%%%%%%%%%%%%%%%%

For $\beta_u$ = 0 as it was noted in section \ref{competitionSkHel}, the conical phase is the globally stable state from zero field to the saturation field ($0< h < 0.5$),\cite{Bak80} (Fig. \ref{PDUA} interval $(a-d)$). skyrmion lattices and helicoids are metastable solutions: skyrmions exist in the interval of magnetic fields from negative critical field with $H_H/H_D = \pi^2/16 = 0.3084$ (Fig. \ref{PDUA} point $b$) to positive critical field with $ H_S/H_D = 0.4006 $ (Fig. \ref{PDUA} point $c$); helicoids exist below the critical fields $H_H/H_D$ (Fig. \ref{PDUA} point $b$).

A sufficiently strong uniaxial anisotropy $\beta_u$ suppresses the conical states. Cones can exist only in the triangular region $(a-d-f)$: within the region $(a-d-B-A)$ they are thermodynamically stable and flip into the saturated state by the second-order phase transition at the critical line $(d-B)$ when the conical structure closes. Within the region $(a-A-B-f)$ the conical phase is a metastable state, at the lines $(a-A)$ and $(A-B)$ it discontinuously transforms into helicoids and skyrmions, respectively.

Modulated states with the propagation vectors perpendicular to the applied field (helicoids and skyrmion lattices) can exist even for larger values of uniaxial anisotropy (up to the point $e$): helicoids occupy the area $(a-b-D-e)$ with the line $(b-D-e)$ of unwinding into homogeneous state, while skyrmions have the existing area $(a-c-B-D-e)$ for positive fields and $(a-b-e)$ for negative fields. The skyrmion lattice is the only  modulated state that can exist in the triangular region $(B-E-D)$, and only helicoids exist in the region $(D-F-e)$ (see inset (i) of Fig. \ref{PDUA}).

By comparing energies of corresponding modulated phases (Fig. \ref{curves} (a)) one can conclude that skyrmions can be stabilized only with simultaneous influence of positive magnetic field and easy-axis uniaxial anisotropy. For easy-plane uniaxial anisotropy, the conical phase is always the global minimum of the system.

The skyrmion states are thermodynamically stable within a  curvilinear triangle $(A-B-D)$ with vertices $(A) = (0.0166, 0.1197)$, $(B) = (0.0907, 0.3187)$, and $(D) = (0.47, 0.05)$) (Fig. \ref{PDUA}). The phase diagram from present rigorous solutions very slightly differs from the calculations within the circular-cell approximation. Only point $(A) = (0.0125, 0.1079)$ has the slightly different coordinates [XI].

The solutions for helicoids exist within area ($a-A-D-e$) with the line $A-D$ of first-order phase transition into skyrmion lattice.

Thus, critical points $A$, $B$, $D$ separate the phase diagram (Fig. \ref{PDUA}) into three distinct regions with thermodynamical stability of each of considered phases.
Moreover, one can introduce different regimes of uniaxial anisotropy:

(I) In the low anisotropy regime ($\beta_u < \beta_{uA} = 0.0166$) only helical states are realized as thermodynamically stable phases: at the line $a-A$ helicoids transform into cones;
%The magnetization curve 
%(Inset (b)) includes the first-order
%transition between the helicoid and conical
%phases and the second-order transition 
%from the conical to the saturated state.

(II) For $\beta_{uA} < \beta_u < \beta_{uB} = 0.0907$ the skyrmion lattice becomes absolutely stable in a certain range of the applied field: at the line $A-D$, first, helicoid flips into skyrmion phase and then at the line $A-B$ skyrmions transforms into cones;
%
%The skyrmionic phase is separated from the helicoidal and conical states by first-order transitions.(Inset(c)).

(III) For $ \beta_{uB} < \beta_u < \beta_{uD} = 0.47$ there is only phase transition between helicoids and skyrmions at the line $A-D$;

(IV) Finally for ( $\beta_{uD} < \beta_u < \beta_{ue}=\beta_{cr}$) the helicoids are thermodynamically stable in the whole region where  modulated states exist.

\subsection{Magnetization curves}

\begin{figure}
\centering
\includegraphics[width=12cm]{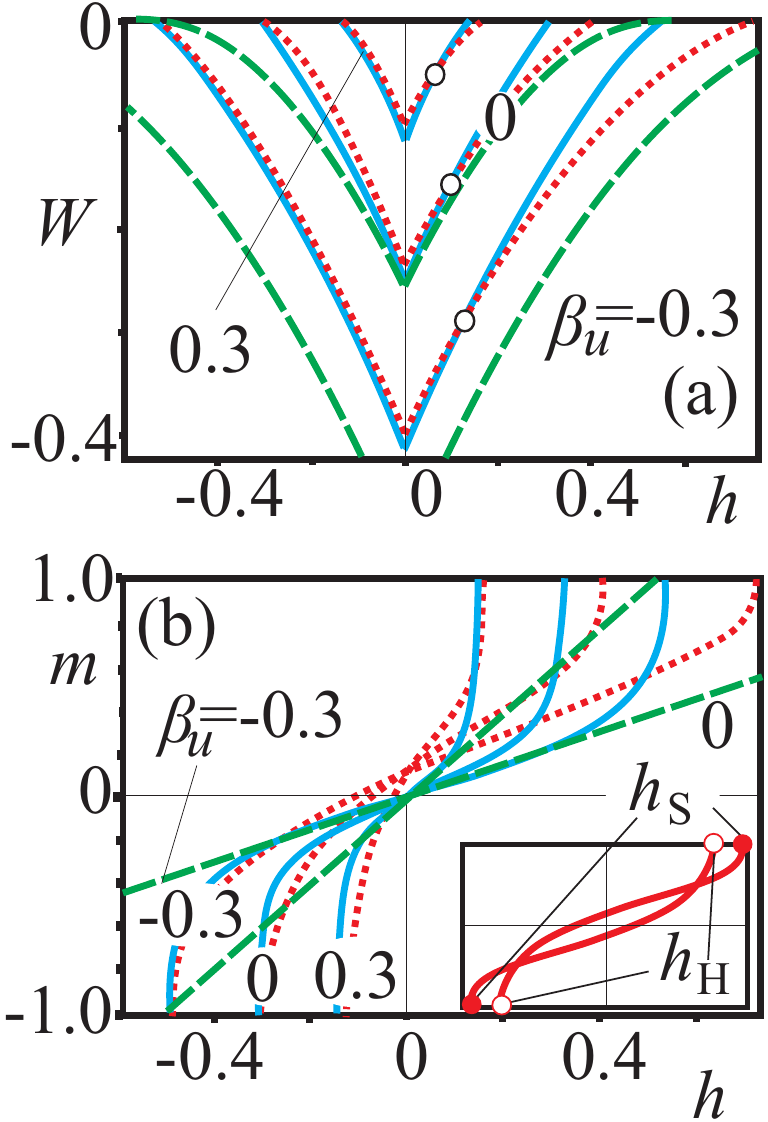}
\caption{
\label{curves}
(a) The energies of the skyrmion lattice (red dotted line), cone (dashed green lines) , and helicoid (solid blue lines) with respect to the homogeneous state plotted as functions of magnetic field for different values of uniaxial anisotropy $\beta$. For $\beta>0.25$ only helicoids and skyrmions can be realized. (b) Magnetization curves of all modulated states for different values of uniaxial distortions $\beta$: green dashed lines for conical phase, red dotted lines for skyrmion lattice, and blue solid lines for helicoid.   Inset shows hysteretic magnetization process for skyrmion lattice (see text for details). 
}
\end{figure}

Fig. \ref{curves} (b) shows the magnetization curves of all considered modulated structures for different values of uniaxial anisotropy.

For conical and helical phases, magnetization curves represent anhysteretic lines symmetric with respect to the field direction.
In the region of helicoid existence, the magnetization changes linearly almost for all values of the applied magnetic field (except drastic increase near the field of saturation), but remains smaller in comparison to the linear magnetization increase of the conical phase.

Magnetization curves for skyrmion lattices bear pronounced hysteretic character with the mutual conversion of two critical fields $h_H$ and $h_{S}$ (points $b$ and $c$ in the Fig. \ref{PDUA}).
For instance, in large negative magnetic fields far beyond  the disappearence of the honeycomb skyrmion texture, isolated skyrmions with the magnetization along $z$ axis  (with $\theta(0)=0$) can be nucleated.
These skyrmions condense into a lattice  in accordance with the physical principles described previously.
In positive magnetic field this skyrmion lattice becomes a honeycomb structure and transforms into the homogeneous state.
Thus, exemplified magnetization curve is composed from three subloops with remanent magnetization in zero magnetic field (Fig. \ref{curves} (b), inset).

%%%%%%%%%%%%%%%%%%%%%%%%%%%%%%%%%%%%%%%%%%%%%%%%%%%%%%%%%%%%%%%%%%%%%%%%%%%%%%%%%%%%%%%%%%%%%%%%%%%%%%%%%%%%%%%%%%%%%%%%%%%%
%old
%\section{Stabilization effect of exchange anisotropy on skyrmion states. Phase diagram of states \label{exchangeAnisBHPhase}}

%new
\section[Stabilization effect of exchange anisotropy on skyrmion states]{Stabilization effect of exchange anisotropy on skyrmion states. Phase diagram of states \label{exchangeAnisBHPhase}}
%%%%%%%%%%%%%%%%%%%%%%%%%%%%%%%%%%%%%%%%%%%%%%%%%%%%%%%%%%%%%%%%%%%%%%%%%%%%%%%%%%%%%%%%%%%%%%%%%%%%%%%%%%%%%%%%%%%%%%%%%%%%

\begin{figure}
\centering
\includegraphics[width=14cm]{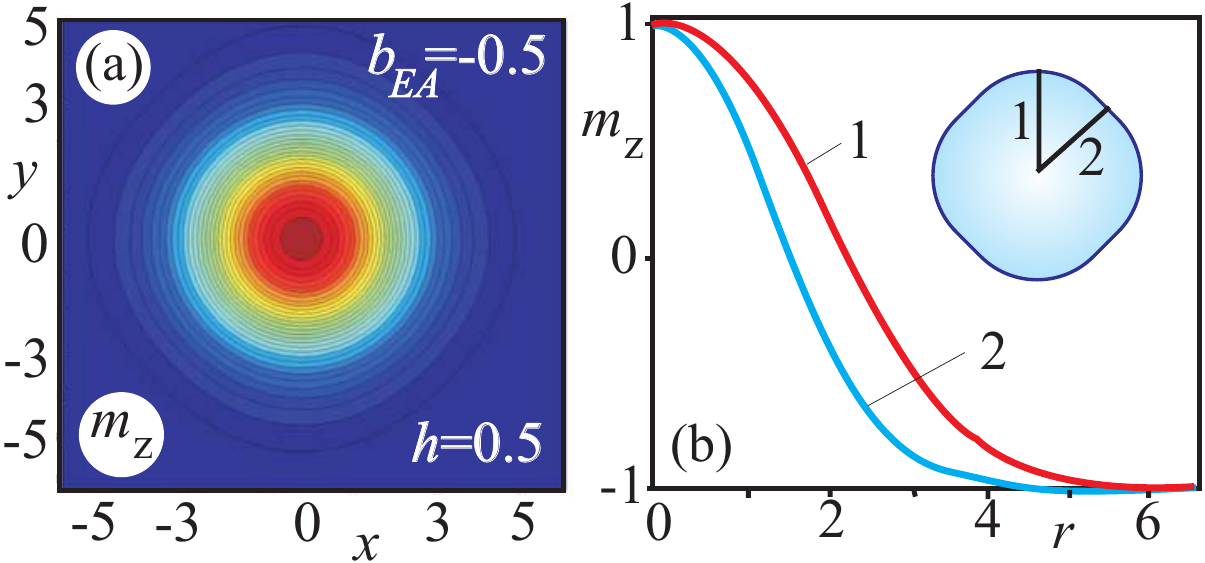}
\caption{
\label{EAIsolated}
(a) The contour plot for $m_z$-component of the magnetization in the isolated skyrmion for $h=0.5,\,b_{EA}=-0.5$. The isolated skyrmion aquires a square shape under influence of cubic exchange anisotropy. (b) Profiles $m_z$ plotted in dependence on the spatial coordinate $r$ in two cross-sections of the isolated skyrmion shown in inset.
}
\end{figure} 

From the numerical investigation of Eq. (\ref{DMdens1}) with an exchange anisotropy (Eq. (\ref{additional})) I now show that skyrmion textures can be stabilized over conical phases even for relatively small values of $b_{EA}$.

The exchange anisotropy on the contrary to uniaxial anisotropy (section \ref{StabilizationUA}),  does not affect the one-dimensional conical phase, but deforms significantly the skyrmion states. With 
\begin{equation}
b_{EA}<0 
\end{equation}
it supplies the skyrmions with additional negative energy density. For some critical value of $b_{EA}^{(crit)}$ (I will distinguish between two values: $b_{EA}^{(crit1)}$ is the critical value of exchange anisotropy when the skyrmion lattice can be stabilized in an applied magnetic field; $b_{EA}^{(crit2)}$ is the value of EA when even in zero field the skyrmion lattice is the global minimum of the system; see phase diagram in Fig. \ref{EA}) the amount of the additional energy is sufficient to make the skyrmions the global minimum of the system. In the following I will consider exactly this mentioned situation. The cones and skyrmions will be considered in the field applied along $<001>$ crystallographic direction. 

\begin{figure}
\centering
\includegraphics[width=15cm]{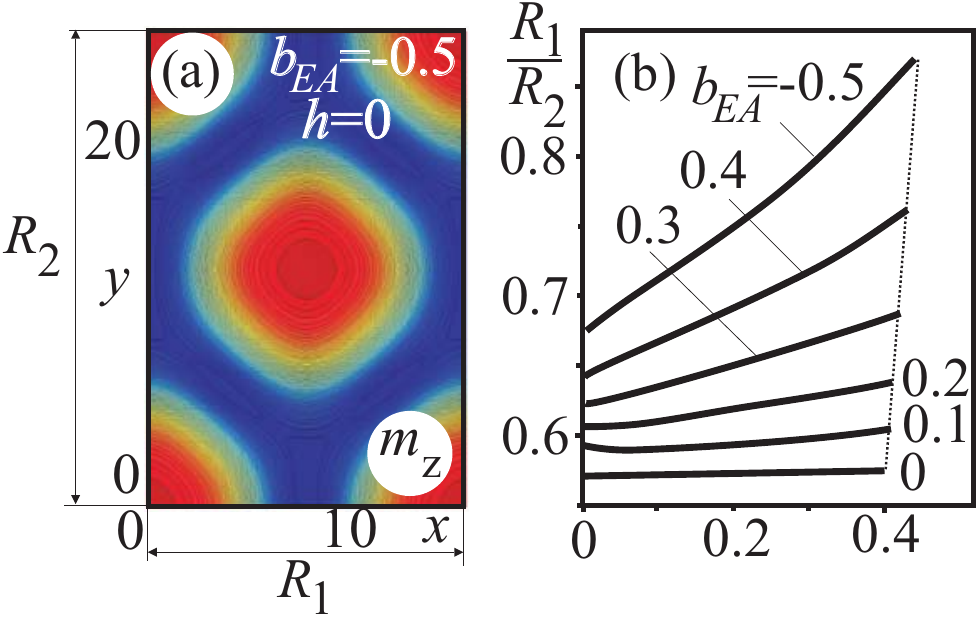}
\caption{
\label{EALattice}
(a) The contour plot for $m_z$-component of the magnetization in the skyrmion lattice for $h=0,\,b_{EA}=-0.5$. (b) The dependences of the ratios $R_1/R_2$ on the applied magnetic field for different values of the exchange anisotropy $b_{EA}$. The perfect hexagonal lattice corresponds to the ratio $R_1/R_2=0.5773$ (as for example for $b_{EA}=0$). The square lattice is characterized by $R_1/R_2$ and can be realized for large values of $b_{EA}$ in the field (as for example for $b_{EA}=-0.5$). 
%The isolated skyrmion aquires the square shape under action of exchange anisotropy. (b) Profiles $m_z$ plotted in dependence on the spatial coordinate $r$ in two cross-sections of the isolated skyrmion shown in inset.
}
\end{figure}

\begin{figure}
\centering
\includegraphics[width=15cm]{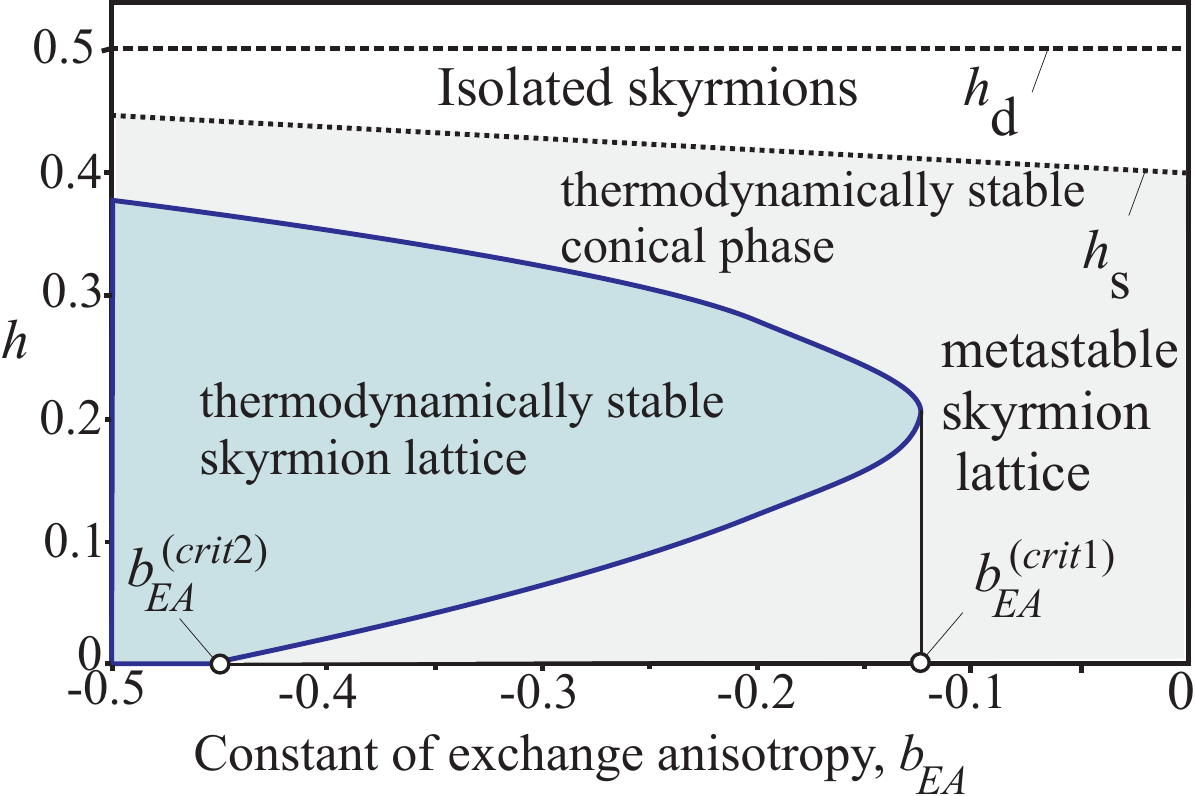}
\caption{
\label{EA}
The phase diagram for conical and skyrmion states in the plane $(b_{EA},h)$. For $b_{EA}<b_{EA}^{(crit1)}$ the skyrmions are energetically favoured over cones in the interval of the applied magnetic field. For $b_{EA}<b_{EA}^{(crit2)}$ the skyrmions are energetically advantageous over cones even for $h=0$. %of states in the presence of exchange anisotropy contains the pocket with thermodynamically stable skyrmion lattice. 
}
\end{figure}

Isolated skyrmions in the presence of exchange anisotropy assume a special character of the magnetization distribution: the double-twisted core retains its circular symmetry, but the boundary region is distorted into a square shape.
%
%These low-symmetry solutions may be induced by the rectangular constraint of numerical method. 
It is clear that the numerical method in those cases, by the restriction to rectangular unit cells, is unable to reproduce the correct energy minimum if the lattice cell undergoes a distortion into parallelogram shape. I neglect this effect in the numerical calculations, because it is small. Thus, the solutions in Figs. \ref{EAIsolated}, \ref{EALattice} and phase diagram in Fig. \ref{EA} have to be considered as semi- quantitative approximations.  
 In Fig. \ref{EAIsolated} (a) such a square-like distribution of the magnetization is shown by contour plot of $z$-component of the magnetization for $b_{EA}=-0.5,\,h=0.5$. In Fig. \ref{EAIsolated} (b) the profiles $m_z=m_z(r)$  are clearly different along the two cuts of the isolated skyrmions (see inset in Fig. \ref{EAIsolated} (b)). 

When isolated skyrmions condense into the lattice with the decreasing magnetic field, they are subject to the influence of two opposite mechanisms: from one side, they tend to form the densely packed lattice, from the other side  however,  the skyrmions try to keep this square symmetry imposed by the exchange anisotropy.  
As a result, the lattice of skyrmions is highly distorted. Rectangular lattices of this type have been calculated and relaxed according to the principles of section \ref{NumericalRecipes}. %Therefore, artificial effects of the chosen elementary cell may %Such a lattice can be characterized as  the intermediate state between square and hexagonal packing of skyrmions (Fig. \ref{EALattice} (a)). 

In Fig. \ref{EALattice} (b) I plotted the ratio $R_1/R_2$ ($R_1$ and $R_2$ are the sizes of the elementary cell along two perpendicular directions $x$ and $y$ shown in Fig. \ref{EALattice} (a)) versus magnetic field for different values of the constant $b_{EA}$. As for perfect hexagonal lattice
\begin{equation}
 \frac{R_1}{R_2}=0.5773 
\end{equation}
(Fig. \ref{EALattice} (b) stright line), the skyrmion lattice in the applied magnetic field shows the  tendency of the deformation toward  the square lattice with 
\begin{equation}
\frac{R_1}{R_2}=1
\end{equation}
(especially for large values of $b_{EA}$, see the last curve in Fig. \ref{EALattice} (b)). With increasing constant of exchange anisotropy the saturation field of the skyrmion lattice (that is the field when the lattice releases the free isolated skyrmions) also increases (dotted line in Fig. \ref{EALattice} (b) and the line $h_S$ in Fig. \ref{EA}). 

%Strong exchange anisotropy leads to the redistribution of the magnetization in two-dimensional skyrmion states making the isolated skyrmions of square form (Fig. \ref{EA} (c)). Nonetheless, the double-twisted core retains its circular symmetry.

In Fig. \ref{EA} I plotted the phase diagram for cones and skyrmions depending on the constant of exchange anisotropy $b_{EA}$. For 
\begin{equation}
b_{EA}<b_{EA}^{(crit1)}=-0.13
\end{equation}
the spacious pocket shows up in the applied magnetic field  with the thermodynamically stable skyrmions. For
\begin{equation}
b_{EA}<b_{EA}^{(crit2)}=-0.45
\end{equation}
even in zero magnetic field the skyrmions have the lowest energy of all modulated phases considered in this chapter. 

%%%%%%%%%%%%%%%%%%%%%%%%%%%%%%%%%%%%%%%%%%%%%%%%%%%%%%%%%%%%%%%%%%%%%%%%%%%%%%%%%%%%%%%%%%%%%%%%%%%%%%%%%%%%%%%%%%%%%%%%%%%%%%
\section{Stabilization of skyrmion textures by cubic anisotropy  \label{StabilizationCubic}}
%%%%%%%%%%%%%%%%%%%%%%%%%%%%%%%%%%%%%%%%%%%%%%%%%%%%%%%%%%%%%%%%%%%%%%%%%%%%%%%%%%%%%%%%%%%%%%%%%%%%%%%%%%%%%%%%%%%%%%%%%%%%%%%

\begin{figure}
\centering
\includegraphics[width=18cm]{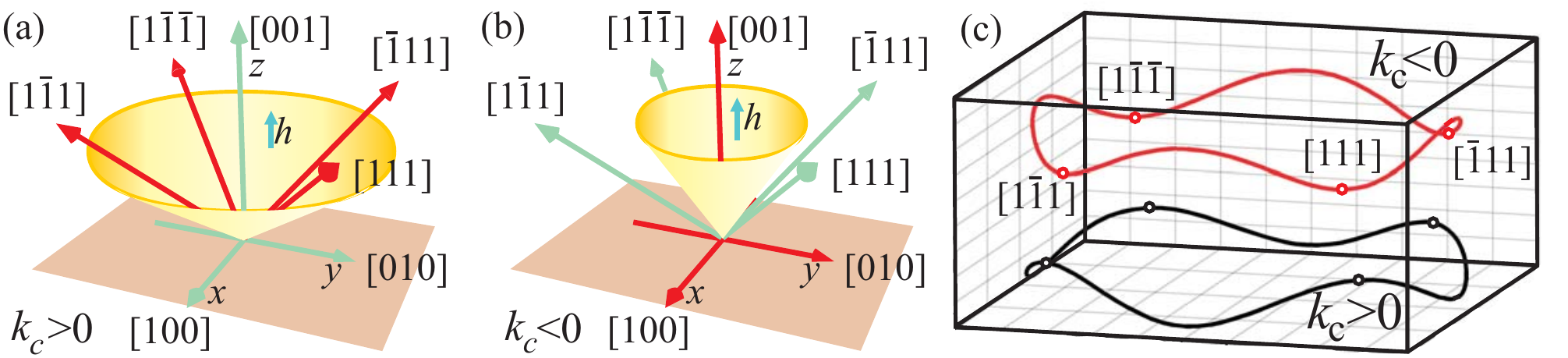}
\caption{
\label{ConusSketch} The sketches of the magnetization rotation in the conical phase in the presence of cubic anisotropy with $k_c>0$ (a) and $k_c<0$ (b) are shown together with the schematic representation of the magnetization traces in a space (c). Depending on the orientation of the cone propagation direction ($z$) with respect to the easy (green arrows) and hard (red arrows) anisotropy axes and the applied magnetic field ($h||[001]$) the energy density of the cone can be increased or  reduced (see text for details). 
}
\end{figure}

In the present section I explicitly refer to the cubic anisotropy that can favour skyrmions over conical phases for suitable  orientation of the applied magnetic field and skyrmion axes (as well as propagation direction of the cones and helicoids) with respect to the easy anisotropy axes. 
Results of this section give straightforward recommendations how to make skyrmionic spin textures the thermodynamically stable state of the system. %I give the qualitative recipes supported by the quantitative calculations how to achieve this goal.
 Calculations for all modulated phases have been obtained rigorously using methods of section \ref{NumericalRecipes}.

The detailed analysis of chiral modulations in the presence of cubic anisotropy offers also practical recommendations to experimentalists under which circumstances to look for  stable skyrmion states in bulk cubic helimagnets. % under the influence of uniaxial stresses 

\subsection{Distorted conical phase in the presence of cubic anisotropy \label{ConeCubicDistorted}}

%In the present section I consider the question of the suppression of the conical phase for the appropriate choice of the magnetic field direction for the given 

As it was noted in section \ref{DistortionsBHphase}, uniaxial anisotropy along the propagation direction suppresses  the conical phase for the values of the anisotropy coefficient $\beta_{u}$ much smaller than it does for the skyrmion and helical phases (Fig. \ref{PDUA}). After cones have been suppressed, skyrmions may become the thermodynamically stable state of the system in the applied magnetic field [XI].

Uniaxial anisotropy does not affect the ideal single-harmonic type of the magnetization  rotation in the cone state, but just leads to the gradual closing of the cone.  Cubic anisotropy, on the contrary, violates the ideal spin configuration of the conical phases:  %The magnetization in the presence of cubic anisotropy rotates along the curvilinear surface of the distorted cone; 
the magnetization deviates from the ideal conical surface trying to embrace the easy axes and to avoid the hard directions (Fig. \ref{ConusSketch}).

Depending on the mutual  arrangement of easy anisotropy axes and propagation direction of the cone the cubic anisotropy can either increase the energy of this phase or decrease it. 
Therefore, the rotation of the magnetization in the conical phase must be in tune with a complex landscape of the cubic anisotropy with various global and local minima.

The homogeneous states in a system with the cubic anisotropy in the applied magnetic field are described by the behaviour of the following energy functional:
\begin{equation}
\Phi(\Theta,\Psi)=k_c(m_x^2m_y^2+mx^2m_z^2+m_y^2m_z^2)-\mathbf{h}\cdot\mathbf{m},\,\mathbf{m}=(\sin\Theta\cos\Psi,\sin\Theta\sin\Psi,\cos\Theta)
\label{CubicHomo}
\end{equation}
where angles $\Theta$ and $\Psi$ define the orientation of the magnetization in the spherical coordinate system. I introduce angles $\Theta$ and $\Psi$ for the magnetization in the homogeneous state to distinguish them from the angles $\theta$ and $\psi$ characterizing the distribution of the magnetization in skyrmion states.

Depending on the values of the coefficient $k_c$ and the components of magnetic fields different spatially homogeneous phases can be realized in the system. %In a general case the obtained phase diagram is three-dimensional surface 
The basic principles how to handle such a type of functionals and to define the manifold of extrema in the applied magnetic field are given explicitly in chapter 2. Here, I will refer to the results of that chapter, while dealing with the modulated phases.

In the forthcoming calculations, the magnetic field $h$ is considered to be applied along $[001]$ crystallographic direction. The cases with $k_c>0$ and $k_c<0$ are discussed separately. Certainly, these two examples cannot address the problem of skyrmion stabilization over cones for random orientation of the magnetic field. But they represent the auxiliary cases consideration of which is instructive in the following. % and helps to chose intuitively the  very But from the point of suppression of the conical phase they allow to 

\vspace{3mm}
\textit{A. Solutions for conical phase with $k_c>0,\,h||[001]$.}
\vspace{3mm}

\begin{figure}
\centering
\includegraphics[width=18cm]{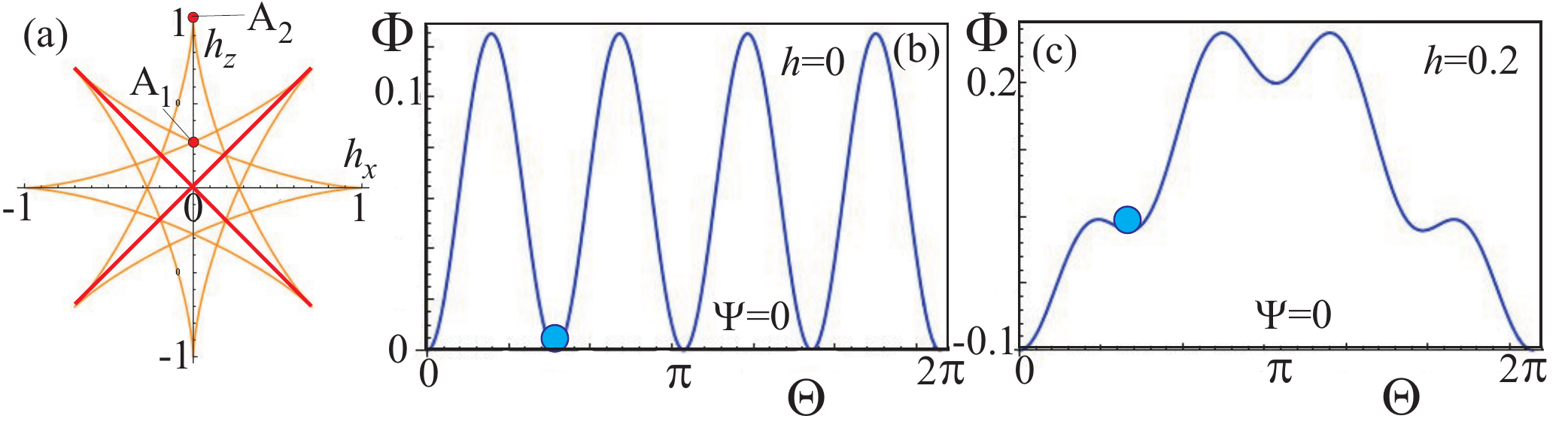}
\caption{
\label{Homo100v1} $k_c>0,\,h||[001]$. (a) The phase diagram in internal field components obtained according to the conventions  of chapter 2. The orange lines are the lability lines bounding the region with plenty of local minima of energy functional (\ref{CubicHomo}). On the red lines two different angular phases have the same energy density, i.e. these are lines of the first-order phase transitions. (b), (c) show the energy density $\Phi=\Phi(\Theta)$ (\ref{CubicHomo}) for fixed value of the azimuthal angle $\Psi=0$ and different values of the applied magnetic field $h$.
%
%The energy density $\Phi$ (\ref{CubicHomo}) plotted as two-dimensional surfaces in dependence on the angles $\Theta$ and $\Psi$ (b) and  three-dimensional polar plots $A+B \Phi$ (c) where $A$ and $B$ are suitable scaling factors.   The cuts of panels (b) and (c) with $\Psi=0$ are shown in panels (a) and (d), correspondingly. The first plot in the panel (a) represents the astroid obtained according to the rules of chapter \ref{astroids}. Path of the rotating magnetization in the conical phase is imaged by the yellow lines with yellow circles being the minima of Eq. (\ref{CubicHomo}). In the applied magnetic field the states of the magnetization in the cone correspond only to the local minima. 
}
\end{figure}

\begin{figure}
\centering
\includegraphics[width=18cm]{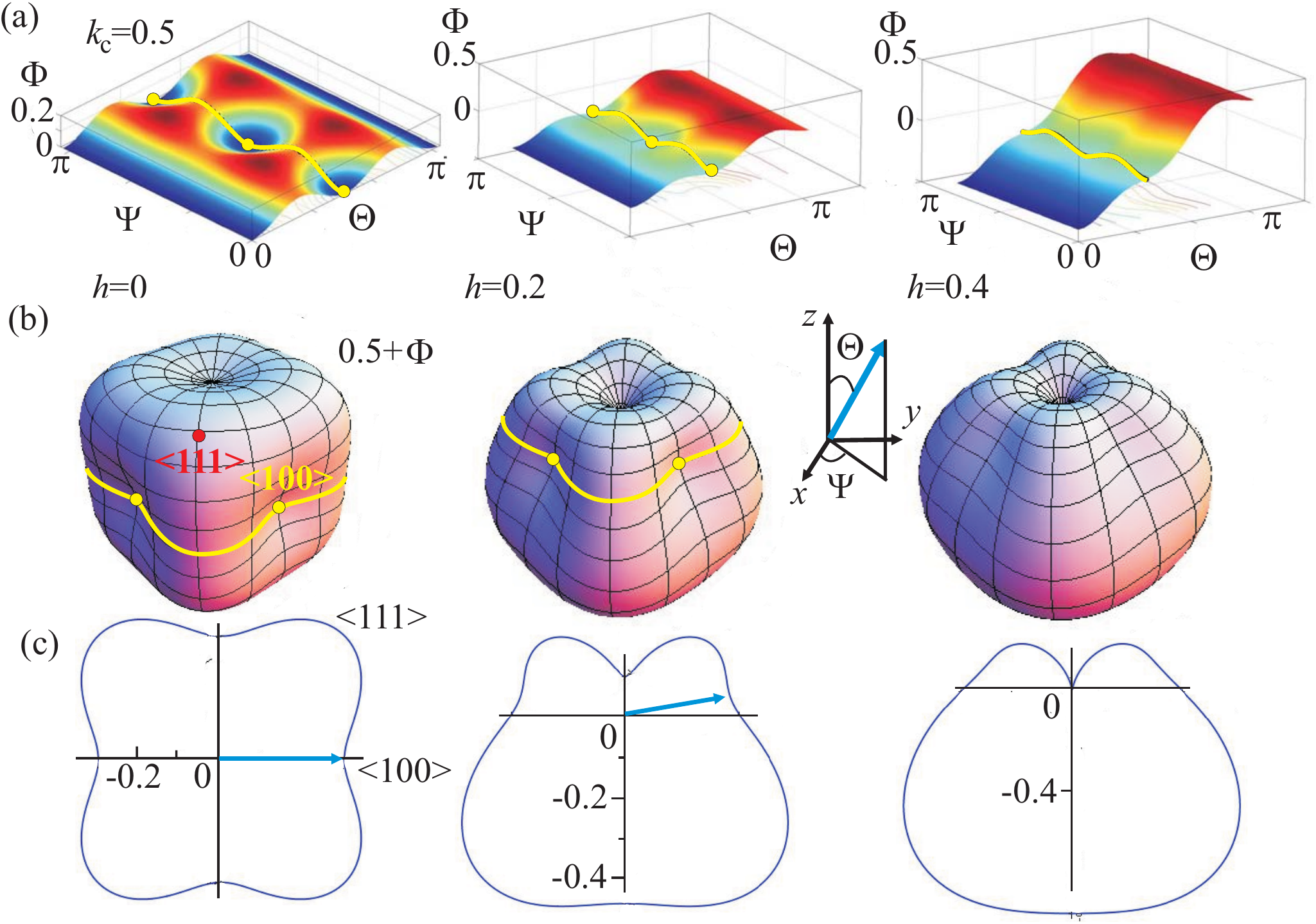}
\caption{
\label{Homo100v2} $k_c>0,\,h||[001]$. The energy density $\Phi$ (\ref{CubicHomo}) plotted as two-dimensional surfaces in dependence on the angles $\Theta$ and $\Psi$ (a) and  three-dimensional polar plots $A+B \Phi$ (b) where $A$ and $B$ are suitable scaling factors. In the present case $A=0.5,\,B=1$.   The cuts of graphs in panel (b) with $\Psi=0$ are shown in panel (c). Path of the rotating magnetization in the conical phase is imaged by the yellow lines with yellow circles being the minima of Eq. (\ref{CubicHomo}). In the applied magnetic field the states of the magnetization in the cone correspond only to the local minima. 
}
\end{figure}

In Fig. \ref{Homo100v2} I have plotted the energy density $\Phi$ of Eq. (\ref{CubicHomo}) for some values of the applied magnetic field as surfaces in dependence on the angles $\Theta$ and $\Psi$ (Fig. \ref{Homo100v2} panel (a)), three-dimensional polar plots  (Fig. \ref{Homo100v2} panel (b)), and their two-dimensional cuts  (Fig. \ref{Homo100v2} panel (c)). Two-dimensional cuts of energy surfaces of panel Fig. \ref{Homo100v2} (a) are plotted in Fig. \ref{Homo100v1} (b), (c) as $\Phi=\Phi(\Theta)$ for $\Psi=0$ and $h=0,\,0.2$. Fig. \ref{Homo100v1} (a) represents the astroid plotted according to the conventions of the chapter 2. The lability (orange) lines have been obtained by solving the system of equation, $\Phi_{\Theta}=0,\,\Phi_{\Theta\Theta}=0$. The red lines are the lines of first-order phase transitions - the solutions with different orientations of the magnetization (global minima of the functional (\ref{CubicHomo})) have equal energies along these lines. The present astroid corresponds to the two-dimensional case with $\Psi=0$. The field applied along [001] crystallographic direction has only $z$-component, i.e. in the following $h=h_z$. Although, any other directions of the applied magnetic  field can be considered. 
From all these graphs the comprehensive analysis of the magnetization rotation in the conical phase can be carried out.

For $k_c>0$ and $h=0$ the equilibrium states of the magnetization correspond to the easy axes of cubic anisotropy oriented along $<001>$ crystallographic directions (green arrows in Fig. \ref{ConusSketch} (a) and blue circle in FIg. \ref{Homo100v1} (b) marking the global minimum of $\Phi(\Theta,\Psi)$). Maxima of the functional (\ref{CubicHomo}) are $<111>$ directions - the hard axes of cubic anisotropy (red arrows in Fig. \ref{ConusSketch} (a)). The equilibrium states of the magnetization in the homogeneous state have the orientations with $\Theta=k\pi/2,\,k=0,1,2...$. In Fig. \ref{Homo100v2} (a), (b) these minima are marked by the yellow circles. In the panel (c) of Fig. \ref{Homo100v2} the orientations of the magnetization are shown by the blue arrows.

For $h=0$ the magnetization in the conical phase rotates in the plane (001) (brown plane in Fig. \ref{ConusSketch} (a)). While rotating, the magnetization leaves one  energy minimum corresponding to $<001>$ directions and, rotating  through the saddle point between hard axes $<111>$, gets into another energy minimum with $<001>$ direction. The trace of the magnetization in the conical phase is shown by thick yellow line in Fig. \ref{Homo100v2} (a),(b).

In the applied magnetic field $h||[001]$ the energy functional (\ref{CubicHomo}) has a global minimum corresponding to the state along the field and local minima for the states of the magnetization deflected from the plane (001). These minima disappear in the point $A_1$ of the astroid (Fig. \ref{Homo100v1} (a)). The local minimum for the magnetization pointing against the field, $\Theta=\pi$, vanishes in the point $A_2$. 

\begin{figure}
\centering
\includegraphics[width=18cm]{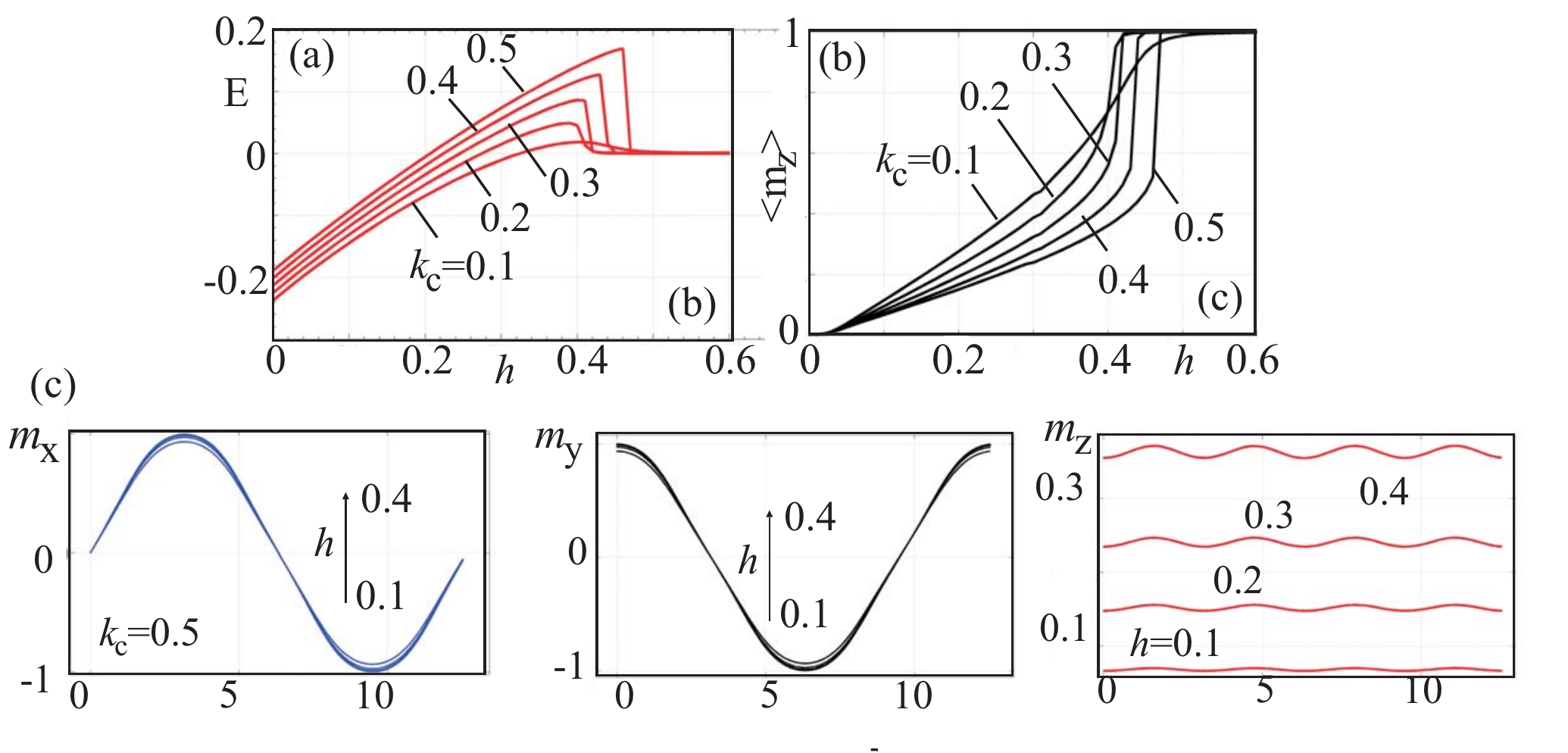}
\caption{
\label{Conus100} $k_c>0,\,h||[001]$. Solutions for the conical phase: (a) energy density of the conical phase with respect to the homogeneous state in dependence on the applied magnetic field $h$ for different values $k_c$ of the cubic anisotropy; (b) magnetization curves showing the component of the magnetization along the field, $m_z$. After reaching the hard cubic axes $<111>$ the magnetization jumps into the homogeneous state. (c) the components of the magnetization plotted along the propagation direction for different values of the applied magnetic field.
}
\end{figure}

Rotation of the magnetization in the conical phase around the field sweeps the metastable states of Eq. (\ref{CubicHomo}) for $h<h(\mathrm{A}_1)$ and saddle points for $h>h(\mathrm{A}_1)$. The conical phase becomes the metastable solution in comparison with the skyrmion lattice (see the phase diagram of states in Fig. \ref{PDCubic} (a)). In a critical field the conical phase by a first-order phase transition flips into the homogeneous state.  In Fig. \ref{Conus100} (a) I plotted the energy density of conical spiral with respect to the homogeneous state. The line $\varepsilon$ in Fig. \ref{PDCubic} (a)  signifies the first-order phase transition when the energy difference of two phases is zero. The fields corresponding to  jump of metastable conical phase with positive energy into homogeneous state are not shown on the phase diagram. 

The magnetization curves for conical phase are depicted in Fig. \ref{Conus100} (b). From the behavior of all components of the magnetization (Fig. \ref{Conus100} (c)) in the applied magnetic field  it can be concluded  that the cones become more distorted in the high magnetic fields - the rotating magnetization tries to avoid the hard anisotropy axes $<111>$.

\vspace{3mm}
\textit{B. Solutions for conical phase with $k_c<0,\,h||[001]$.}
\vspace{3mm}

\begin{figure}
\centering
\includegraphics[width=18cm]{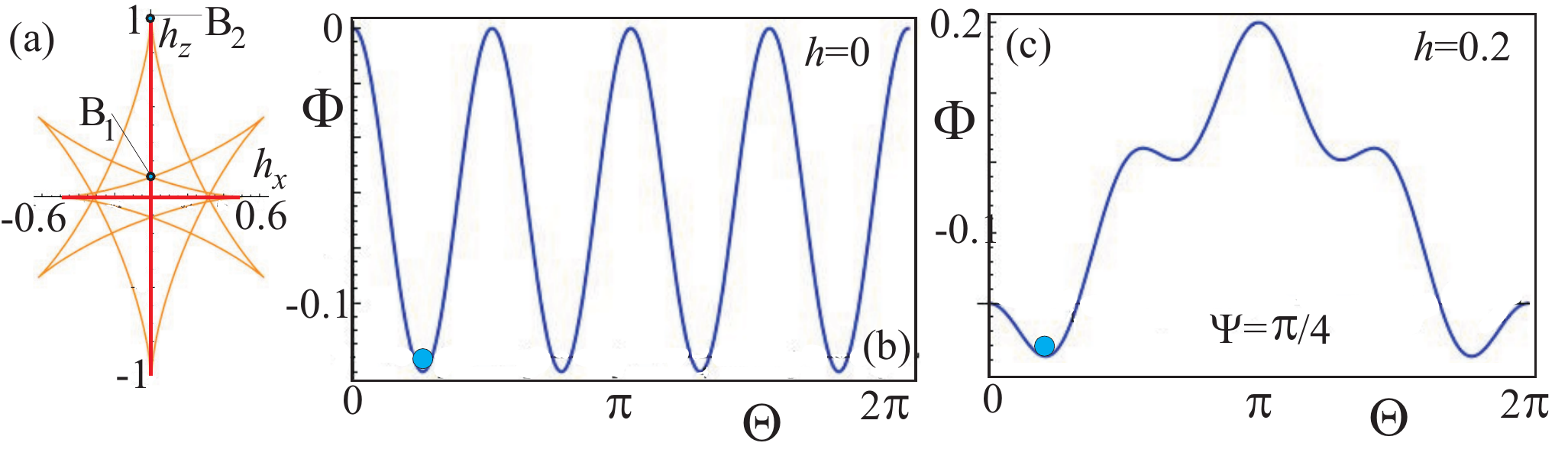}
\caption{
\label{Homo111v1} $k_c<0,\,h||[001]$.  (a) Astroid obtained according to the rules of chapter 2. The red lines are the lines of the first-order phase transitions. The orange lines (lability lines) bound the region with a multitude of local minima. Magnetic field is applied along $z$-direction, i.e. $h=h_z$. (b), (c) functional $\Phi=\Phi(\Theta)$ plotted with fixed azimuthal angle $\Psi=\pi/4$. The homogeneous states (global minima of Eq. (\ref{CubicHomo})) swept by the rotating magnetization in the conical state are marked by blue circles in (b) and (c). %In the applied magnetic field the magnetization in the cones rotates to sweep the  (blue circles in (a)). %the states of the magnetization in the cone correspond only to the local minima. 
}
\end{figure}

\begin{figure}
\centering
\includegraphics[width=18cm]{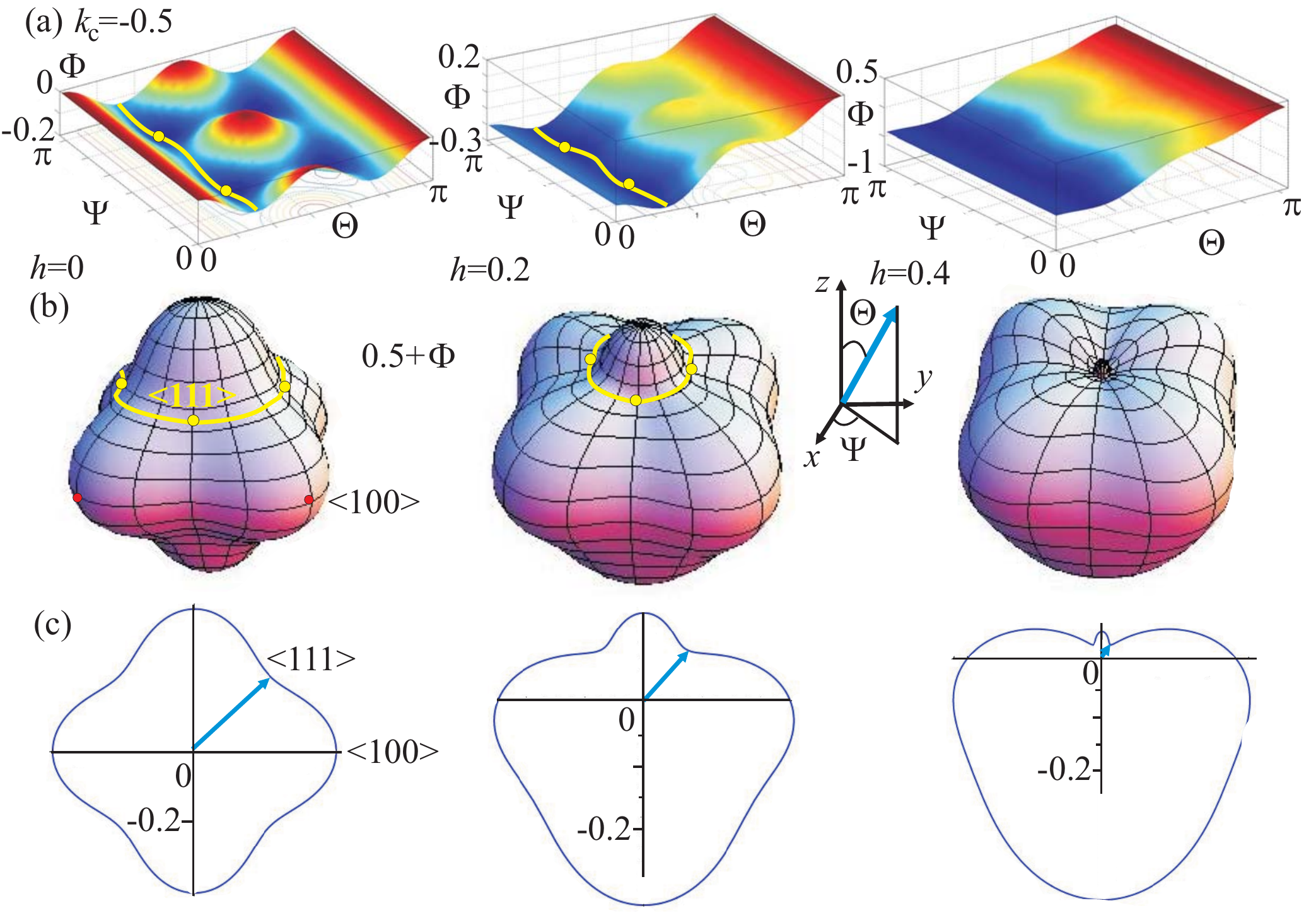}
\caption{
\label{Homo111v2} $k_c<0,\,h||[001]$. The energy density (\ref{CubicHomo}) plotted as two-dimensional surfaces in dependence on the angles $\Theta$ and $\Psi$ (a) and  three-dimensional polar plots (b).   The cuts of panel (a) with $\Psi=\pi/4$ are shown in panel (c).  Path of the rotating magnetization in the conical phase is shown by the yellow lines with yellow circles being the minima of Eq. (\ref{CubicHomo}). In the applied magnetic field the magnetization in the cones rotates to sweep the global minima of Eq. (\ref{CubicHomo}). %the states of the magnetization in the cone correspond only to the local minima. 
}
\end{figure}

For $k_c<0$ and $h=0$ the equilibrium states of the energy functional (\ref{CubicHomo}) correspond to the easy axes of cubic anisotropy oriented along $<111>$ crystallographic directions (green arrows in Fig. \ref{ConusSketch} (b)). The angle of these easy direction with respect to the field $h||[001]$ is $70.5^{\circ}$. Maxima of the functional (\ref{CubicHomo}) are $<001>$ directions - the hard axes of cubic anisotropy (red arrows in Fig. \ref{ConusSketch} (b)). %The equilibrium states of the magnetization in the homogeneous state have the orientations with $\Theta=k\pi/2,\,k=0,1,2...$. In Fig. \ref{Homo100} (b), (c) these minima are marked by the yellow circles. In the panel (d) the orientation of the magnetization is shown by the blue arrow. 

The magnetization in the conical phase performs such a rotation to sweep the easy directions $<111>$ (Fig. \ref{Homo111v2} (a), (b)). Even in zero field the conical phase has non-zero component of the magnetization along the field (Fig.  \ref{ConusSketch} (b)). 

In the applied magnetic field the global minima of Eq. (\ref{CubicHomo}) gradually approach the field direction. For $h>h(\mathrm{B}_2)$ (see the astroid in Fig. \ref{Homo111v1} (a)) only the states with the magnetization along the field can exist. In the intermediate point $\mathrm{B}_1$ the local minima of those $<111>$ axes dissappear that make angles more than $90^{\circ}$ with the field, i.e. the easy axes under the brown plane in Fig. \ref{ConusSketch} (b). 

\begin{figure}
\centering
\includegraphics[width=18cm]{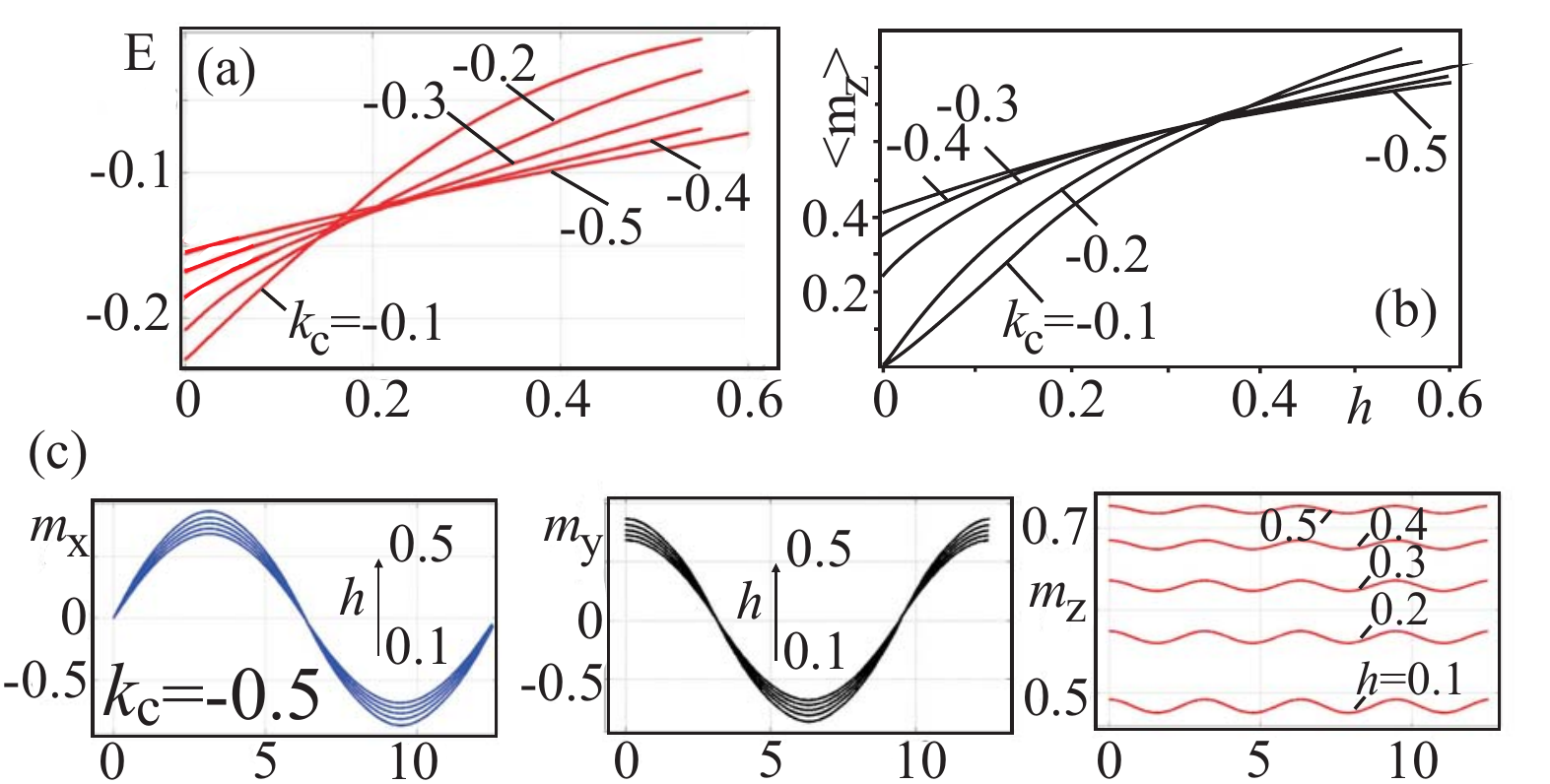}
\caption{
\label{Conus111} $k_c<0,\,h||[001]$. Solutions for the conical phase: (a) energy density of the conical phase with respect to the homogeneous state in dependence on the applied magnetic field $h$ for different values $k_c$ of the cubic anisotropy; by the second-order phase transitions cones transform into the homogeneous state with the magnetization along the field. (b) Magnetization curves showing the component of the magnetization along the field, $m_z$. Even in zero magnetic field the magnetization has non-zero $z$- components.  (c) The components of the magnetization plotted along the propagation direction for different values of the applied magnetic field.
}
\end{figure}

In Fig. \ref{Conus111} (a) I plotted the energy density of the conical phase with respect to the homogeneous state for different values of cubic anisotropy $k_c$ in dependence on the field $h$. 
%The cones transform into the homogeneous state by the second-order phase transition. 
The magnetization curves for conical phase (Fig. \ref{Conus111} (b)) display the non-zero $m_z$-component of the magnetization in zero field as described earlier. 
 From the behavior of all components of the magnetization (Fig. \ref{Conus111} (c)) in the applied magnetic field  it can be concluded  that the cones undergo the greatest deformation of their $m_z$-component in zero magnetic field. %With increasing the field high magnetic fields - the rotating magnetization tries to avoid the hard anisotropy axes <111>. 

\subsection{Distorted helicoid in the presence of cubic anisotropy \label{HelicoidDistortedCubic}}

In the section \ref{helicoidsBH} I introduced the following definition for the helicoid: helicoid is a spiral state with the propagation direction perpendicular to the applied magnetic field.  (Fig. \ref{spirals1} (a)). In the presence of cubic anisotropy such a definition  must be generalized to include arbitrary orientations of the applied magnetic field and propagation directions of helicoids. 

Both the spin arrangements and the corresponding propagation directions in the helicoid are found to be extremely sensitive to the orientation and strength of the applied magnetic field as well as to the sign and value of the anisotropic constant $k_c$. Perturbations of the uniform rotation for the helicoid are related to the shape of potential profiles of homogeneous states (Figs. \ref{Homo100v2}, \ref{Homo111v1}, \ref{Homo111v2}). 

In a general case, there is a multitude of solutions for helicoids characterized by  variable directions of propagation vectors. In the following, helicoids are defined as states with vectors $\mathbf{k}$ oblique to the field. 
%
%I will look for different orientations of the wave vectors $\mathbf{k}$  with respect to the field and chose the helical state with the lowest energy. 
Conical phase (see section \ref{ConeCubicDistorted}) in the present definition can be considered as one of the helicoids with the propagation direction along the field. 

%In the following, I will imply under helicoids such states with the propagation direction oblique to the field. In the present definition the conical phase considered in the section \ref{ConeCubicDistorted} is one of the helicoids with the propagation direction along the field. 

\vspace{3mm}
\textit{A. Solutions for helicoids with $k_c>0,\,h||[001]$.}
\vspace{3mm}

\begin{figure}
\centering
\includegraphics[width=18cm]{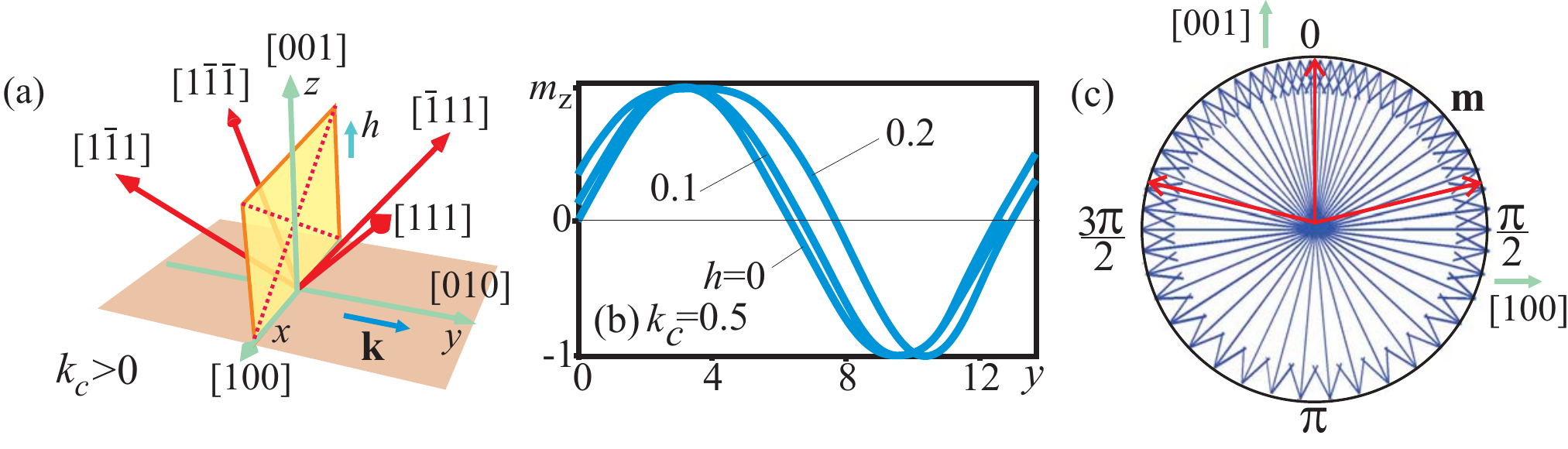}
\caption{
\label{Helicoid100} $k_c>0,\,h||[001]$. Solutions for the helicoid: (a) sketch showing the plane of the magnetization rotation (yellow plane), easy (green arrows) and hard (red arrows) directions of the cubic anisotropy, the directions of the applied magnetic field $h$ and the helicoid propagation direction $\mathbf{k}$. The projections of hard anisotropy axes onto the plane of rotation are shown by dotted red lines. (b) $m_z$-component of the magnetization in dependence on the coordinate $y$ for different values of the field. With increasing magnetic field parts of the curves with the magnetization along the field widen. (c) Polar plot for the component of the magnetization perpendicular to $\mathbf{k}$ for $h=0.2$. The densest distribution of the magnetic vectors corresponds to $\theta=0$ although the local minima can be distinguished also for $\theta=\pm\pi/2$.
}
\end{figure}

For the case with  $k_c>0,\,h||[001]$ the helicoids propagate in the plane (001) along one of the easy axes $<001>$ of cubic anisotropy (Fig. \ref{Helicoid100}). Magnetic field  is applied perpendicularly to the vector $\mathbf{k}$ as it was considered for helicoids of isotropic functional $W_0$ (section \ref{helicoidsBH}). Deviation of the propagation direction from the plane (001) as well as from the easy cubic axis in the plane increases the energy of the helicoid.

The plane of the magnetization rotation in the helicoid contains the easy anisotropy axes $<001>$ (green arrows in Fig. \ref{Helicoid100} (a) including the easy axis $[001]||h$)  and the projections of the hard anisotropy direction $<111>$ onto this plane (red dotted lines in Fig. \ref{Helicoid100} (a)). Already for $h=0$ the helicoid accomplishes an inhomogeneous rotation disturbed by anisotropic interactions. 

Increasing magnetic field leads to slow  rotation of the magnetization in the vicinity of the axis $[001]$ and acceleration of the rotation for the directions opposite to the field. In Fig. \ref{Helicoid100} (c) the polar plot  is shown for the component of the magnetization perpendicular to the vector $\mathbf{k}$. Distribution of the magnetic moments is densest for $\theta=0$. The angular phases obtained by deflecting the inplane magnetization with $\theta=\pm\pi/2$ correspond to the local minima of Eq. (\ref{CubicHomo}) and also lead to the denser distribution of the magnetic moments (see Fig. \ref{Homo100v2}). Described solutions are depicted in Fig. \ref{Helicoid100} (c) by red arrows. %The magnetic vectors 

In Fig. \ref{Helicoid100} (b) I have plotted $m_z$ component of the helicoid in dependence on the spatial coordinate $y$ along the propagation direction for different values of the applied magnetic field. For some critical value of the field the helicoid transforms into the homogeneous state with the magnetization along the field. This transition is signalled by an unlimited growth of the period for the helicoid and leads to the set of isolated domain walls with infinite separation between them. %In Fig. \ref{Helicoid100} (b) the period of the helicoid for $h=0.2$ is
%
%Nevertheless, the finite stiffness of the exchange interaction prevents a complete annihilation of the domain walls and they may exist with finite thickness within homogeneous state. 
%
The line of these critical fields on the phase diagram (Fig. \ref{PDCubic} (a)) has a label $\eta$ (green dashed line). The energy density of the helicoid with respect to the homogeneous state is shown in Fig. \ref{PDCubic} (b) (solid blue line).

% additional the distribution it has the local minimum which have lower energy in comparison with the local minimum with $\theta=\pi$ (see Fig. \ref{Homo100}). % The local minimum with $\theta=\pi$  has the larger energy than for 

\vspace{3mm}
\textit{B. Solutions for helicoids with $k_c<0,\,h||[001]$.}
\vspace{3mm}

For the case $k_c<0,\,h||<001>$ the energy of the helicoid must be minimized with respect to the orientation of the wave vector $\mathbf{k}$ relative to the applied magnetic field. The equilibrium solutions for helicoids are directly related to the energy landscape of cubic anisotropy (see Fig. \ref{Homo111v2}). %Therefore, for each value of the field $h$ all possible directions of the helicoid propagation direction must be considered, their energy calculated, and the directions with the global energy minimum defined. 

\begin{figure}
\centering
\includegraphics[width=18cm]{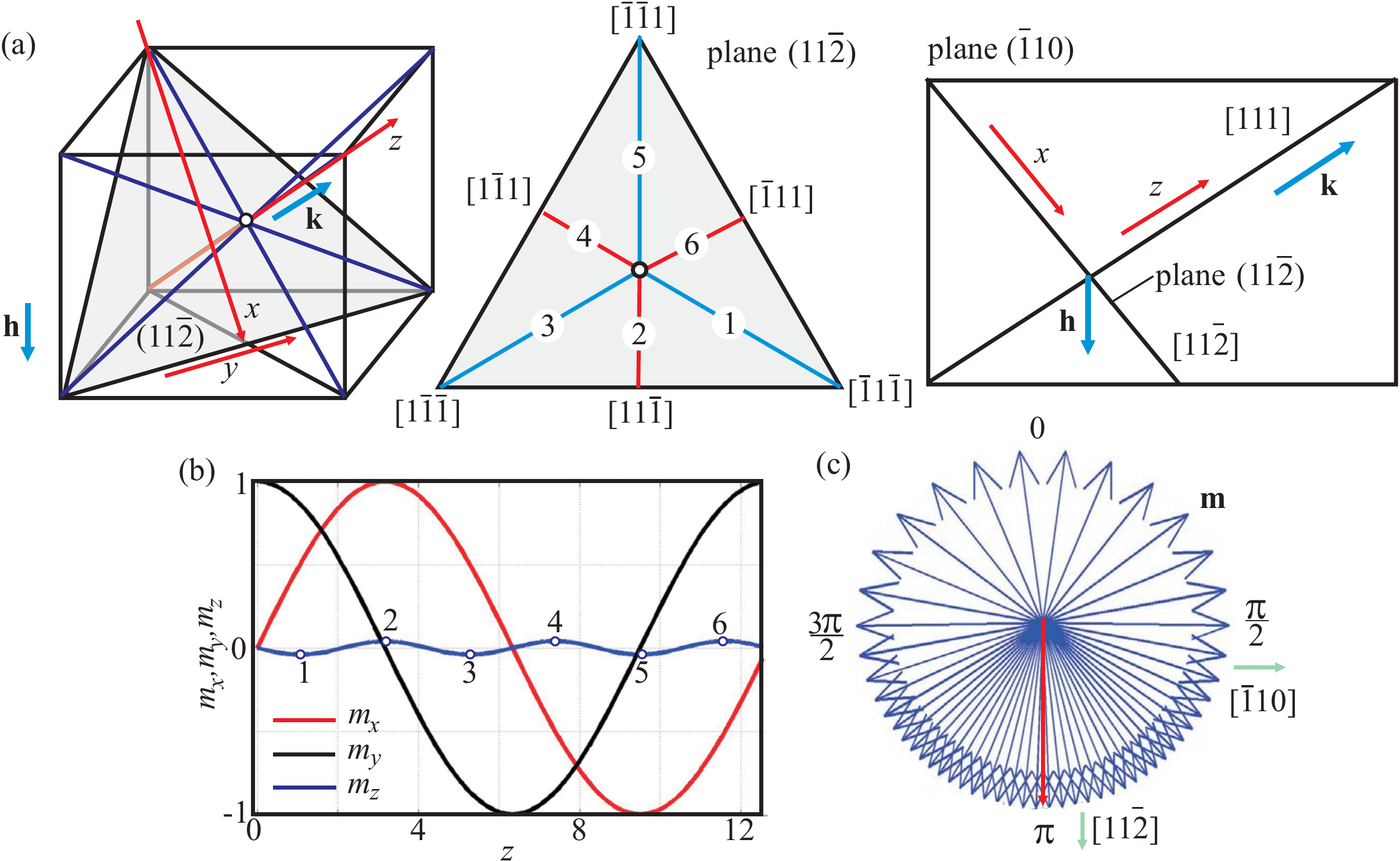}
\caption{
\label{Helicoid111} $k_c<0,\,h||[00\overline{1}]$. Solutions for the helicoid with $\mathbf{k}||[111]$: (a) sketch showing the plane of the magnetization rotation (grey triangular plane), easy  directions of the cubic anisotropy (blue straight lines), the directions of the applied magnetic field $h$ and the helicoid propagation direction $\mathbf{k}$, and the coordinate axes (red arrows) related to the helicoid. The projections of the easy anisotropy axes onto the plane of the magnetization rotation are numbered and marked by blue  (the axes are above the plane) and red (the axes are under the plane) color.
 (b) $m_x$ (red line), $m_y$ (black line), and $m_z$ (blue line) components of the magnetization in dependence on the coordinate $z||\mathbf{k}$ for $h=0$. The maxima and minima of the $m_z$-component correspond to the deviations toward easy anisotropy directions.  (c) Polar plot for the component of the magnetization perpendicular to $\mathbf{k}$ for $h=0.2$. The densest distribution of the magnetic vectors corresponds to the lower part of the plot: the rotating magnetization spans the easy direction in the direct vicinity of the applied magnetic field. 
}
\end{figure}

For $h=0$, the propagation direction of a helicoid  was found to  point to $<111>$ crystallographic directions. For definiteness in the following calculations,  I assume  $\mathbf{k}||[111]$, $\mathbf{h} || [00\overline{1}]$ (see sketch in Fig. \ref{Helicoid111} (a)).  The magnetization in the helicoid rotates in the plane $(11\overline{2})$ (grey shaded triangle in Fig. \ref{Helicoid111} (a)). The plane of the magnetization rotation contains the projections of easy anisotropy axes $<111>$. I marked the easy directions under the plane of rotation  by blue color  and above the plane - by red. Also all easy directions have been numbered. Rotating magnetization deviates from the plane $(11\overline{2})$ and sweeps these easy exes. The $m_z$-component of the magnetization shown in Fig. \ref{Helicoid111} (b) is negative for the crystallographic directions $[\overline{1}1\overline{1}[$ (marked as 1), $[1\overline{1}\overline{1}]$ (3), $[\overline{1}\overline{1}1]$ (5), and positive for $[11\overline{1}]$ (2), $[1\overline{1}1]$ (4), $[\overline{1}11]$ (6).

For $h=0$ the helicoid has lower energy in comparison with the cones with $\mathbf{k}||[00\overline{1}]$. While conical phase is able to sweep four easy anisotropy axes (see section \ref{ConeCubicDistorted} \textit{B}), the rotating magnetization in the helicoid meets six easy anisotropy directions on its way (Fig. \ref{Helicoid111} (a)).

Applied magnetic field $\mathbf{h}||[00\overline{1}]$ leads to the significant distortions of the helicoid structure. The field destroys the degeneracy of energy minima of cubic anisotropy: easy axes 1,2, and 3 have lower energy in comparison with metastable directions 4, 5, and 6. During this complex magnetization process the wave vector $\mathbf{k}$ of the helicoid is directed along the metastable minimum $[111]$ with slight change of its orientation in the applied magnetic field.  The distribution of the magnetic vectors in the plane perpendicular to the propagation direction becomes denser in the lower part of the polar plot in Fig. \ref{Helicoid111} (d). 

In Fig. \ref{PDCubic} (d) I plotted the energy density of the cone (blue line) and helicoid (green line) versus $h$. For some critical value of the magnetic field (point $\beta$ in Fig. \ref{PDCubic} (d)) conical phase becomes energetically more favourable than the helicoid. This point indicates the first-order phase transition between these one-dimensional modulated phases. The critical fields  have been plotted in the phase diagram (Fig. \ref{PDCubic} (c)) for different values of the cubic anisotropy constant $k_c$. 
Note, that in the present geometry the cone sweeps easy axes 1,2,3 as well as the axis $[\overline{1}\overline{1}\overline{1}]$ perpendicular to the plane of rotation of the helicoid. As soon as the metastable minima 4,5,6 dissappear, the helicoid transforms into the conical phase (see the third plot in panel (a) of Fig. \ref{Helicoid111}).

\subsection{Transformation of the hexagonal skyrmion lattice in the presence of cubic anisotropy \label{SkyrmionCubicTransformation}}

\begin{figure}
\centering
\includegraphics[width=15cm]{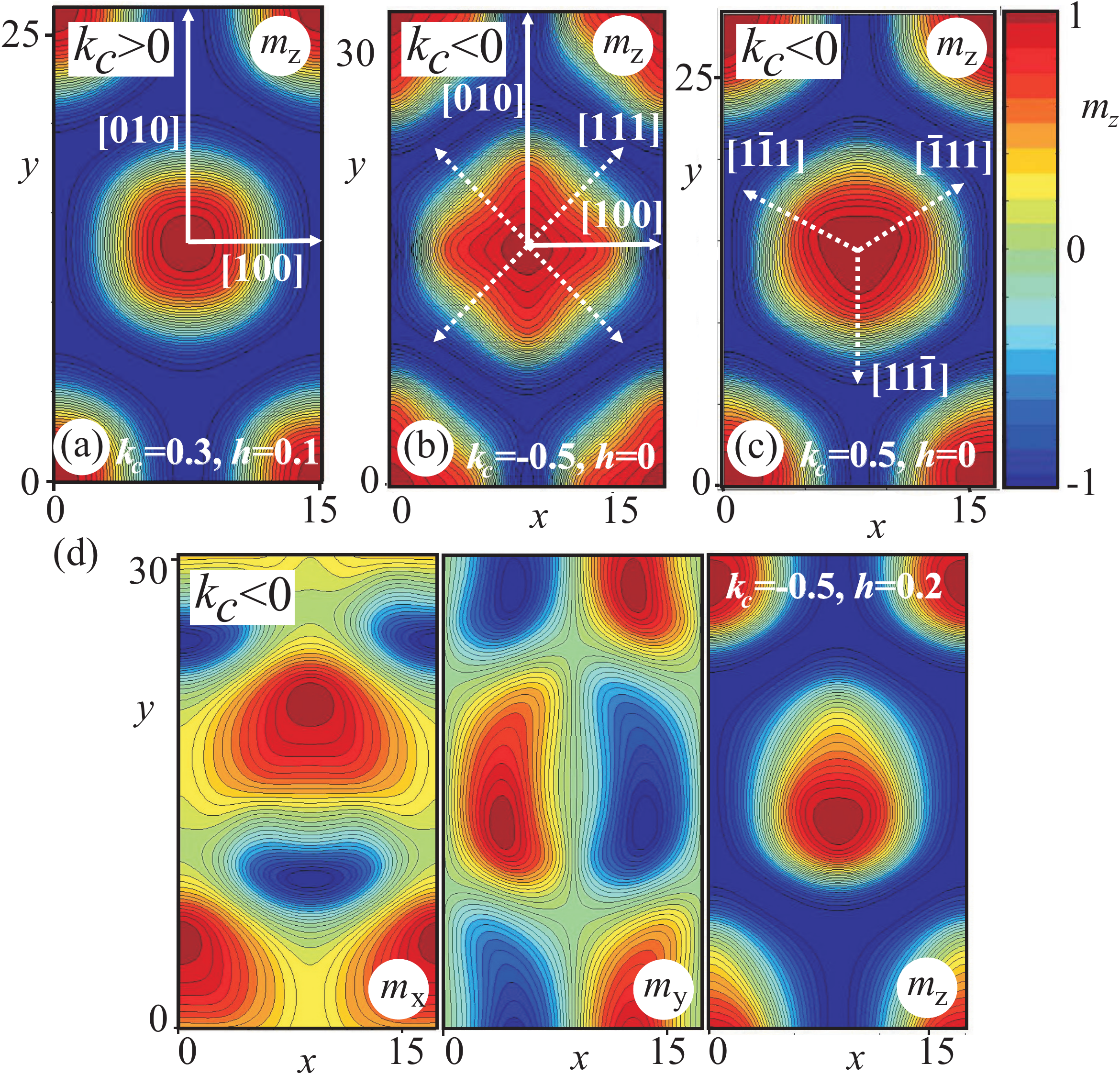}
\caption{
\label{SkyrmionCubic} Contour plots for components of the magnetization in skyrmion lattices for both signs of the cubic anisotropy constant $k_c$: (a), (b) contour plots for $m_z$-components of the magnetization with the axis of the skyrmion lattices directed along the field, i.e. along [001]. Axes $<100>$ of cubic anisotropy are shown as white arrows. In (a) with $k_c>0$, these axes are easy directions, whereas in (b) with $k_c<0$ - they are hard anisotropy axes.  (c) Contour plot for $m_z$-component of the magnetization in the skyrmion lattice with the axis along [111] crystallographic direction. Dotted white arrows indicate the projections of anisotropy axes $<111>$ onto the plane of skyrmion lattice. (d) Contour plots for $m_x$, $m_y$, and $m_z$ components of the magnetization in a case when axes of skyrmions do not point to the equilibrium state of the functional (\ref{CubicHomo}). The cores of the skyrmions are shifted from the center of the lattice cell, but the lattice retains the stability against transformation into helicoids. 
}
\end{figure}

Alongside with the drastic influence on the conical phases and helicoids (see sections \ref{ConeCubicDistorted} and \ref{HelicoidDistortedCubic}), cubic anisotropy distorts significantly skyrmion states: the symmetry of the skyrmion cores reflects the underlying energy landscape of the cubic anisotropy (Figs. \ref{Homo100v2}, \ref{Homo111v2}) and undergoes the respective  transformation (Fig. \ref{SkyrmionCubic}). 

In the calculations according to the methods of section \ref{NumericalRecipes}, the direction of the skyrmion axes have been tuned with respect to the field and easy anisotropy axes in the search of the states with the lowest energy (the same minimization of the energy had been done in the section \ref{HelicoidDistortedCubic} \textit{B} for helicoids). 
%
%As it was done for helicoids, the direction of the skyrmion axes with respect to the field and easy anisotropy axes must be tuned, and the states with the lowest energy must be found. 
%
For $k_c>0$ the equilibrium position of the skyrmion axis is codirectional with the applied magnetic field. For $k_c<0$ the skyrmion axis has been found to follow the global minimum of the cubic anisotropy (Eq. (\ref{CubicHomo})), i.e. for $h=0$ the skyrmion axis is directed along the easy  cubic axes $<111>$, but in the field it starts to move toward the field. 

In Fig. \ref{SkyrmionCubic} (a)-(c) I have plotted the contour plots for $m_z$-components of the magnetization in skyrmion lattices for both signs of the cubic anisotropy $k_c$. 
%
%The white (solid) arrows show the axes of the cubic anisotropy: for $k_c>0$ these axes are easy directions, for $k_c<0$ - hard axes. Dotted white arrows in Fig. \ref{SkyrmionCubic} (c) are projections of easy anisotropy axes $<111>$ onto the plane of skyrmions. As it is seen, the perfect circular symmetry of the skyrmion cores is impaired by the corresponding hard and easy anisotropy axes. 
%
In Fig. \ref{SkyrmionCubic} (a) and (b) the axes of skyrmion lattices are directed along the field; the cores of skyrmions become square shaped with the tendency either  to elongate or to shorten along particular directions.  

%It is necessary to note, that 

The skyrmion lattice of Fig. \ref{SkyrmionCubic} (a) can be stabilized only in the applied magnetic field. The field localizes the skyrmion cores and prevents skyrmions from the transformation   into helicoids. For the field lower than some critical value (in Fig. \ref{PDCubic} (b) this field is marked by $\gamma_0$) the easy axes of the cubic anisotropy in the plane of the skyrmion lattice induce the instability of skyrmions with respect to  helicoids. By numerical means used in the present thesis it is hardly possible to obtain the solutions of the skyrmion lattice for zero and small magnetic fields.  The easy directions of the cubic anisotropy are shown by the white arrows in Fig. \ref{SkyrmionCubic} (a). 

For another critical value of the field the skyrmion lattice releases the free isolated skyrmions. In Fig. \ref{PDCubic} (a) these fields are depicted as the line $\gamma$ (dashed red line). The energy density of skyrmion lattice is plotted in Fig. \ref{PDCubic} (b) (red solid line for $k_c=0.5$). With increasing value of the constant $k_c$ the interval between two critical fields (i.e. between the points $\gamma_0$ and $\gamma$) decreases. For the constants $k_c>k_c(E)$ (Fig. \ref{PDCubic} (a)) the skyrmion lattice is highly unstable. % and is hardly achieved in the numerical simulations. 

The skyrmion lattice plotted in Fig. \ref{SkyrmionCubic} (b) has the larger energy in comparison with the lattice directed along the $<111>$ crystallographic directions (Fig. \ref{SkyrmionCubic} (c)). White (solid) arrows in this case with $k_c<0$ indicate hard axes of the cubic anisotropy. White dotted arrows are projections of the easy axes $<111>$ onto the plane $(001)$ of the skyrmion lattice. In the present calculations such a lattice was used as a cross-check of calculations: the energies of two skyrmion lattices coincide, when the angular phases of the functional (\ref{CubicHomo}) are aligned along the field. %different angular phases of cubic anisotropy dissappear in the field. %with equilibrium direction of its axis, i.e. along the $<111>$ crystallographic directions. 

 %In zero and small magnetic fields the lattice transforms into the helicoid. For For the large values of the cubic anisotropy $k_c$ the interval of the fields between the field of instability with respect to the helicoid and the field when isolated skyrmions are released decreases. For the magnetic field $h>h(E)$ (Fig. \ref{PDCubic} (a)) the skyrmion lattice is highly instable and hardly achieved in the numerical simulations. 

In Fig. \ref{SkyrmionCubic} (c) the axis of the skyrmion lattice points to the crystallographic direction $[111]$. In this case the dotted white arrows show the projections of the cubic easy axes $<111>$ onto the plane $(11\overline{2})$. The core of the lattice acquires the shape of a curvilinear triangle. With increasing magnetic field such a lattice gradually rotates keeping its axis parallel to an equilibrium state of the energy functional (\ref{CubicHomo}). In the panel (d) of Fig. \ref{SkyrmionCubic} I plotted the skyrmion lattice in the applied magnetic field, but with the axis directed along $[111]$ axis. The figure shows that the skyrmion lattice is essentially robust against the transformation into the helicoid even if its core is displaced from the central position in the lattice cell. Energy density of the skyrmion lattice is plotted in Fig. \ref{PDCubic} (d) (red solid line). The skyrmions are only metastable states in comparison with cones and helicoids. %The first-order phase transitions between the skyrmions and cones (point $\alpha_1$), and skyrmions and helicoids (point $\alpha_2$ in Fig. \ref{PDCubic} (d)) are hidden by the state with global energy minimum. 

\subsection{The phase diagrams of states in the presence of magnetocrystalline cubic anisotropy  \label{CubicPDAll} } 

\begin{figure}
\centering
\includegraphics[width=18cm]{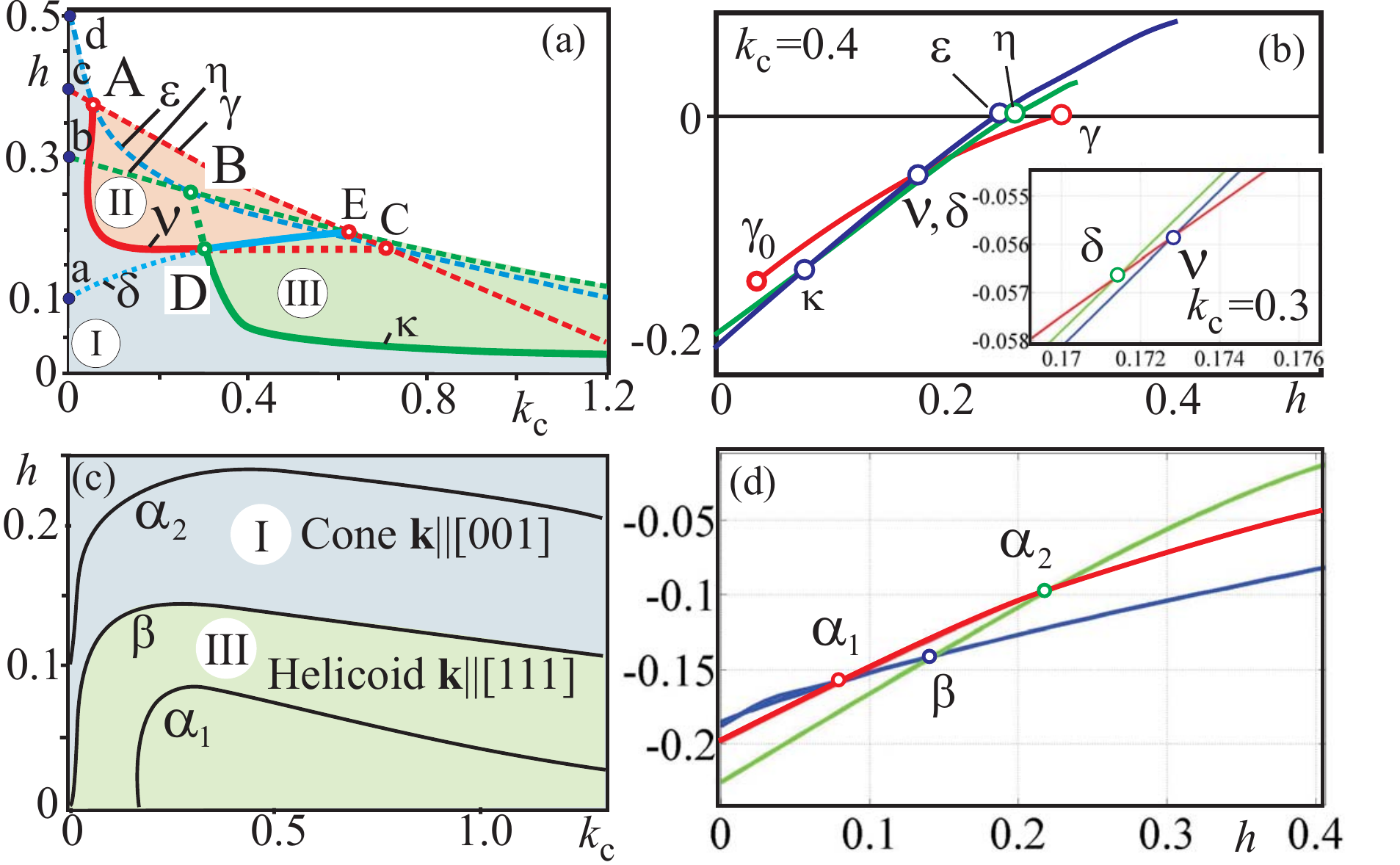}
\caption{
\label{PDCubic} Phase diagrams of states for $k_c>0$ (a) and $k_c<0$ (c). Magnetic field $\mathbf{h}$ is applied along [001]. The regions of the thermodynamical stability are colored by red (skyrmions), blue (cones), and green (helicoids). The detailed description of the phase transitions between modulated phases is given explicitly in the text. (b), (d) The energy densities of the modulated phases in dependence on the field for fixed value of $k_c$ (see section \ref{CubicPDAll} for details).
}
\end{figure}

The phase diagrams of states for both signs of cubic anisotropy constant $k_c$ and the field applied along $[001]$ are plotted in Figs. \ref{PDCubic} (a), (c). Further, I analyse each of these phase diagrams and give some qualitative recommendations as far the thermodynamical stability of skyrmions with respect to conical phases is concerned. 

\vspace{3mm}
\textit{A. Phase diagram of states for $k_c>0,\,\mathbf{h}||[001]$.}
\vspace{3mm}

Cones as modulated states with negative energy relative to the homogeneous state exist below the line d-A-B-C (dashed blue line $\varepsilon$ in Fig. \ref{PDCubic} (a)). At this line cones flip into the saturated state by a first-order phase transition as described in section \ref{ConeCubicDistorted} \textit{A}. Nevertheless, above this line cones still exist as states with positive energy. Only within the region filled with a blue color (0-d-A-D or region I) they are thermodynamically stable. In the remaining part the cones are metastable states. At the lines $\nu$ (red solid line A-D in Fig. \ref{PDCubic} (a)) and $\kappa$ (green solid line) cones transform discontinuously into skyrmions (red-colored area II) and helicoids (green colored region III), respectively. Dotted lines mark the phase transitions between the metastable states: $\delta$ is the line of first-order phase transition between skyrmions and helicoids (blue line a-D), for $k_c=0$ the point 'a' corresponds to this transition in the isotropic case; line D-B (green dotted line) stands for the transition between metastable cones and helicoids in the region where skyrmions are thermodynamically stable states; line D-C (dotted red line) is the line of the first-order phase transition between skyrmions and cones in the region of stability of helicoids. 

As it is seen from the phase diagram (Fig. \ref{PDCubic} (a)), the skyrmion states are thermodynamically stable within a curvilinear triangle (A - D - E) with vertices (A) = (0.047, 0.379), (D) = (0.233, 0.264), and (E) = (0.613, 0.203)). At the line D-E they transform into helicoids, and at c-A-E-C (red dashed line $\gamma$) - into the homogeneous state with the magnetization along the field.  (C)=(0.651,0.176) is the point of intersection of the lability lines for cones (blue dashed lines) and skyrmions; in the point E the lability lines for skyrmions and helicoids cross each other. Point B has the coordinates (0.269,0.253). For $k_c>k_c(\mathrm{C})$ helicoids and cones can exist for much larger values of the applied magnetic field than skyrmions; skyrmions undergo the elongation into the helicoids in this region (see also section \ref{SkyrmionCubicTransformation}).

The solutions for helicoids exist below the line b-B-E (green dashed line) where they turn into the homogeneous state. In the region III of the phase diagram (Fig. \ref{PDCubic} (a)) helicoids are the thermodynamically stable states of the system. Due to the strong influence of the cubic anisotropy on the conical phase, helicoids can exist in  higher fields than cones for $k_c>k_c(\mathrm{B})$. 

The present phase diagram has been built by comparing energies of corresponding modulated phases. In Fig.  \ref{PDCubic} (b) the energies of the skyrmion lattice (red line), helicoid (green line), and cone (blue line) are plotted in dependence on the applied magnetic field $h$ for $k_c=0.4$. In zero field the conical phase is the state with the minimal energy. Then, in the point $\kappa$ the cone transforms into the helicoid. Points $\delta$  and $\nu$ indicate the transitions to skyrmions from helicoids and to skyrmions from cones, respectively. These transitions demand the closer look at them. For $k_c=0.3$, $h(\delta)<h(\nu)$, but for $k_c=0.4$, $h(\delta)<h(\nu)$ (see inset of Fig. \ref{PDCubic} (b)). For $h<h(\gamma_0)$ the skyrmion lattice may elongate into the helicoid (see section \ref{SkyrmionCubicTransformation} for details). The skyrmion cores become instable with respect to elliptic distortions. Numerically, such a transformation is accompanied by the drastic increase of the grid spacings ($\Delta_y>>\Delta_x$ or $\Delta_x>>\Delta_y$) along one of the spatial directions $y$ or $x$ (see section \ref{NumericalRecipes} for the introduction into the numerical recipes of the present calculations).  %increases significantly () of the %in the case of skyrmion lattice 

\vspace{3mm}
\textit{B. Phase diagram of states for $k_c<0,\,\mathbf{h}||[001]$.}
\vspace{3mm}

For $k_c<0$ and the field $\mathbf{h}||[001]$ only one-dimensional chiral modulations are present in the phase diagram as thermodynamically stable states of the system (Fig. \ref{PDCubic} (c)). At zero field the helicoid (green line) has lower energy in comparison with the cone (blue line in Fig. \ref{PDCubic} (d)). The reason of this is explicitly explained in the section \ref{HelicoidDistortedCubic} \textit{B}: rotating magnetization in the helicoid sweeps 6 easy axes of cubic anisotropy, while the conical phase  - only 4. The situation is drastically changed in the applied magnetic field: point $\beta$ signifies transformation of the helicoid into the cone by the first-order phase transition. 

Skyrmions are metastable solutions. Points $\alpha_1$ and $\alpha_2$ of the transitions to skyrmions from cones and to skyrmions from helicoids are characterized by the higher energy densities comparing with the energy of the global helical and conical phases, respectively. Therefore, these transitions are hidden. In this connection the influence of the higher-order anisotropy terms may have significant influence on the skyrmion states. 

\vspace{3mm}
\textit{C. General remarks how to stabilize skyrmion states in the presence of cubic anisotropy}
\vspace{3mm}

Considered phase diagrams of states (Figs. \ref{PDCubic} (a), (c))  allow to deduce some qualitative recommendations how to stabilize skyrmions over conical phases in the presence of cubic anisotropy. Such phase diagrams, however, cannot be considered as complete, since I did not considered possible three-dimensional states realized in the system. Moreover, competition of different small anisotropic contributions will also distort the stability regions of different modulated phases.
 %e determined by subtle additional, but important, effects. For this reason, it is technically not feasible to calculate the magn phase diagram in full detail for a real, 3D system. In particular near magn ordering there is not clear hierarchy of magnetic couplings because of the nucleation of localised soltonic states.
 %The influence   

(i) As it was concluded in section \ref{ConeCubicDistorted}, cubic anisotropy effectively suppresses  conical phases for $k_c>0$ and $\mathbf{h}||[001]$. Rotating magnetization of the conical phase in this case sweeps the metastable directions of the energy functional (\ref{CubicHomo}). The same effect may be achieved for the field $\mathbf{h}||<111>$ and $k_c<0$. In this case,  the hard axes $<100>$ of the cubic anisotropy impair the ideal harmonic rotation of the magnetization in the conical phase. The phase diagram of states looks qualitatively similar to the phase diagram in Fig. \ref{PDCubic} (a), but with slightly different coordinates for all critical points. 
Therefore, the suppression of the cone depends on the sign of the cubic anisotropy constant $k_c$: for $k_c>0$ the field must be applied along $<001>$, for $k_c<0$ - along $<111>$. The cubic anisotropy $k_c$ must be larger than some threshold value corresponding to the point A in Fig. \ref{PDCubic} (a).

For $k_c<0$, $\mathbf{h}||[001]$, and  $h>h(\beta)$, the conical phase is the thermodynamically stable state of the system (Fig. \ref{PDCubic} (c)). The rotating magnetization in a cone  sweeps the global minima of energy functional (\ref{CubicHomo}). The same situation will be also realized for $k_c>0$ and $\mathbf{h}||<111>$, when the magnetization spans easy anisotropy axes <100>. The skyrmions will form only metastable states in these situations.

(ii) At the same time the constant $k_c$ of cubic anisotropy must not be larger than the critical value $k_c(\mathrm{E})$. Otherwise, the skyrmions will tend to elongate into spirals (see for details section \ref{SkyrmionCubicTransformation}). Such an instability of skyrmions is related to the easy anisotropy axes <100> in the plane of the skyrmion lattice (Fig. \ref{SkyrmionCubic} (a)). Skyrmions can be stabilized only in the applied magnetic field $h>h(\gamma_0)$ (Fig. \ref{PDCubic} (b)). For $k_c<0,\,\mathbf{h}||<111>$ the skyrmions will suffer from instability toward helicoids with $\mathbf{k}||<111>$ as easy cubic axes $<111>$ make some angle with the skyrmion plane.

\section{Candidate materials for experimental observation of skyrmion textures}

From the general phenomenological point of view, the choice of materials that will show skyrmion lattices as low-temperature states in applied fields is dictated only  by symmetry requirements and a magnetic ordering transition.
Therefore, many different magnetic crystals from classes D$_{2d}$ and C$_{nv}$ could be listed as promising objects of pointed searches for skyrmion lattice phases  in their magnetic phase diagrams.
Here, I mention only very few of them, where clear indications of non-collinear magnetic states are known 
from early experiments.

{\textit{Defect spinel structure magnets}}
GaM$_4$X$_8$. The magnetic phase diagram of GaMo$_4$X$_8$ shows a clear intermediate
phase between ground-state and field saturated state \cite{Rastogi87}.
The material behaves as almost a ferromagnet, and Rastogi and Wohlfahrt \cite{Rastogi87} pointed out the similarity with the behavior of MnSi and the possibility of a twisted non-collinear spin-structure.

{\textit{CeTMSn magnets.}}
 The examplary CeCuSn exists in two modifications, $\alpha$- and $\beta$-CeCuSn. Only the $\beta$-variant is non-centrosymmetric with space-group P6$_3$mc  belonging to Laue class C$_{6v}$ and displays a series of marked anomalies in the magnetization data $M(H)$, while the centrosymmetric variant behaves apparently as a simpler
magnetic systems. The direct comparison of the two different crystallographic states suggests an important
role of chiral DM interactions \cite{Sebastian06}.

In Ref.\cite{JMMM94} two tetragonal materials Tb$_3$Al$_2$ and Dy$_3$Al$_3$ were proposed as suitable candidates for the observation of chiral skyrmions and spiral structures. 
Both crystals belong to the space group C$_{4v}^4$ ; Tb$_3$Al$_2$ has a Curie temperature $T_C$ = 100 K, and for Dy$_3$Al$_3$, $T_C$ = 190 K has been measured \cite{Barbara68,Barbara71}. 
At high temperatures they have rather complicated easy-axis type magnetic structures.
At a transition temperature $T_t$ they seem to switch to an easy-plane structure.
This transition was found at $T_t$ = 10 K for Tb$_3$Al$_2$ and at $T_t$ = 20 K for Dy$_3$Al$_3$. 
Up to now modulated magnetic structures have not been identified in these materials, but in view of the magnetic symmetry they could be present, particularly near T$_t$, where the uniaxial anisotropy constant is small and conditions for the
realization of a magnetic mixed state are favourable.

The cubic materials MnSi, FeGe, Fe$_x$Co$_{1-x}$Si, and Co$_x$Mn$_{1-x}$Si belong to another group where skyrmions are believed to induce anomalies of phase diagram near the ordering transition. 
In these compounds spiral structures related to the Dzyaloshinsky-Moriya interactions are well known \cite{Ishikawa76,Wilkinson76,Beille83}.
The helix pitch $L_D=2\pi/|\mathbf{k}|$ is in general large in these compounds: it is 18 $\mathrm{nm}$ in the case of MnSi or even larger (>230 nm) for Fe$_{0.8}$Co$_{0.2}$Si (see also some examples in the Table \ref{table1}).
%
%$L_D$ is proportional to the ratio of the counter-acting exchange and Dzyaloshinskii constants (\ref{helix0}) and introduces a fundamental \textit{length} characterizing a magnitude of chiral modulations in non-centrosymmetric magnets.

\begin{table}[h]
\caption{N\'eel temperatures ($T_N$), helix periods ($L_D$), and saturation fields ($H_D$) \\
for some cubic helimagnets, data from Ref. \cite{Lebech89}. \label{table1}}
\begin{center}
\begin{tabular}{|p{2cm}||p{1.5cm}|p{2.2cm}|p{2.2cm}|p{2.2cm}|p{2cm}|}%{llllllllll}
%\br
\hline
Compound \quad \quad & MnSi \quad \quad & FeGe & Fe$_{0.3}$Co$_{0.7}$Si \quad \quad
 & Fe$_{0.5}$Co$_{0.5}$Si \quad \quad & Fe$_{0.8}$Co$_{0.2}$Si \\ \hline
 %\mr
 \hline
$T_N$ [K] & 29.5& 278.7& 8.8 & 43.5& 32.2\\ \hline
$L_D$ [nm] & 18.0 & 68.3- 70.0 & 230&  90.0 & 29.5 \\ \hline
$H_D$ [T] & 0.62 & 0.2 & (6.0 $\pm$ 1.5)$\cdot 10^{-3}$ &  (4.0 $\pm$ 0.5)$\cdot 10^{-2}$  & 0.18 \\
%\br
\hline
\end{tabular}
\end{center}
% {\small{\quad \quad The data are from Refs. \cite{Lebech89}.}}
\end{table}
The skyrmion states in the present magnets can be easily stabilized by the small cubic anisotropy with easy axes along $<001>$ crystallographic directions and the applied magnetic field directed along $<001>$ as shown in section \ref{StabilizationCubic}. The main effect of the cubic anisotropy in this case is suppression of  the conical phase. The phase diagram depicted in Fig. \ref{PDCubic} (a) will be qualitatively the same for easy axes of cubic anisotropy along $<111>$ and the field parallel to one of these directions. %Note, that in the compounds with the cubic anisotropy 

The skyrmion states may be stabilized over cones and helicoids also by the uniaxial anisotropy.
The results of section \ref{StabilizationUA} may help to clarify the role of small uniaxial distortions in high-pressure experiments in MnSi -  an important and unsolved problem, which continues to attract widespread attention.
For example for MnSi, earlier experiments \cite{FranusMuir84} and analysis \cite{PlumerWalker82} of magnetoelastic couplings
allow a quantitative estimate of strain-induced uniaxial anisotropy. % showing that the effects predicted here can be achieved in experiment.
The magnetoelastic coupling with uniaxial strains $u_{zz}$ is given by \(w_{me}=b \,u_{zz}\,(M_z/M_S)^2\), where $M_S=$~50.9~A/m is the saturation magnetization \cite{Bloch75} and $b=$~7.4~GPa  is a magnetoelastic coefficient 
derived from the magnetostriction data in Ref. \cite{FranusMuir84}.
Using exchange constant $A=$~0.11~pJ/m, as estimated from the spin-wave stiffness reported in Ref. \cite{Grigoriev05},
and $D=2\,q_0\,A=$~0.86~$\mu$J/m$^2$ for MnSi \cite{Nature06} one has $K_0\simeq$~17~kJ/m$^3$ and a dimensionless 
scale $b/K_0\simeq$~44 for the induced anisotropy.
Thus, a modest strain $u_{zz}=$~0.0024 is sufficient to reach an induced anisotropy $K/K_0=$~0.1 which is enough to stabilize the skyrmion lattice in magnetic field. % well within the region for stable  skyrmion lattices in the phase diagram Fig.~\ref{PD}.
This strain corresponds to a tensile stress $\sigma_{zz}$=~680 MPa for MnSi by using the elastic constant $c_{11}$=~283~GPa.\cite{Stishov08}
The rather low uniaxial stress necessary to stabilize the skyrmion lattice is particularly relevant for pressure experiments
with a uniaxial disbalance of the applied stresses, but it could also be achieved in epitaxial films.

Additional uniaxial anisotropy may be also of surface-induced nature. In magnetic nanosystems surface/interface interactions provide additional stabilization mechanism of skyrmion states. As recently the skyrmion states were observed in thin magnetic layers of Fe$_{1-x}$Co$_x$Si and FeGe \cite{Yu10,Yu10a}, this might have the significant contribution. 
%
%

%To stabilize skyrmion states, an additional uniaxial anisotropy must be present.
%
%External stresses or growth anisotropies may be used to create an easy axis.
%
%Then these materials become examples for the symmetry class D$_n$.

%\section{Parts of good text}

%This experimental work has been based on our earlier theoretical predictions [8-10] and is now fruitful with the break-through reported in Ref. [1]. (In experimental part) 

\section{Conclusions}

%An overview of recent theoretical and experimental results is presented on chiral modulated and localized states 
In non-centrosymmetric magnetic materials, Dzyaloshinskii-Moriya exchange based on the relativistic spin-orbit couplings stabilizes helical one- dimensional modulations,  as well as solitonic textures, i.e., localized topologically non- trivial baby- skyrmions - repulsive particle-like spin textures imbedded into homogeneously magnetized "parental" state.
These isolated skyrmion excitations can be manipulated as particle-like entities. Their relevant length scale can be tuned by the competition between direct and chiral DM exchange and may range from few atomic spacings up to microns. 
Theoretical results for the basic phenomenological continuum theory of chiral magnets demonstrate that localized spin-textures with constant value of the magnetization modulus may form extended regular states.
The formation of skyrmionic textures is determined by the stability of the localized solitonic cores and their geometrical incompatibility that frustrates homogeneous space-filling.
On the contrary to the circular-cell approximation, used as a method of choice to obtain approximate solutions for skyrmion lattices in early papers of A. N. Bogdanov et al. \cite{JMMM94,pss94,JMMM99}, the rigorous solution for skyrmion states in this chapter are derived by the direct energy minimization for phenomenological models of non-centrosymmetric helimagnets from different crystallographic classes. 
These numerical results provide a comprehensive description of the structure of the skyrmion lattice and its evolution in the applied magnetic field directed either opposite or along the magnetization in the center of the skyrmion cell. Differences of lattice parameters from circular- cell approximation and from numerical calculations lie within 2\% and, therefore, demonstrate that CCA is a good approximation.

It is shown that for crystals from Laue classes D$_{2d}$ and C$_{nv}$ skyrmion lattices are stable with respect to one-dimensional helices in the applied magnetic field. As the transition between spiral and skyrmion states is a first- order phase transition, domains of coexisting phases may be formed. For cubic helimagnets and other systems with Lifshitz invariants attached to three spatial directions, additional anisotropic contributions suppressing conical phase must be considered. Skyrmion lattices can be stabilized in a broad range of thermodynamical parameters in the presence of uniaxial anisotropy. Skyrmion stability demands the combined effect of uniaxial anisotropy and magnetic field. These findings demonstrate that distorted cubic helimagnets are very promising objects for investigations of skyrmion states. On the other hand, skyrmion states may be stabilized over cones by small cubic anisotropy itself. To achieve this goal the applied magnetic field must point along particular crystallographic directions strongly deforming the conical state. 


\begin{thebibliography}{00}

\bibitem{JETP89}   A. N. \ Bogdanov and D. A. \ Yablonsky,
%Theormodynamically stable vortices in magnetically 
%ordered crystals. Mixed state of magnetics.
 Zh. Eksp. Teor. Fiz. {\textbf 95}, 178 (1989) [Sov. Phys. JETP 68, 101 (1989)].
 
\bibitem{JMMM94} A. Bogdanov, A. Hubert, J. Magn. Magn. Mater. {\textbf{138}}, 255 (1994).
 
 \bibitem{Nature06} 
U. K. R\"o\ss ler, A. N. Bogdanov, C. Pfleiderer,
% Spontaneous skyrmion ground states in magnetic metals 
Nature \textbf{442}, 797 (2006).

\bibitem{pss94} A. Bogdanov, A. Hubert, phys. stat. sol. (b)  {\textbf{186}}, 527 (1994).

\bibitem{Dz64}  I.\ E.\ Dzyaloshinskii, J.\ Sov.\ Phys.\ JETP-USSR {\textbf{19}}, 960 (1964).

\bibitem{Yu10} X. Z. Yu, Y. Onose, N. Kanazawa \textit{et al.}, Nature, {\textbf{465}}, 901 (2010).

\bibitem{Yu10a} X. Z. Yu, N. Kanazawa, Y. Onose \textit{et al.}, Nature Mater. \textbf{10}, 106 (2011).












\bibitem{Dz57} I.\ E.\ Dzyaloshinskii, Sov.\ Phys.\ JETP {\textbf{5}}, 1259 (1957).

\bibitem{Moriya60}  T.\ Moriya, Phys.\ Rev.\ {\textbf{120}}, 91 (1960).

\bibitem{LandauLifshitz} L. D. Landau and E. M. Lifshitz, \textit{Statistical Physics. Course of Theoretical Physics} (Pergamon, Oxford, 1997), Vol. V.

\bibitem{Bode07} M. Bode, M. Heide, K. von Bergmann, P. Ferriani \textit{et al.}, %, S. Heinze, 
%G. Bihlmayer, A. Kubetzka, O. Pietzsch, S. Bl\"ugel, R. Wiesendanger,
%Chiral magnetic order at surfaces driven by inversion asymmetry.
Nature {\textbf{447}}, 190 (2007).


\bibitem{Heinze11} S. Heinze \textit{et al.}, accepted to Nature Physics  (2011) (see also APS March Meeting 2010, March 15-19,2010, abstract L34.014).

\bibitem{Butenko09} A. B. Butenko et al. Phys. Rev. B \textbf{80},  134410 (2009).









\bibitem{Izyumov84}  Yu.\ A.\ Izyumov, Sov.\ Phys.\ Usp.\ {\textbf{27}}, 845 (1984).
% Modulated, or long-periodic, magnetic structures of crystals

\bibitem{Bogdanov02k} A. N. Bogdanov, U. K. R\"o\ss ler, M. Wolf, and K. -H. M\"uller, Phys. Rev. B \textbf{66}, 214410 (2002).

\bibitem{books} P. G. De Gennes and J. Prost, \textit{The Physics of Liquid Crystals} (Oxford University Press, Oxford, 1993), 2nd ed. %; A. Hubert, R. Sch{\"{a}}fer, \textit{Magnetic Domains} (Springer, Berlin 1998).

\bibitem{Brandt95} E. H. Brandt, Rep. Prog. Phys. \textbf{58}, 1465 (1995).

\bibitem{Brandt03} E. H. Brandt, Phys. Rev. B \textbf{68}, 054506 (2003).

\bibitem{JMMM99} A. Boganov, A. Hubert, J. Magn. Magn. Mater.  {\textbf{195}}, 182 (1999).


\bibitem{Hubert98} A. Hubert, R. Sch{\"{a}}fer,  \textit{Magnetic Domains} 
% , The Analysis of Magnetic Microstructures,
(Springer, Berlin 1998).

\bibitem{JETP95} A. Bogdanov,
% New localized solutions of the nonlinear field-equaitons
JETP Lett. {\textbf 62}, 247 (1995).

\bibitem{DeGennes75}  P. G. de Gennes, in \textit{Fluctuations, Instabilities, and Phase transitions}, ed. T. Riste, NATO ASI Ser. B, vol.~2 (Plenum, New York, 1975).

\bibitem{BP75} A. A. Belavin, and A. M. Polyakov, Pis'ma Zh. Eksp. Teor. Fiz. \textbf{22}, 503 (1975) [JETP Lett. \textbf{22}, 245 (1975)]. 

\bibitem{Metropolis} N. Metropolis, A. W. Rosenbluth, M. N. Rosenbluth, A. H. Teller, and E. Teller, J. of Chem. Phys. \textbf{21}, 1087 (1953). 

\bibitem{MC} K. Binder, D. W. Heerman, \textit{Monte Carlo Simulation in Statistical Physics} (Springer, Berlin 1992), 2nd ed.

\bibitem{Muhlbauer09} S. M\"uhlbauer, B. Binz, F. Jonietz \textit{et al.},
%Skyrmion lattice in a Chiral Magnet
 Science, {\textbf{323}}, 915 (2009).
 
 
 \bibitem{Bak80} P.\ Bak and M.\ H.\ Jensen, J.\ Phys. C: Solid State Phys.\ {\textbf 13}, L881 (1980).
 
 \bibitem{Rastogi87} A. K. Rastogi, E. P. Wohlfarth, phys. stat. sol. (b) \textbf{142}, 569 (1987).
% GaMo4S8_GaMo4Se8_F-43m_magnetic-phsae-diagram

\bibitem{Sebastian06} C. P. Sebastian \textit{et al.}, arXiv: 0612225 (2006).
% CeCuSn_P6_3mc_alpha_vs_beta

\bibitem{Barbara68} B. Barbara, C. B\'ecle, J.-L. Feron, R. Lemaire, and R. Pauthenet, C.R. Acad. Sci. Paris \textbf{267}B, 244 (1968).

\bibitem{Barbara71} B. Barbara, C. B\'ecle, R. Lemaire, and D. Paccard, J. Physique \textbf{32}, C1-299 (1971).

\bibitem{Ishikawa76} Y.\ Ishikawa, K.\ Tajima, D.\ Bloch, and M.\ Roth,
Solid State Comm.\ {\textbf{19}}, 525 (1976).

\bibitem{Wilkinson76} C.\ Wilkinson, F.\ Sinclair, and J.\ B.\ Forsyth, 5th Conf.\ on Solid Compounds of Transition
Elements. Extended Abstracts, Uppsala 1976, p.\ 158.

\bibitem{Beille83} J. Beille, J. Voiron, M. Roth, 
% "Long period helimagnetism in the cubic-B20 FexCo1-xSi and CoxMn1-xSi alloys"
Sol. State Comm. \textbf{47}, 399  (1983).


\bibitem{Lebech89}
%FeGe 
B.\ Lebech, J.\ Bernhard, and T.\ Freltoft, J.\ Phys.: Condens.\ Matter {\textbf{1}}, 6105 (1989).
% Magnetic structures of cubic FeGe
% studied by small-angle neutron scattering


\bibitem{FranusMuir84} E. Franus-Muir, M. L. Plumer, and E. Fawcett, J. Phys. C \textbf{17}, 1107 (1984).
% Magnetostriction in the spin-density-wave phase % of MnSi

\bibitem{PlumerWalker82} M. L. Plumer and M. B. Walker, J. Phys. C \textbf{15}, 7181 (1982).
% Magnetoelastic effects in the spin-density-wave phase of MnSi


\bibitem{Bloch75} D. Bloch, J. Voiron, V. Jaccarino, and J. H. Wernick, Phys. Lett. A \textbf{51}, 259 (1975).
% The high-field-heigh pressure magnetic properties of MnSi

\bibitem{Grigoriev05} S. V. Grigoriev, S. V. Maleyev, A. I. Okorokov, Y. O. Chetverikov, R. Georgii, P. B\"oni, D. Lamago, H. Eckerlebe, and K. Pranzas, Phys. Rev. B \textbf{72}, 134420 (2005).

\bibitem{Stishov08}  S. M. Stishov, A. E. Petrova, S. Khasanov \textit{et al.},
% A. E. Petrova, S. Khasanov, G. K. Panova
J. Phys.: Condens. Matter. \textbf{20}, 235222 (2008).

\end{thebibliography}
\end{document}